\documentclass[aps,twocolumn,showpacs,amsmath,amssymb,superscriptaddress,longbibliography,pra]{revtex4-1}
\usepackage{physics}
\usepackage{graphicx}
\usepackage{epsf}
\usepackage{graphicx}
\usepackage{dcolumn}
\usepackage{amsmath}
\usepackage{cases}
\usepackage{color}
\usepackage[colorlinks, linkcolor=blue, citecolor=red]{hyperref}
\usepackage{soul}
\usepackage[capitalise]{cleveref}
\usepackage[dvipsnames]{xcolor}
\makeatletter
\let\old@makecaption=\@makecaption
\usepackage{subcaption}
\let\@makecaption=\old@makecaption
\makeatother
\usepackage{blindtext}
\usepackage{ulem}
\usepackage{siunitx}

\newcommand{\bsigma}{\boldsymbol{\sigma}}

\renewcommand{\b}[1]{{\boldsymbol{#1}}}

\newcommand{\sgn}{\mathop{\mathrm{sgn}}}
\newcommand{\rcol} {\textcolor{red}}
\newcommand{\bcol} {\textcolor{blue}}
\newcommand{\gcol} {\textcolor{ForestGreen}}

\newcommand{\nn}{\nonumber}
\newcommand{\half}{{\textstyle{\frac{1}{2}}}}

\usepackage{cases}

\begin{document}

\title{Majorana-Weyl cones in ferroelectric superconductors}
\author{Hennadii Yerzhakov}
\affiliation{Department of Physics, Bar-Ilan University, Ramat Gan, Israel}
\author{Roni Ilan}
\affiliation{Department of Physics, Tel Aviv University, Tel Aviv, Israel}
\author{Efrat Shimshoni}
\affiliation{Department of Physics, Bar-Ilan University, Ramat Gan, Israel}
\author{Jonathan Ruhman}
\affiliation{Department of Physics, Bar-Ilan University, Ramat Gan, Israel}

\date\today

\begin{abstract}
Topological superconductors are predicted to exhibit outstanding phenomena, including non-abelian anyon excitations, heat-carrying edge states, and topological nodes in the Bogoliubov spectra. Nonetheless, and despite major experimental efforts, we are still lacking unambiguous signatures of such exotic phenomena.  
In this context, the recent discovery of coexisting superconductivity and ferroelectricity in lightly doped and ultra clean SrTiO$_3$ opens new opportunities. 
Indeed, a promising route to engineer topological superconductivity is the combination of strong spin-orbit coupling and inversion-symmetry breaking. 
Here we study a three-dimensional parabolic band minimum with Rashba spin-orbit coupling, whose axis is aligned by the direction of a ferroelectric moment. 
We show that all of the aforementioned phenomena naturally emerge in this model when a magnetic field is applied. 
Above a critical Zeeman field, Majorana-Weyl cones emerge regardless of the electronic density. 
These cones manifest themselves as Majorana arcs states appearing on surfaces and tetragonal domain walls. Rotating the magnetic field with respect to the direction of the ferroelectric moment tilts the Majorana-Weyl cones, eventually driving them into the type-II state with Bogoliubov Fermi surfaces. 
We then consider the consequences of the orbital magnetic field. First, the  single vortex is found to be surrounded by a topological halo, and is characterized by two Majorana zero modes: One localized in the vortex core and the other on the boundary of the topological halo. 
Based on a semiclassical argument we show that upon increasing the field above a critical value the halos overlap and eventually percolate through the system, causing a bulk topological transition that always precedes the normal state. Finally, we propose concrete experiments to test our predictions.
\end{abstract}

\maketitle

\section{Introduction}
\label{sec: Intro}
Finding robust experimental realizations of topological superconductivity is an important goal, both for fundamental research of topological matter and for possible applications to quantum technology~\cite{alicea2012new,ando2015topological,lutchyn2018majorana}. However, materials which naturally host such exotic ground states are scarce. Moreover, measuring non equivocal signatures of topological superconductivity is   an outstanding experimental challenge~\cite{mackenzie2017even,Yu2021,frolov2022believe}, because such signatures are often obscured by imperfections in the sample or probe. Most candidate materials also realize low-dimensional topological superconducting states.
Thus, new candidate bulk superconductors might help overcome such challenges.

Over a decade ago Fu and Kane have shown how strong spin-orbit coupling combined with the obstruction of time-reversal symmetry on the surface of a topological insulator converts proximity $s$-wave superconductivity to a topological state~\cite{fu2008superconducting}. Indeed, the combination of spin-orbit coupling, the lack of an inversion center and the breaking of time reversal symmetry are key ingredients in a variety of exotic theoretical predictions and phenomena, including Majorana zero modes~\cite{Sau2010,Lutchyn2011,Oreg2011,Potter2012}, the Fulde–Ferrell–Larkin–Ovchinnikov (FFLO) state~\cite{agterberg2003novel,Dimitrova2007,Michaeli2012,loder2015route}, Majorana-Weyl cones~\cite{Sato2009,Sato2010,Gong2011,Jiang2011,Seo2012,Seo2013}, 
and Ising superconductivity~\cite{xi2016ising,hsu2017topological,mockli2018robust,wickramaratne2020ising}.   

The coexistence of  superconductivity and ferroelectricity  in low-density systems~\cite{Rischau2017,fei2018ferroelectric,russell2019ferroelectric,tomioka2022superconductivity,Scheerer2020,Tuvia2020} opens new opportunities in this context. A ferroelectric crystal breaks inversion symmetry spontaneously and therefore can be easily manipulated. Moreover, such systems are often close to their ferroelectric transition, where the dielectric constant is huge~\cite{WEAVER1959274,Muller1979}. As a consequence, the influence of disorder is dramatically suppressed~\cite{ambwani2016defects,Collignon2019}. These properties make low-density superconductors close to a ferroelectric quantum critical point prime candidates for engineering unconventional superconducting states.

The paradigmatic example of such polar superconductors is lightly doped SrTiO$_3$ (STO)~\cite{Collignon2019,Gastiasoro2020b}. In its natural form however, STO is paraelectric~\cite{WEAVER1959274,Muller1979,rowley2014ferroelectric}. By doping it with Ca or Ba~\cite{Collignon2019,tomioka2022superconductivity}, substituting $^{16}$O with  $^{18}$O~\cite{Stucky2016} or by applying  epitaxial strain~\cite{salmani2020order} one can drive STO to the polar phase, where inversion is spontaneously broken. It has been shown that low-density superconductivity exists in the ferroelectric phase~\cite{Stucky2016,sakai2016critical,Rischau2017,Herrera_2019,engelmayer2019ferroelectric,wang2019charge,enderlein2020superconductivity,Salmani2021} and is even enhanced~\cite{ahadi2019enhancing,tomioka2022superconductivity}.

Motivated by the physics in ferroelectric STO, we revisit the problem of a Rashba spin-orbit coupled superconductor subject to a magnetic field~\cite{Kanasugi2018,Kanasugi2019}, where we focus on the case of three spatial dimensions. 
Rashba spin-orbit coupling originates from the combination of inversion breaking by the ferroelectric moment and atomic spin-orbit coupling~\cite{PETERSEN200049,khalsa2012theory}. Therefore, the axis of the Rashba spin-orbit coupling can vary in space and may also be externally manipulated. 

\begin{figure*}[!htbp]
    \centering
    \begin{subfigure}[t]{0.29\textwidth}
    \centering
    \includegraphics[width=\linewidth]{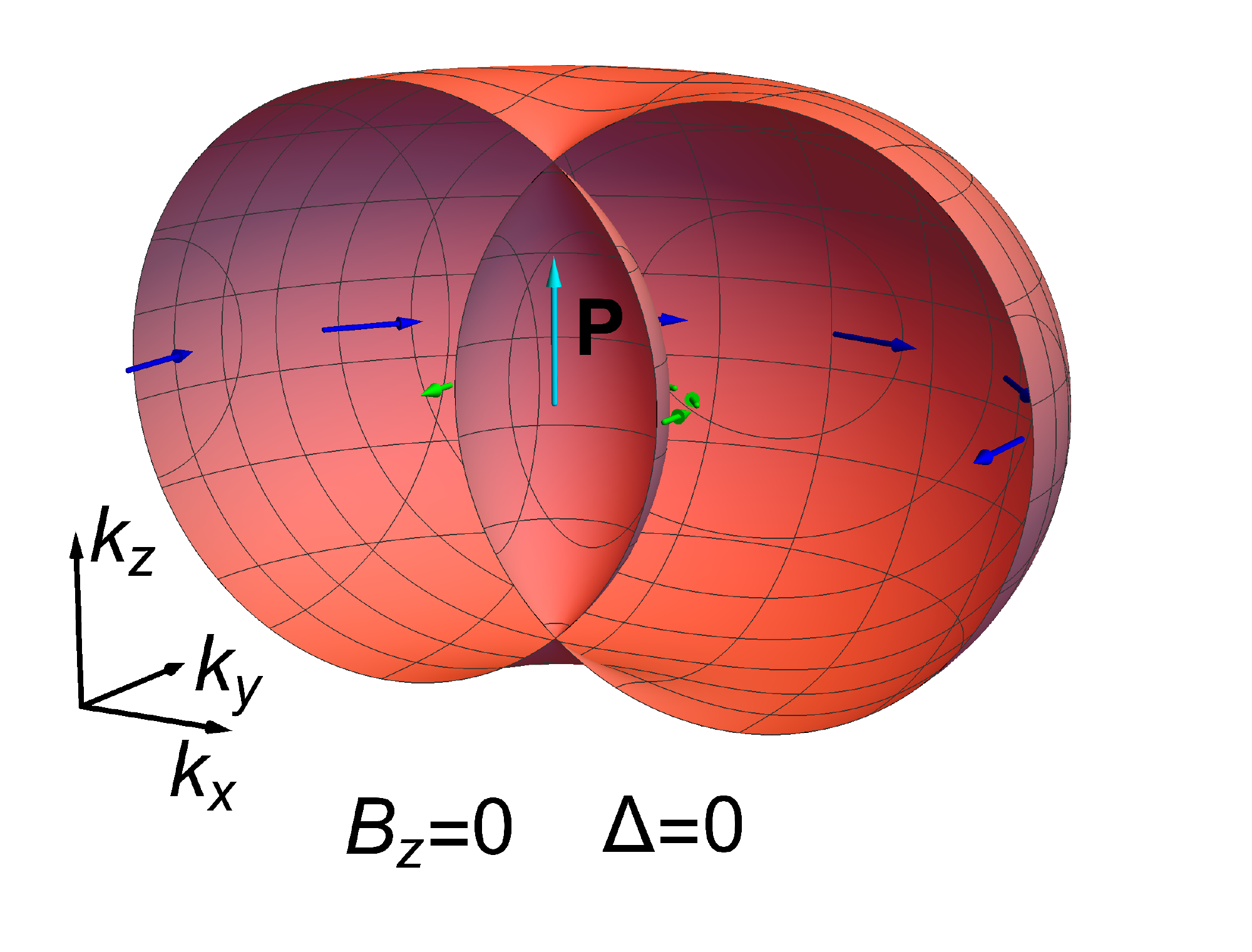}
    \caption{}
    \label{fig: FS1}
    \end{subfigure}
    \begin{subfigure}[t]{0.29\linewidth}
    \centering
    \includegraphics[width=\textwidth]{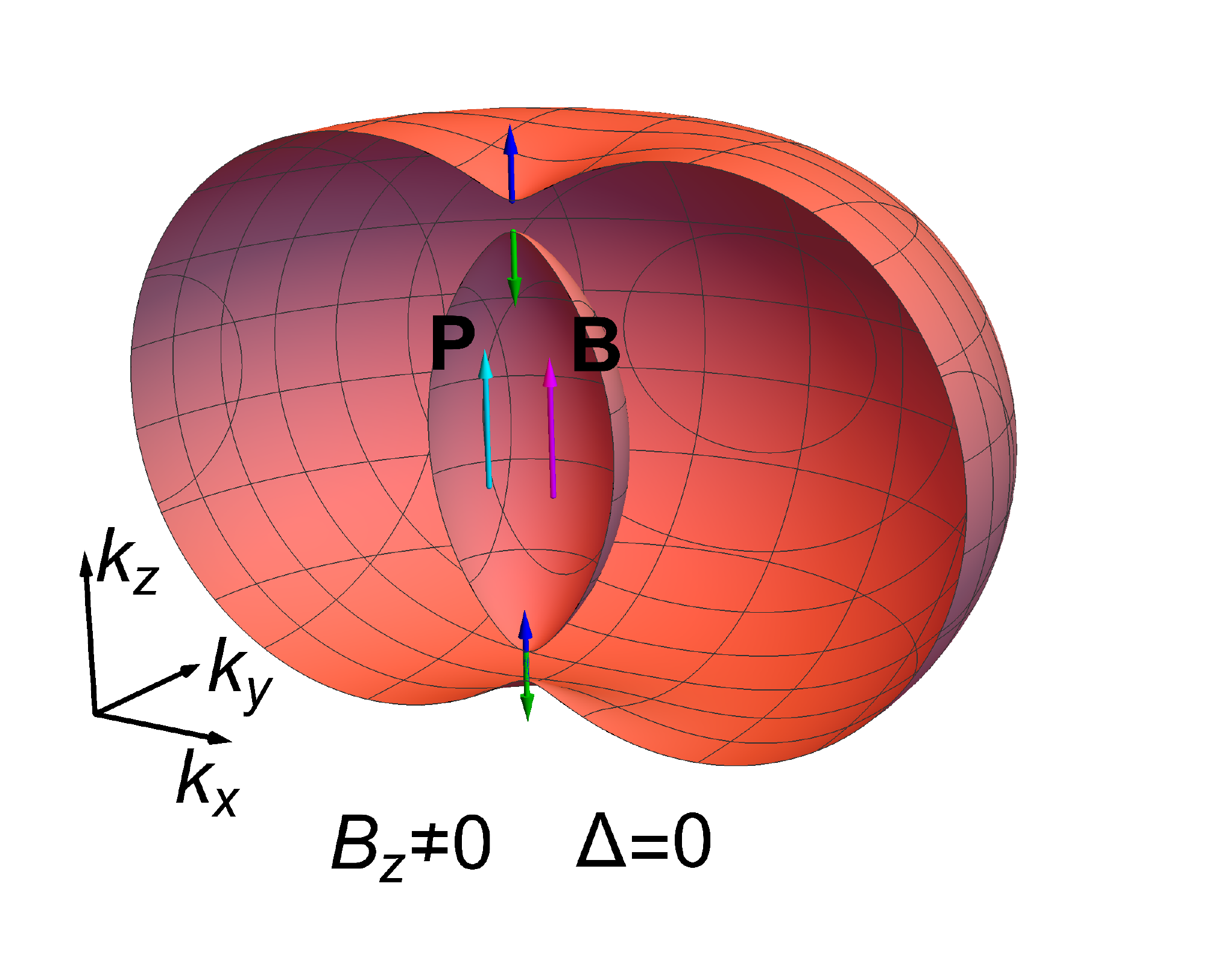}
    \caption{}
    \label{fig: FS2}
    \end{subfigure}
    \begin{subfigure}[t]{0.1\textwidth}
    \centering
    \includegraphics[height=2.5\linewidth]{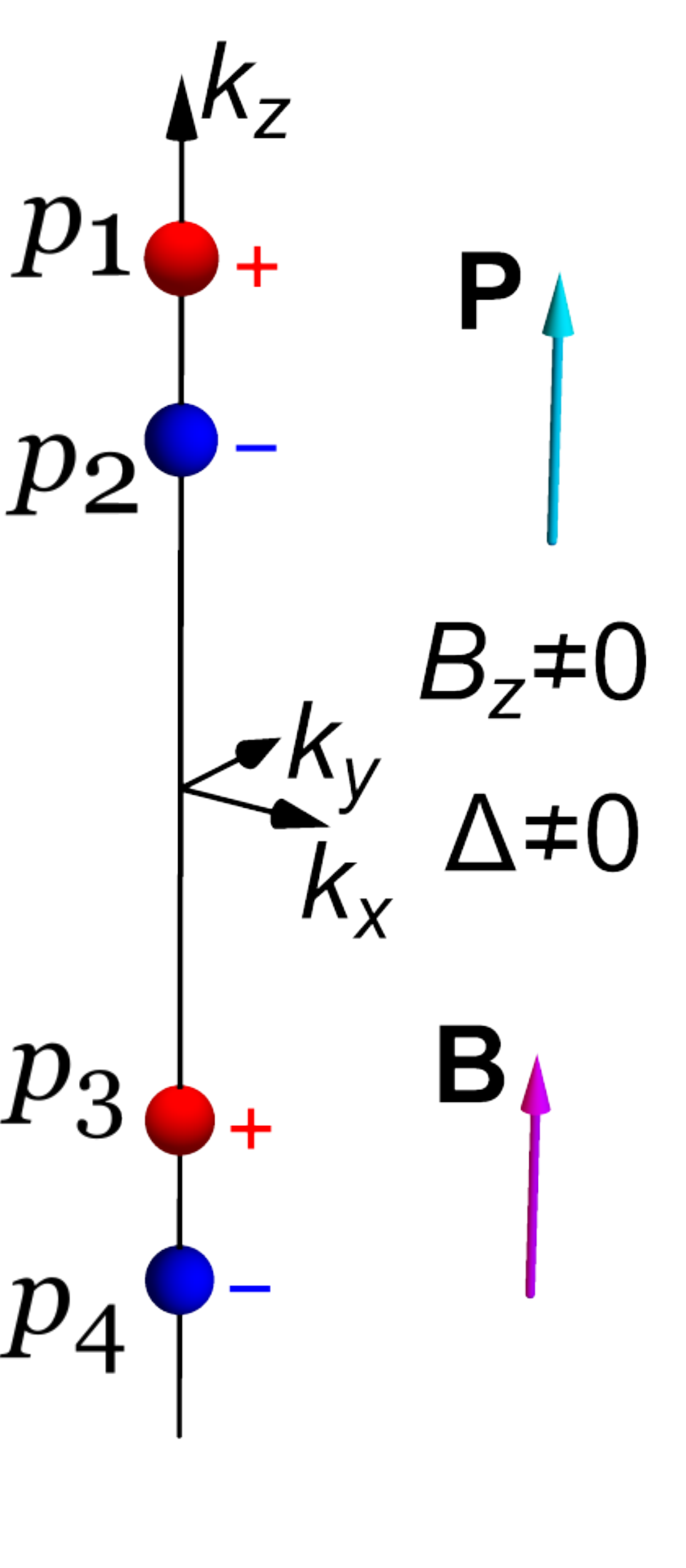}
    \caption{}
    \label{fig: FS3}
    \end{subfigure}
    \begin{subfigure}[t]{0.21\textwidth}
    \centering
    \includegraphics[width=\linewidth]{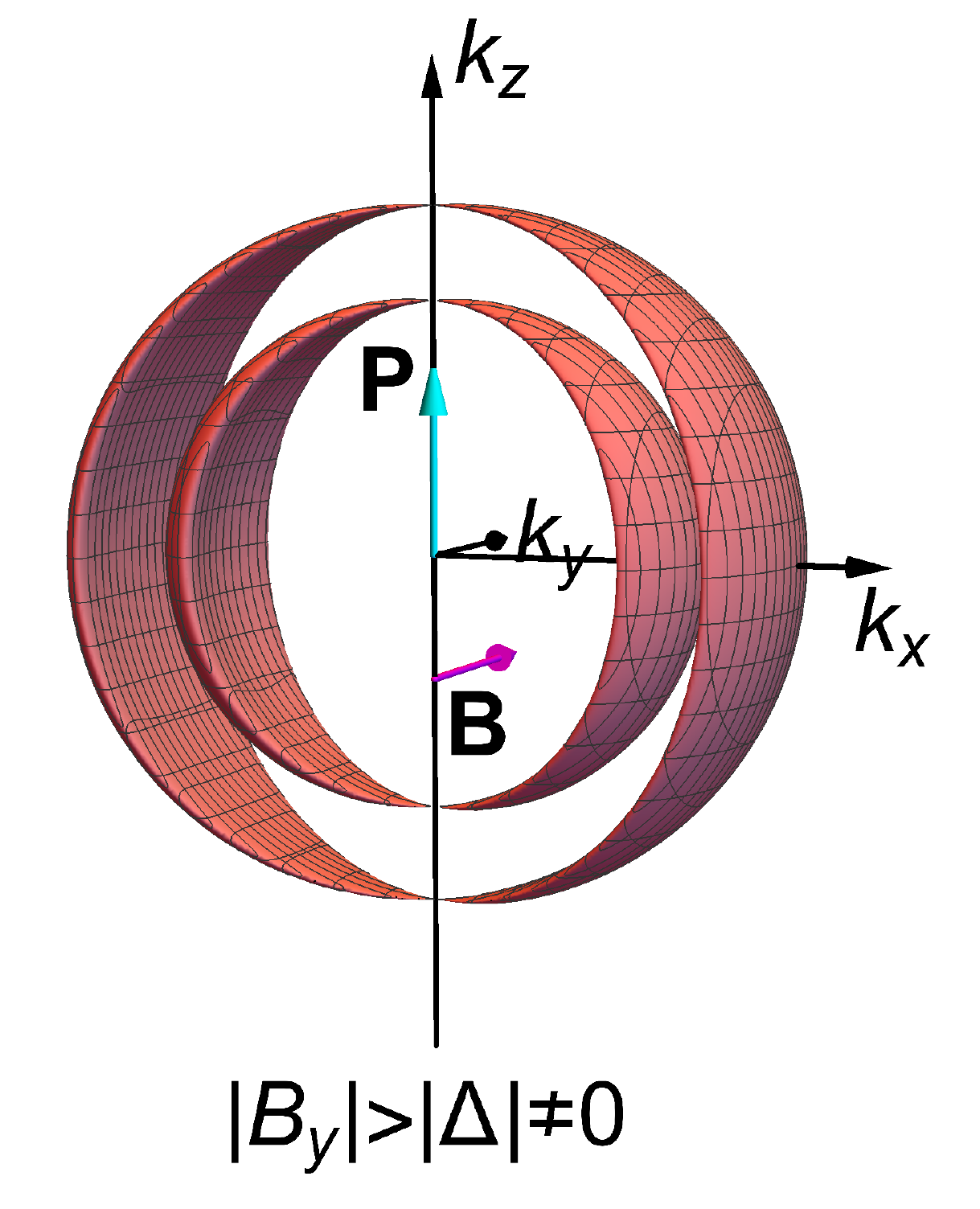}
    \caption{}
    \label{fig: FS4}
    \end{subfigure}
    \caption{(a) The Fermi surface of the free Rashba gas (in the ferroelectric phase). Blue and green arrows denote spin texture of the outer and inner sheet of the FS, respectively.  (b) In the presence of non-zero magnetic field parallel to the polarization, the Fermi surface develops a gap at the points of overlap of the two sheets of the Fermi surface of the Rashba gas. Blue and green arrows denote the direction of the spins corresponding to the momenta on the FS lying on $k_z$-axis. (c) Depairing effect of the Zeeman field is maximal along $k_z$, and sufficiently strong magnetic field destroys superconductivity locally in momentum space, making the spectrum gapless at the specific momenta along $k_z$ - the Weyl points denoted by the red and blue spheres, which colors signify positive and negative chiralities, respectively. (d) Bogoliubov Fermi surfaces appear when the magnetic field is perpendicular to the polarization direction. 
    }
\end{figure*}

In the absence of superconductivity and magnetic fields, the Fermi surfaces are spin split everywhere in momentum except for two pinching points, which lie along the axis of the polar vector (Fig.~\ref{fig: FS1}). Consequently, in the superconducting state, pair breaking is strongest at the vicinity of these points. When the magnetic field exceeds a critical threshold, the gap closes along this polar axis causing four Majorana-Weyl points to emerge, accompanied by surface Majorana arcs. We then show that the Majorana-Weyl cones can be tilted by tuning the angle between the polar moment and field, such that the  superconductor becomes type-II-Weyl with Bogoliubov Fermi surfaces~\cite{agterberg2017bogoliubov,venderbos2018pairing} above a certain critical angle. We also study the Fermi arcs forming on domain walls between different polarization directions. We find that chiral surface states do not appear for all angles of the magnetic field.  

Finally, we turn to the more realistic scenario, where the field is non-homogeneous and penetrates the sample through line vortices. We first study the single vortex problem, where we show that the magnetic field always exceeds the critical threshold close enough to the center, forming a topological halo surrounding the vortex. We show that each such vortex has a single zero mode in its core with a counterpart at the boundary of the halo, yielding corresponding signatures in the tunneling density of states. 
Then, 
as the magnetic field is increased towards $H_{c2}$ the density of vortices increases and the halos begin to overlap, forming larger topological regions. As a consequence we predict that the trivial-superconducting and normal states are always separated by a {putative} topological phase in any polar superconductor. The topological phase is characterized by percolation of the halos, akin to a transition between integer quantum Hall states.  

The rest of this paper is structured as follows. In~\cref{sec: Model}, we describe the model and show in the mean-field picture, neglecting the orbital effects of the magnetic field, that Majorana-Weyl superconductivity develops when the magnetic field exceeds certain threshold. In~\cref{sec: Domain interfaces}, we discuss the Fermi arcs on surfaces and interfaces between the ferroelectric domains. In~\cref{sec: Vortex}, we consider a more realistic model taking into account orbital effects of the magnetic field. We show that in addition to a Majorana string located in the core, an isolated vortex is surrounded by a chiral Majorana mode with the wavefunction peaked at a finite distance from the core. {Based on semiclassical considerations, we propose}
that with the increase of the magnetic field towards $H_{c2}$, there is always a percolation-type phase transition to a bulk Majorana-Weyl superconductivity, at which chiral modes going around each vortex overlap. In this section we also calculate contribution from the Majorana modes to the tunneling density of states. Finally, we give our conclusions with emphasis on experimental consequences caused by the physics considered in~\cref{sec: Conclusions}. Throughout the paper we work in units in which $\hbar=k_B=1$.


\section{Majorana-Weyl superconductivity in the presence of a Zeeman field}
\label{sec: Model}

We now describe the microscopic model.
We start with the coupling between the optical phonon displacement $\b{P}$
and the conduction electrons~\cite{KoziiFu2015,ruhman2016superconductivity,Gastiasoro2020,kumar2021spinphonon,Gastiasoro2021}
\begin{align}
\hat{H}_{el-ph} = \sum_{\b{k},\b{q}} \bar{\lambda} \psi_{\b{k}+\b{q/2}}^\dagger (\b{k}\times \b{\sigma}) \psi_{\b{k}-\b{q/2}} \cdot \b{P}_{\b{q}},
\label{el_ph}
\end{align}
where
$\psi_\b{k}$ is an annihilation operator for the electron with momentum $\b{k}$ and $\bar{\lambda}$ is a coupling. This term has its microscopic origin as a consequence of combined effect of spin-orbit coupling and interorbital hybridization allowed by inversion breaking~\cite{PETERSEN200049}.

In the ferroelectric phase
, the displacement field $\b{P}$ develops a non-zero expectation value. 
We emphasize however, that this expectation value does not imply the presence of long-ranged  electric fields, which are always screened by the itinerant electrons beyond the Thomas-Fermi length scale. The term ``ferroelectricity" is commonly used in the literature to describe the polar state even when it is metallic. 
 However, in this case the ``ferroelectric'' phase transition refers to a structural transition, where  inversion symmetry is broken. As a result, the coupling Eq. (\ref{el_ph}) leads to the celebrated Rashba spin orbit coupling
$
\hat{H}_{SOC} = \sum_\b{k} \lambda \,\psi_{\b{k}}^\dagger  [(\b{k}\times \b{\sigma})\cdot \hat{n}] \,\psi_{\b{k}}
$, where $\hat n$ is a unit vector parallel to the ferroelectric order parameter $\langle \b P \rangle$ and $\lambda = \abs{\langle \b P \rangle} \bar{\lambda}$.


Additionally, we consider a Zeeman coupling to an external magnetic field $\b{B}$, and neglect its orbital effects \textit{pro tem} \footnote{this is justified in the regime $\frac{g \mu_B B}{\mu} \ll 1$ (where $\mu$ is the chemical potential, $\mu_B$ is the Bohr magneton and $g$ is the Lande' g-factor).}. Without loss of generality we align the $z$-axis with the local polarization $\langle \b P \rangle$ (hence $\hat{\b n} = \hat {\b z}$), and obtain the dispersion Hamiltonian
\begin{align}
H({\b{k}})=\psi_\b{k}^\dagger \epsilon_\b{k} \psi_\b{k} + \lambda \psi_{\b{k}}^\dagger (\b{k}\times \b{\sigma})_z \psi_\b{k} - \frac{ g \mu_B}{2} \b{B}\cdot \bsigma
\label{Hk0}
\end{align}
where $\b{\sigma}=(\sigma_x,\sigma_y,\sigma_z)$ is a vector of Pauli matrices in spin space and we have assumed the dispersion $\epsilon_\b{k}=\frac{ k^2}{2 m}-\mu$ is spherically symmetric. In the following, we work in units in which $ g \mu_B/2=1$. 

We next add an attractive interaction between electrons, which causes a Cooper instability at low temperature. 
For simplicity we restrict ourselves to $s$-wave superconductivity~\footnote{
In general, inversion breaking leads to a state of mixed singlet-triplet superconducting states~\cite{Gorkov2001}.}, which is also reported in the experiments on paraelectric STO~\cite{Collignon2019}.

Finally, writing the Hamiltonian in BdG form we obtain 
\begin{align} \label{eq: BDG hamiltonian}
    \hat{H} &= \half\sum_{\b k} \Psi^\dagger_{\b k} H_{BdG}(\b{k}) \Psi_{\b k} \\ \nn &= \half\sum_{\b k} \Psi^\dagger_{\b k} \begin{pmatrix}
        H(\b{k})     & \b{\Delta}  \\
        \b{\Delta}^\dagger    & -H^*(-\b{k})
        \end{pmatrix} \Psi_{\b k},
\end{align}
where $\Psi^\dagger_{\b k} = (\psi^\dagger_{\b k},\psi_{-\b k}^T)$ is the Nambu spinor, $\b{\Delta} = i \sigma_y \Delta$ in the $s$-wave BCS channel, and we choose a gauge in which $\Delta$ is real.
The BdG Hamiltonian above enjoys a particle-hole symmetry, implemented by  $\mathcal{P}=\tau_x \mathcal{ C}$, where $\tau_j,\  j=x,y,z$ are Pauli matrices in the particle-hole space and $\mathcal{ C}$ is the complex conjugation operator. Namely, the Hamiltonian obeys $\mathcal{P} H_{BdG}(\b{k}) \mathcal{P}^\dagger = -H_{BdG}(-\b{k})$. Additionally, when $\b B \parallel \b P$ the Hamiltonian has rotational symmetry about the axis {parallel to the polarization}, where the rotation includes both spatial and spin rotation. In the presence of higher order terms due to the lattice, the continuous rotational symmetry is reduced to discrete four-fold rotations about the polarization axis. 

The energy dispersion is determined from the solutions of a quartic equation [see~\cref{eq: Spectrum eq general}], which for a magnetic field parallel to the polarization yields

\begin{align}
\label{eq: Spectrum eq}
    E_{\b{k}}^2 = \epsilon_\b{k}^2+B^2 +\lambda^2 k_{\perp}^2+\Delta^2 \pm 2 \sqrt{\epsilon_\b{k}^2(B^2 +\lambda^2 k_{\perp}^2) + \Delta^2 B^2},
\end{align}
where $\b{k}_{\perp}=(k_x,k_y)$ denotes the projection of the momentum onto the $xy$-plane. 

The 3D Fermi surfaces of the free Rashba gas described by \cref{Hk0} have the shape obtained by rotating two displaced circles around the axis connecting their crossing points 
(see~\cref{fig: FS1}). Consequently, the crossings form pinching points along the $k_z$ axis, where two Fermi sheets with opposite helicities touch. Upon turning on a magnetic field in the $z$-direction, the two sheets separate, and the spins at these points becomes co-linear with the field direction. 
Thus, the depairing effect of the magnetic field in the superconducting phase is expected to be strongest at these pinching points. 

Indeed, a sufficiently strong magnetic field closes the gap at the pinching points on the $k_z$ axis. From~\cref{eq: Spectrum eq}, we see that the gap closes for $B^2>\Delta^2$ at momenta $\b{p}=(0,0,p_z)$, where
\begin{align}
B^2=\Delta^2 + \epsilon_{p_z}^2\; .
\end{align}
This equation is satisfied at four points
\begin{align}
{p}_{j}=\pm\sqrt{2m (\mu \pm \sqrt{B^2 - \Delta^2})}\; ,
\end{align}
with $j=1,\ldots,4$ labeled in descending order along the $k_z$-axis (see~\cref{fig: FS3}). 
The closing of the gap at these momenta can be viewed as a topological phase transition in the two-dimensional Hamiltonian $H_{BdG}(p_x,p_y,p_z)$, where $p_z$ is a tuning parameter.  Indeed, for $p_{2}<p_z<p_1$ and $p_4<p_z<p_3$ the two dimensional Bloch bands have non-zero Chern numbers $\pm 1$ (of equal sign), signaling that the Weyl nodes are monopoles of Berry charge.
It is worth noting that in the low density limit there are only two Weyl nodes $p_1$ and $p_4$, in accord with the finding of previous literature~\cite{Gong2011,Jiang2011,Seo2012,Seo2013}. 

Rotation of $\b{B}$ with respect to the ferroelectric moment $\b{P}$ profoundly changes the quasiparticle
spectrum. Due to the rotational symmetry, the dispersion is symmetric for both $\b{k} \rightarrow -\b{k}$ and $E \rightarrow -E$ separately, when $\b B \parallel \b P$. However, in the presence of a perpendicular component, the spectrum is invariant only under the combined action of these two operations.
This means that when the angle is large enough, the Weyl cones over tilt and become type II~\cite{Soluyanov2015,volovik2018}, which is accompanied by the development of the Fermi surface of zero-energy Bogoliubov quasiparticles~\cite{agterberg2017bogoliubov,venderbos2018pairing} (see~\cref{fig: FS4}). This mechanism is analogous to the one described in Ref.~\cite{Yuan2018} for the surface of a 3D topological insulator and 2DEG Rashba spin-orbit gases with the proximity induced superconductivity and applied in-plane magnetic field. For more details see~\cref{app: Appendix_Bogoluibov FS}. 

To make these observations more concrete, we derive the low-energy effective Hamiltonian in the vicinity of the Weyl nodes by projecting to the low-energy subspace. 
This yields  the $2\cross2$ Hamiltonian 
\begin{align}
\label{eq: 2D Weyl Hamiltonian}
    H_{eff}(\b k) = C_0^j k_j  \sigma_0 + A_i^j k_j \sigma_i,
\end{align}
where
\begin{align} \label{eq: Weyl Hamiltonian coeffs}
    C_0^x & = - \frac{\lambda}{B} \frac{\epsilon_\b{p}}{B} B_y,\ C_0^y = \frac{\lambda}{B} \frac{\epsilon_\b{p}}{B} B_x,\ C_0^z=0, \\ \nn
    A_x^x & = - \lambda \Delta \frac{ B_x B_y}{B^2(B+B_z)},\ A_x^y=-\lambda \Delta \frac{B^2+B_z B - B_x^2}{B^2(B+B_z)}, \\ \nn
    A_y^x & =  \lambda \Delta \frac{B^2+B_z B - B_y^2}{B^2(B+B_z)},\ A_y^y =  \lambda \Delta \frac{ B_x B_y}{B^2(B+B_z)}), \\ \nn
    A_z^z & = \frac{p_z}{m}  \frac{\epsilon_\b{p}}{B},
\end{align}
and all other components of the matrix $\b{A}$ are equal to zero. 
The chiralities of the Weyl nodes are determined by 
\begin{align}
    \chi = \sgn ( \det \b{A} ) = \sgn ( \lambda^2 \Delta^2 \frac{p_z}{m} \frac{\epsilon_{p_z}}{B} \frac{B_z}{B^3}),
\end{align}
and are controlled by $B_z$, which is the projection of the magnetic field, $\b B$, on the polarization vector $\b P$. 


The $\sigma_0$-term in~\cref{eq: 2D Weyl Hamiltonian}, which is proportional to the components of $\b B$ that are perpendicular to $\b P$, is responsible for tilting the Weyl cones when the magnetic field and polarization are not collinear.
This can be seen by noting the energy spectrum of the Hamiltonian~\cref{eq: 2D Weyl Hamiltonian} 
\begin{align} \label{eq: Weyl cone II dispersion}
    \epsilon(\b{k})=C_0^j k_j \pm \sqrt{\sum_i (A_i^j k_j)^2}\, .
\end{align}
As mentioned above, 
the system can even be driven into a type-II phase, where the cones tilt is so strong they dip below the Fermi energy and form Bogoliubov Fermi surfaces~\cite{agterberg2017bogoliubov}. The condition for Bogoliubov Fermi surfaces to develop is the existence of non-zero $\b{k}$ such that 
$C_0^j {k}_j > \sqrt{\sum_i (A_i^j {k}_j)^2}$. Using the expressions in~\cref{eq: Weyl Hamiltonian coeffs}, we find that this criterion is satisfied when $B_{\perp}^2>\Delta^2$. Close to the cone, the Bogoliubov Fermi surface 
sheet defined by $\epsilon(\b{k})=0$ from~\cref{eq: Weyl cone II dispersion} is a cone with the opening angle in $k_xk_y$-plane $\phi=\pi-2 \arcsin(\frac{\Delta}{B_{\perp}})$. However, inspecting the full Hamiltonian~\cref{eq: BDG hamiltonian} (see~\cref{app: Appendix_Bogoluibov FS}), we find that, in fact, the Bogoliubov Fermi surfaces form the shape of two bananas touching at the Weyl points (see \cref{fig: FS4}).

Before proceeding to the physical consequences of the Weyl nodes, we comment that in our model they appear exactly at zero energy. This is however, not fixed by symmetry, but an artifact of the gap function we chose, which is purely the $A_{1g}$ representation (s-wave). The inversion symmetry breaking renders this representation indistinguishable from $A_{2u}$ ($p_z$, which is triplet). Therefore the gap is in general a mixture of the two, which is characterized by nodes shifted from zero energy, where the sign of the shift for each node depends on the sign of the momentum along $z$. Such a shift will inflate the nodes leading to small Bogoliubov Fermi surfaces (see~\cref{app: Appendix_pz-wave}).

We finally note that the angle between $\b P$ and $\b B$ can be spatially manipulated, for example
across a domain wall separating different ferroelectric domains. This opens a path to control the Weyl nodes, as we discuss in the following section. 

\section{Fermi arcs on surfaces and Domain-walls }
\label{sec: Domain interfaces}

In this section we discuss the Majorana Fermi arcs, which appear on surfaces and domain walls. We first review the well known case of an interface between a single domain and vacuum. We then turn to the case of internal tetragonal domain walls.

\subsection{Majorana arcs on the surface of a single domain}

We first show that Majorana arc states appear on the boundary between a single domain and the vacuum. {Assuming that the ferroelectric moment $\b P$ is tilted with an angle $\theta$ to the interface, we pick a coordinate system such that the $yz$-plane is in the plane of the interface, the $z$-axis aligns with the projection of $\b P$ onto the interface,} and the $x$-axis directs into the domain. The Hamiltonian for the domain  is given by~\cref{eq: BDG hamiltonian} with the replacement $k_x \rightarrow -i \partial_x$, yielding
$H(-i \partial_x, \b{k}_{||})$, where $\b{k}_{||} = (k_y,k_z)$ is a momentum in the plane of interface.
We then seek zero energy eigenstates $\Psi_{\b{k}_{||}}(x) = \left(u_{\b{k}_{||}\uparrow}(x),u_{\b{k}_{||}\downarrow}(x),v_{\b{k}_{||}\uparrow}(x),v_{\b{k}_{||}\downarrow}(x)\right)^T$ satisfying open boundary conditions:

\begin{subnumcases}{}
  H(-i \partial_x, \b{k}_{||}) \Psi_{\b{k}_{||}}(x)  = 0,  \label{eq: vacuum boundary problem 1} \\
  \Psi_{\b{k}_{||}}(0)  =0. \label{eq: vacuum boundary problem 2}
\end{subnumcases}
The Bogoliubov quasiparticles operators are defined as
\begin{align}
    \gamma^\dagger & = \int d \b{r} \sum_{s=\uparrow, \downarrow} \left[ u_{s}(\b{r}) c_s^\dagger(\b{r}) + v_{s}(\b{r}) c_s(\b{r})  \right]  \\ 
    & =\int dx \int d \b{k}_{||} \sum_{s=\uparrow, \downarrow} \left[ u_{s,\b{k}_{||}}(x) c_{\b{k}_{||}s}^\dagger(x) + v_{s,\b{k}_{||}}(x) c_{-\b{k}_{||}s}(x)  \right]. \nn
\end{align}
Thus, the reality condition $\Tilde{\gamma}^\dagger=\Tilde{\gamma}$, where $\Tilde{\gamma}=e^{i \frac{\phi}{2}} \gamma$ with a possibly non-zero phase $\phi$, reads
\begin{align}
\label{eq: reality condition}
    v_{s,\b{k}_{||}}^*(x) = e^{i \phi} u_{s,-\b{k}_{||}}(x).
\end{align}

We now look for the solution of~\cref{eq: vacuum boundary problem 1} in the form $\Psi_{\b{k}_{||}}(x) = \Psi_{0,\b{k}_{||}} e^{-\alpha x}$, where
$\Re(\alpha)>0$ implies decaying solutions as $x \rightarrow \infty$. Plugging this into~\cref{eq: vacuum boundary problem 1}, we obtain the characteristic equation for $\alpha$ [ee~\cref{app: Appendix_Domains}, ~\cref{eq: characteristic for alpha}], the solutions of which for the parallel momentum $\b{k}_{||}$, denoted $\alpha_{\b{k}_{||}}$, obey
$\alpha_{\b{k}_{||}}=\alpha^*_{-\b{k}_{||}}$ in accord with the reality condition~\cref{eq: reality condition}. 
Analysis shows (see~\cref{app: Appendix_Domains}) that for $\abs{B_y}<\Delta$ there are four roots with positive real part. In this case, a general decaying solution for~\cref{eq: vacuum boundary problem 1} is a linear combination of four solutions: $\Psi_{\b{k_{||}}}(x)=\sum_{i=1..4} C_i \Psi_{\b{k_{||}},i}(x)$. Plugging this into the boundary condition Eq.~(\ref{eq: vacuum boundary problem 2}) 
and requiring vanishing of the determinant of the resulting set of the
linear equations with respect to coefficients $C_i$, one obtains $\b{k}_{||}$ for which a non-trivial solution, corresponding to the Majorana-Fermi arc, exists. For $\abs{B_y}>\Delta$, when the Weyl cones overtilt in $x$-direction, we do not find Fermi arcs on the $x=0$ surface.

It is easy to find analytical solution for $B_y=0$. In this case, it is expected that Majorana-Fermi arcs are formed at $k_y=0$. Indeed, in this case~\cref{eq: characteristic for alpha} for $\alpha$ splits into two simpler ones 
\begin{align} \label{eq: characterictic for alpha at By=0}
    &\left(\epsilon_{k_z}-\frac{\alpha^2}{2m} \right)^2 - \lambda^2 (k_z \sin \theta - i \alpha \cos \theta)^2 \\&-B^2+\Delta^2 = 
    - 2 i \eta \lambda \Delta (k_z \sin \theta - i \alpha \cos \theta),\nn
\end{align}
where $\eta=\pm1$, and we find $v_{k_z,s} = \eta u_{k_z,s}$, as required by particle-hole symmetry. Thus, the problem separates into two sectors corresponding to $\eta=\pm1$. 
The number of roots in the right half-plane depends on the sign of the quantity
\begin{align}
\label{eq: rotated topology condition }
    \Xi=\left(\frac{1}{2m} \left(k_z^2+k_z^2 \tan^2 \theta\right) - \mu\right)^2 -B^2 + \Delta^2.
\end{align}
Defining a new (primed) coordinate system, rotated such that its $k_z'$-axis aligns with the ferroelectric moment, $k_z=k_z' \cos \theta$, we see that $\Xi<0$ is just the condition for the momenta $k_z'$ to lie between the two Weyl nodes $\b{p}_1$ and $\b{p}_2$ or $\b{p}_3$ and $\b{p}_4$. Precisely, for $\Xi<0$ ($\Xi>0$), there are three (two) roots in the right half-plane for $\eta=1$, and one (two) roots for $\eta=-1$. The Dirichlet boundary condition~\cref{eq: vacuum boundary problem 2} and the normalization of the wave-function define three conditions to be satisfied. Thus, for $\eta=1$ for momenta on $k_z$-axis lying between the projections of two nearby Weyl nodes, a non-trivial solution corresponding to Majorana-Fermi arc exists, see the dashed lines in~\cref{fig: FermiArcs1}. In~\cref{app: Appendix_Domains}, we show that for non-zero $B_x$ and $\abs{B_y}<\Delta$, the Majorana-Fermi arcs remain to be straight lines connecting the projections of the Weyl nodes.


\subsection{Majorana arcs on domain walls}
In the previous subsection we showed that Majorana zero modes (MZMs) connecting into Fermi arcs appear on the boundary with vacuum.
We now turn to discuss another situation relevant to experiments in STO: Domain walls between different tetragonal domains. 
To understand the nature of such domain walls, we recall that low-temperature STO has spontaneously broken its cubic symmetry into tetragonal structure. In this phase each oxygen octahedra rotates about one of the three cubic axis, clockwise or anticlockwise, alternating from unit cell to unit cell~\cite{cowley1964lattice,Collignon2019}, which is known as  antiferrodistortive (AFD) order. The rotation axis fixes the polarization direction when tuning into the ferroelectric phase. For example, in calcium doped STO, the polarization develops in the $[1,1,0]$ or $[1,\bar{1},0]$ directions~\cite{Bednorz1984,Kleemann1997} if we assume the axis of the AFD rotation is $[0,0,1]$. Without loss of generality we consider this specific case hereafter. 

\begin{figure}[!htbp]
    \centering
    \includegraphics[width=\linewidth]{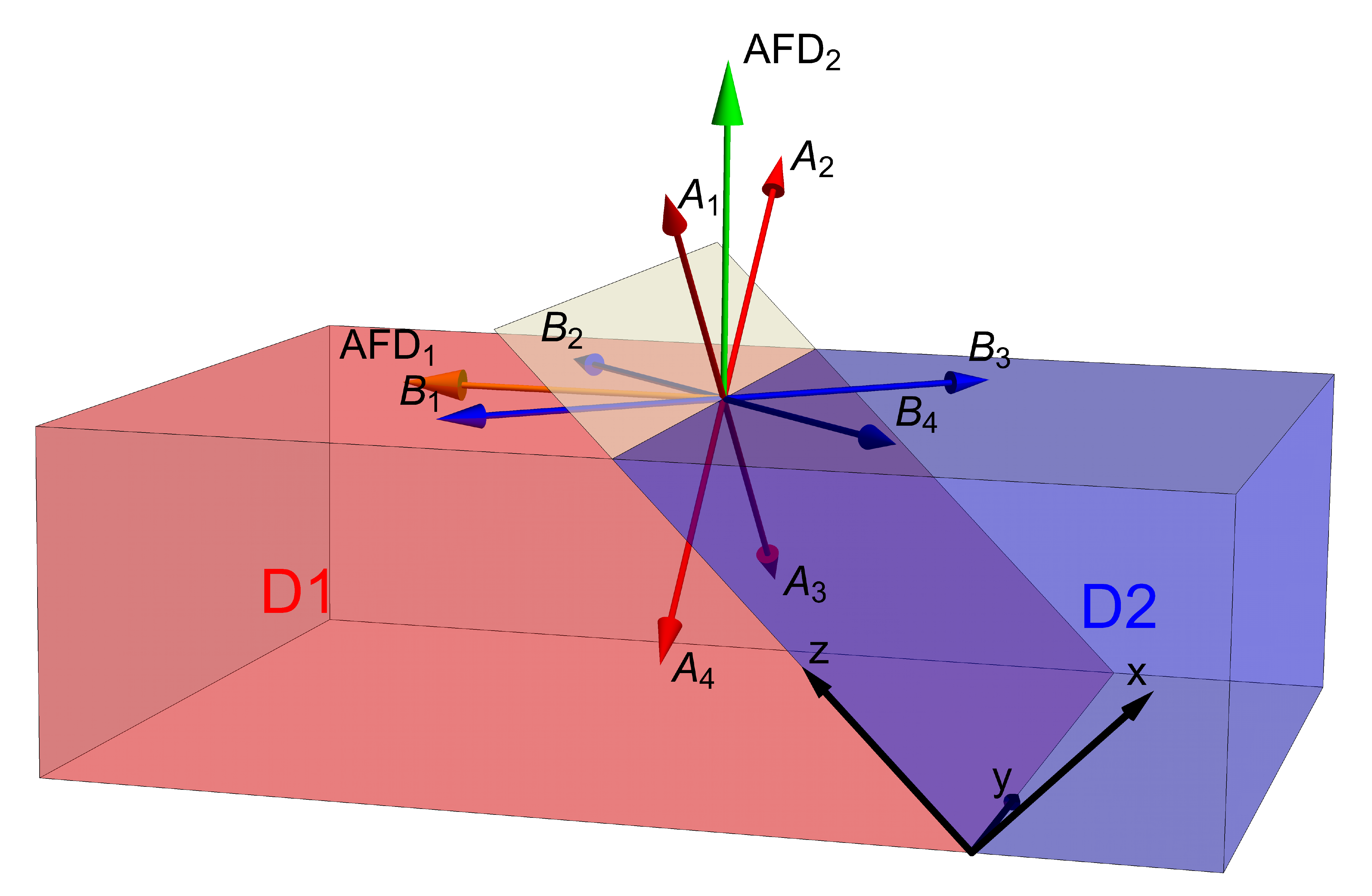}
    \caption{Interface between the two AFD domains, $D_1$ and $D_2$,  with ferroelectric orders. AFD$_{1,2}$ - direction of the AFD distortion in the left (red) and right (blue)) domains, respectively; $A_i, B_i$ - possible directions of the ferroelectric moments in the left (red) and right (blue)) domains, respectively. 
    }
    \label{fig: Domain Setup1}
\end{figure}

The AFD phase is notoriously known to breakout in domains~\cite{kalisky2013locally,honig2013local}, which appear in two types, one endows the system with the reflection symmetry about the wall and the other endows the system with the reflection symmetry about the wall combined with a glide~\cite{Hellberg2019}. The AFD order parameters in neighbouring domains constitute $\pm \frac{\pi}{2}$ angle with each other. In turn, the  ferroelectric polarization in the neighbouring domains will also differ by direction with a relative angle of $\frac{\pi}{3}$ or $-\frac{2\pi}{3}$, see~\cref{fig: Domain Setup1}. 

\begin{figure*}[!htbp]
    \centering
    \begin{subfigure}[t]{0.3\textwidth}
    \centering
    \includegraphics[width=\linewidth]{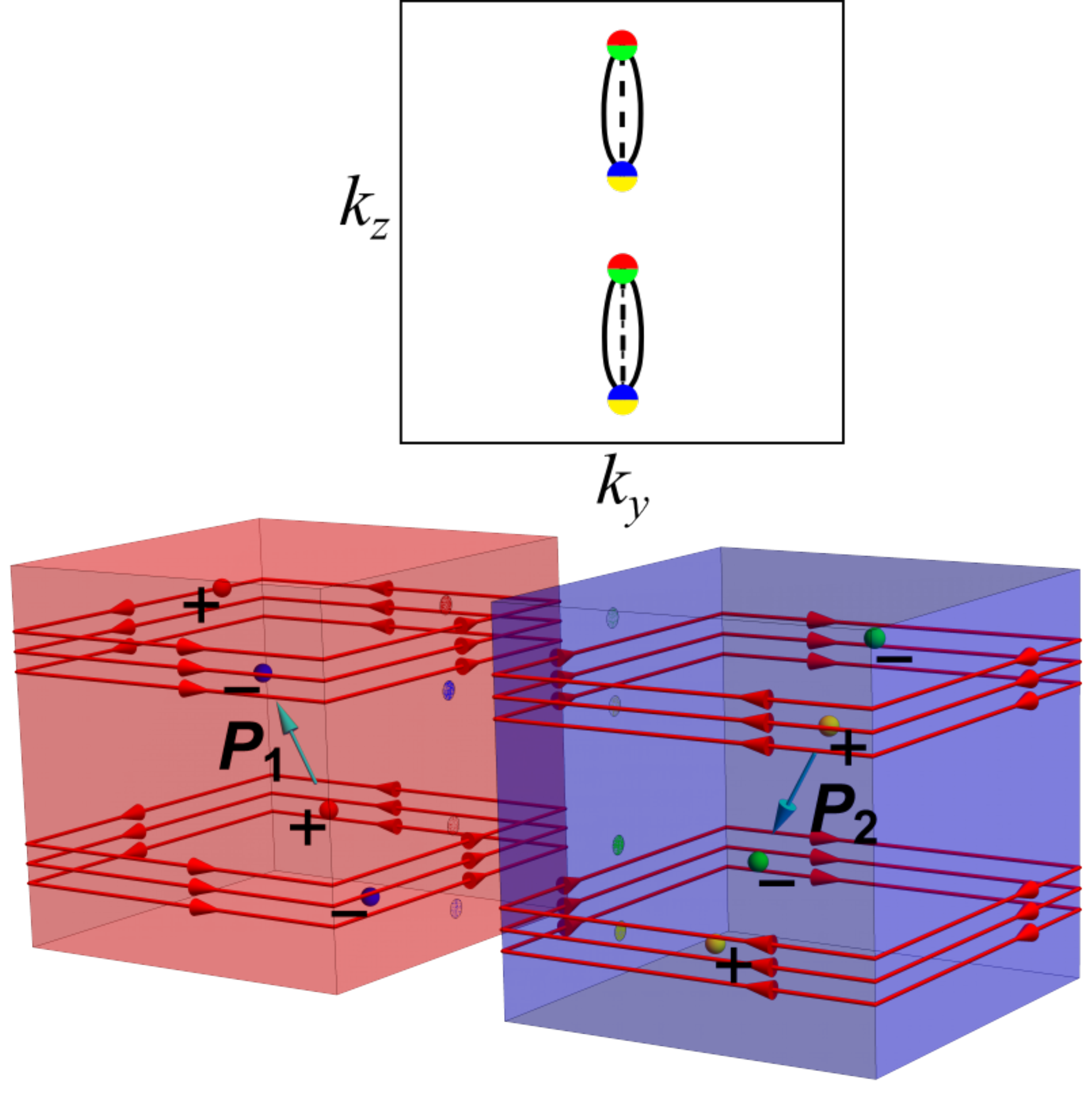}
    \caption{}
    \label{fig: FermiArcs1}
    \end{subfigure}
    \qquad
    \begin{subfigure}[t]{0.3\textwidth}
    \centering
    \includegraphics[width=\linewidth]{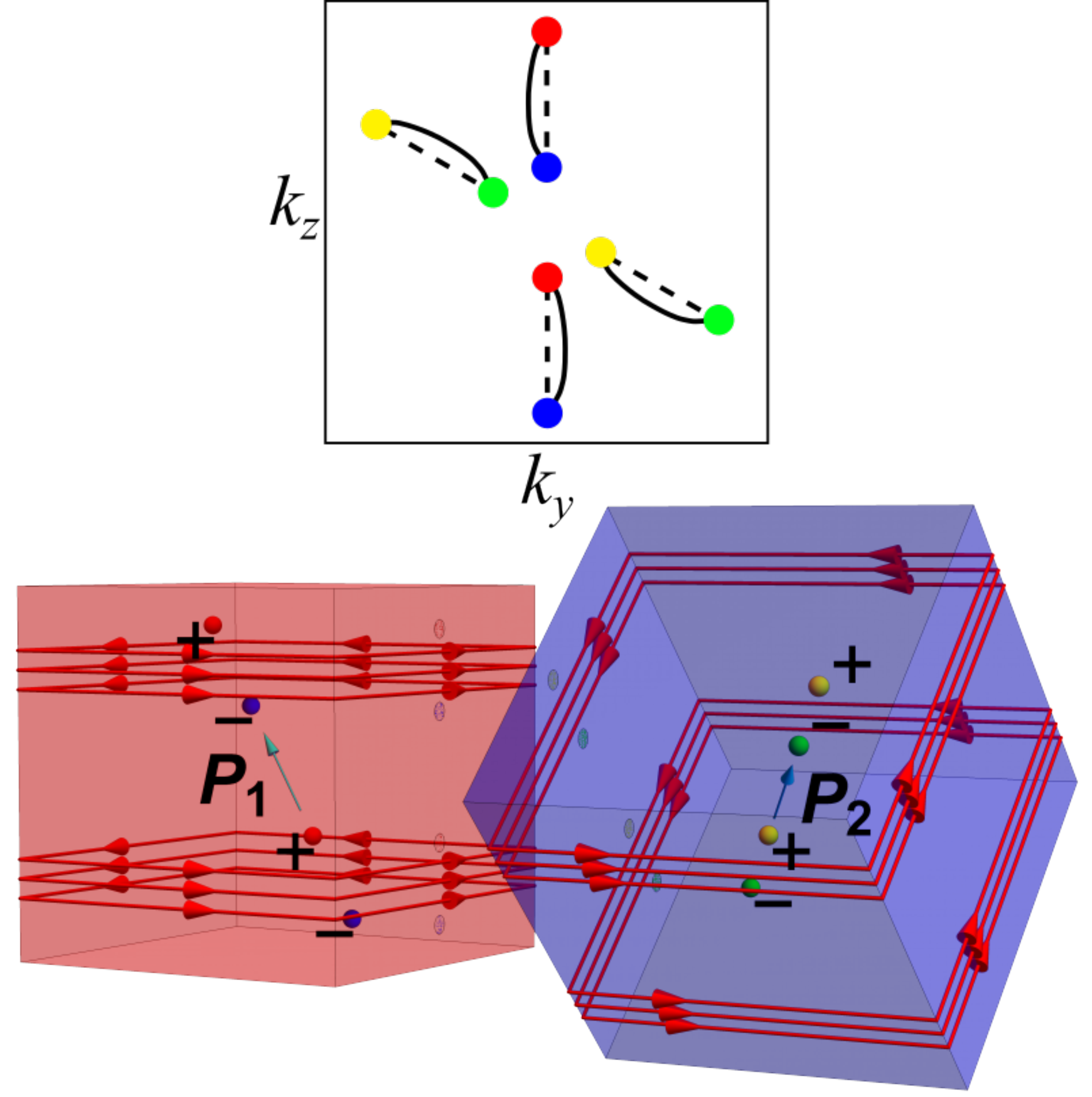}
    \caption{}
    \label{fig: FermiArcs2}
    \end{subfigure}
    \begin{subfigure}[t]{0.3\textwidth}
    \centering
    \includegraphics[width=\linewidth]{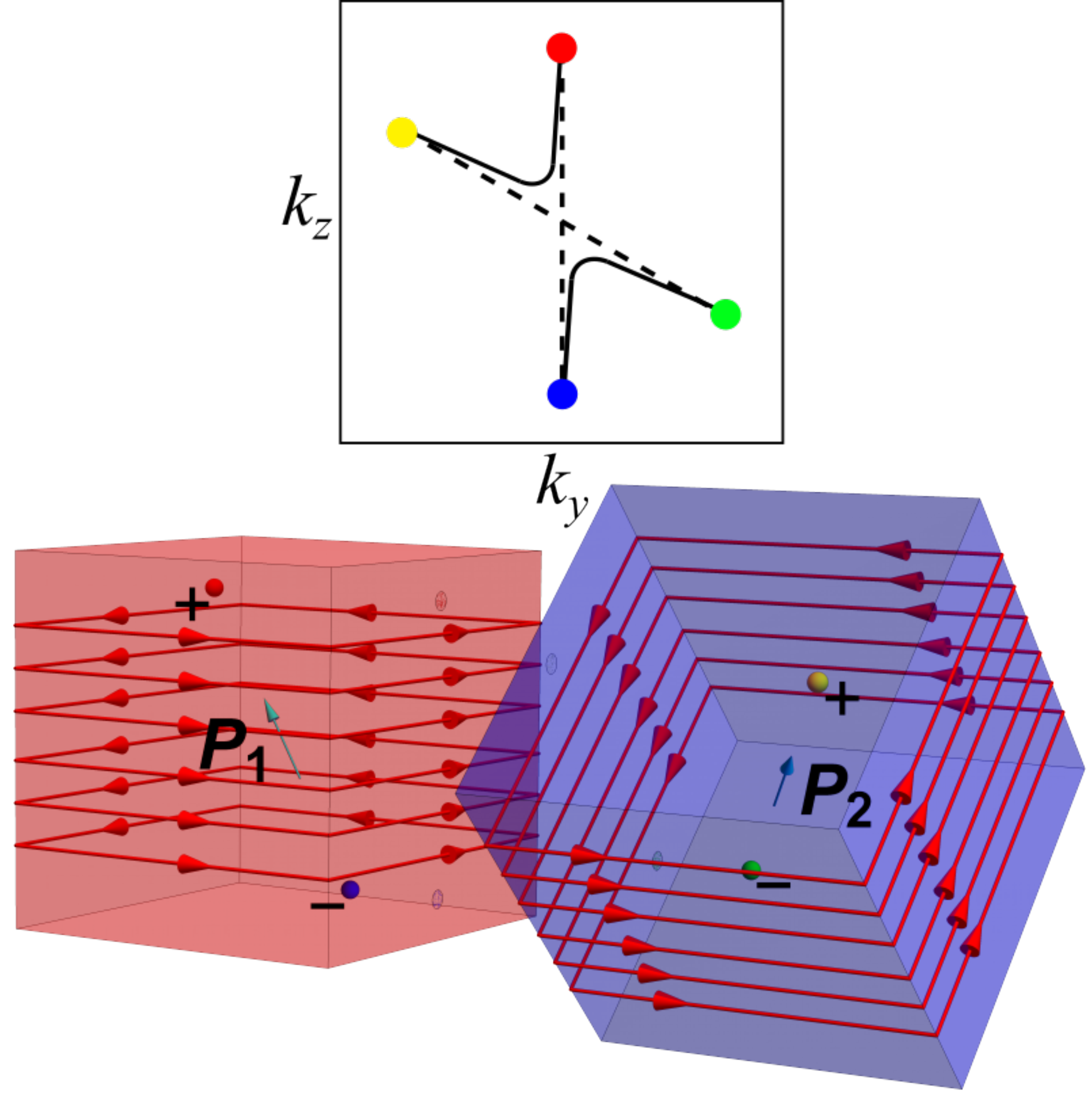}
    \caption{}
    \label{fig: FermiArcs3}
    \end{subfigure}
    \caption{Upper panel: A schematic illustration of Majorana-Fermi arcs at the interface between the domain walls in the surface-momentum space. Dashed lines are Majorana-Fermi arcs of the isolated domains, and solid lines are reconstructed Majorana-Fermi arcs for the interface between the glued domains; colored circles are projections of Weyl nodes onto the surface: red and yellow - of positive chiralities for domains $D_1$ and $D_2$, respectively; blue and green - of negative chiralities for domains $D_1$ and $D_2$, respectively. Lower panel: A schematic illustration of the way the isolated domains are glued. Colors of the Weyl nodes (denoted by spheres with a nearby ``$\pm$" sign denoting the sign of the node's chirality) are in accord with the color scheme of their projections onto the surface; red lines on the domain surfaces with arrows depict zero-energy chiral modes; and $\b{P_{1,2}}$ (cyan-colored vectors) are polarization vectors. \textbf{a}) Projections of Weyl nodes of different chiralities onto the interface coincide: corresponds to the ferroelectric moments along $A_4$ and $B_2$ in the glued domains. \textbf{b}) Projections of Weyl nodes onto the surface do not coincide in the regime of four Weyl nodes (per domain): corresponds to the ferroelectric moments along $A_4$ and $B_1$ in the glued domains. \textbf{c}) As the magnetic field increases, the two Weyl nodes at $p_z=p_2$ and $p_z=p_3$ come closer and eventually annihilate. Here we consider the case when projections of the remaining Weyl nodes do not coincide, which corresponds to the ferroelectric moments along $A_4$ and $B_1$ in the glued domains.
    }
\end{figure*}

We fix the polarization vector in the first domain to be $A_4$ (~\cref{fig: Domain Setup1}). 
When the polarization vector in the second domain is $B_2$ or $B_4$, the Weyl nodes coincide when projected onto the momentum plane parallel to the wall (see~\cref{fig: FermiArcs1}). In contrast, if the polarization vector in the second domain is $B_1$ or $B_3$, the projections of the Weyl nodes from the two domains are at different points (~\cref{fig: FermiArcs2,fig: FermiArcs3}). Below we present a qualitative description of the resulting Fermi arcs for these scenarios. 

In both scenarios, the effective low-energy Hamiltonian is given by~\cite{Dwivedi2018,Murthy2020}
\begin{align}
\label{eq: effective interface Hamiltonian}
    H_{eff}(\b{k}_{||})=\begin{pmatrix}
        \epsilon_1(\b{k}_{||}) & a(\b{k}_{||}) \\
        a^*(\b{k}_{||}) & \epsilon_2(\b{k}_{||})
    \end{pmatrix},
\end{align}
where $\epsilon_{1,2}(\b{k}_{||})$ are the low-energy chiral modes of each of the domains, $D_1$ and $D_2$, and the off-diagonal matrix component $a(\b{k}_{||})$ are the couplings. The eigenvalues of \cref{eq: effective interface Hamiltonian} are given by 
$
    x^2-(\epsilon_1+\epsilon_2)x + \epsilon_1 \epsilon_2 - \abs{a}^2 = 0
$ and therefore, the Fermi arc states obey the equation 
\begin{align}
\label{eq: reconstruction condition}
\epsilon_1 \epsilon_2 = \abs{a}^2\,.
\end{align}

{\it (i) The scenario in which the projections of the Weyl points onto the interface of both domains coincide--} This happens when the  polarization vector in $D_1$ is $A_4$, and the polarization vector in $D_2$ is $B_2$ or $B_4$. We assume the magnetic field $\b{B}$ lies in $xz$-plane for simplicity. We then identify two cases:

\noindent
case \textbf{I}-- The chiralities of the Weyl nodes with coinciding projections are the same. In this case $\epsilon_2=-\epsilon_1$, and the condition becomes $-\epsilon_1^2 = \abs{a}^2$, which can be satisfied only when $\abs{a}^2=0$ for $\b{k}_{||}$ at which $\epsilon_1=0$. However, there is no symmetry that fixes $a(\b k)=0$ for $\b k$ on that line. Therefore, the arcs are gapped out in the general case. In~\cref{app: Appendix_Domains}, we discuss such unprotected zero energy solutions.  

\noindent
case \textbf{II}-- The chiralities of the Weyl nodes with coinciding projections are opposite. Here $\epsilon_1=\epsilon_2$. Consequently,  the arcs are robust and found on the lines for which $\epsilon_1(\b{k}_{||})=\pm \abs{a(\b{k}_{||})}$ (see~\cref{fig: FermiArcs1}).

An important consequence of the scenario of coinciding  Weyl points when projected to the domain wall,  is  that a rotation of the magnetic field $\b B$ about the $y$-axis allows to continuously tune between case \textbf{I} and case \textbf{II}. Then we expect arc states to disappear and reappear as a function of angle.



{\it (ii) The scenario where projections of the Weyl nodes do not coincide--} This happens when the
polarization vector in $D_1$ is $A_4$, and the polarization vector in $D_2$ is $B_1$ or $B_3$. 
For $\Delta^2<B^2<\Delta^2 + \mu^2$ the Majorana Weyl arcs will ``repel" and ``attract" each other as shematically illustrated in~\cref{fig: FermiArcs2}. For $B^2>\Delta^2 + \mu^2$, a more significant reconstruction of the Majorana Fermi arcs happen. For the point close to the crossing point, we can write $\epsilon_1 \approx v_1 k_y$, and $\epsilon_2 \approx - v_2 ( k_z \sin \theta + k_y \cos \theta )$, where $\theta$ is the angle between the polarizations' projections onto the interface. Then, from~\cref{eq: reconstruction condition}, we find
\begin{align}
    k_z=-\frac{\abs{a}^2}{v_1 v_2 k_y \sin \theta} -k_y \cot \theta,
\end{align}
which defines a hyperbola in the vicinity of the crossing point, now connecting the projections of the Weyl nodes of the same chirality (see~\cref{fig: FermiArcs3}).


\section{Weyl-superconductivity in the presence of vortices}
\label{sec: Vortex}
Up to this point we have only considered  the Zeeman coupling to the magnetic field. We now turn to consider the consequence of the orbital coupling. In a type-II superconductor, the field can induce vortices when it exceeds the value $H_{c1}$. We distinguish two limits of interest. In the small magnetic field limit $H_{c1}<B \ll H_{c2}$ the distance between vortices is much greater than the coherence length and each vortex can be treated independently. In the opposite limit, $B\lesssim H_{c2}$ the vortices become densely packed, overlap and significantly reduce the global average value of the order parameter. 

In what follows, we focus on these two limits. We start with the single vortex problem. Using the results of \cref{sec: Model}, we show that individual vortices in ferroelectric superconductors can contain non-trivial Majorana bound states, even when the bulk superconducting state is trivial. Then in the next step, {based on semiclassical considerations} {(namely, assuming strong localization of the Majorana states on a scale much smaller than the coherence length),} we find that there is always a critical magnetic field $B_* < H_{c2}$, marking a percolation transition to a {putative} topological state with Majorana-Weyl nodes in the bulk. 

\subsection{The single vortex problem - Non-trivial bound states}
\label{sec: Single vortex}
\begin{figure}
    \centering
    \includegraphics[width=0.9\linewidth]{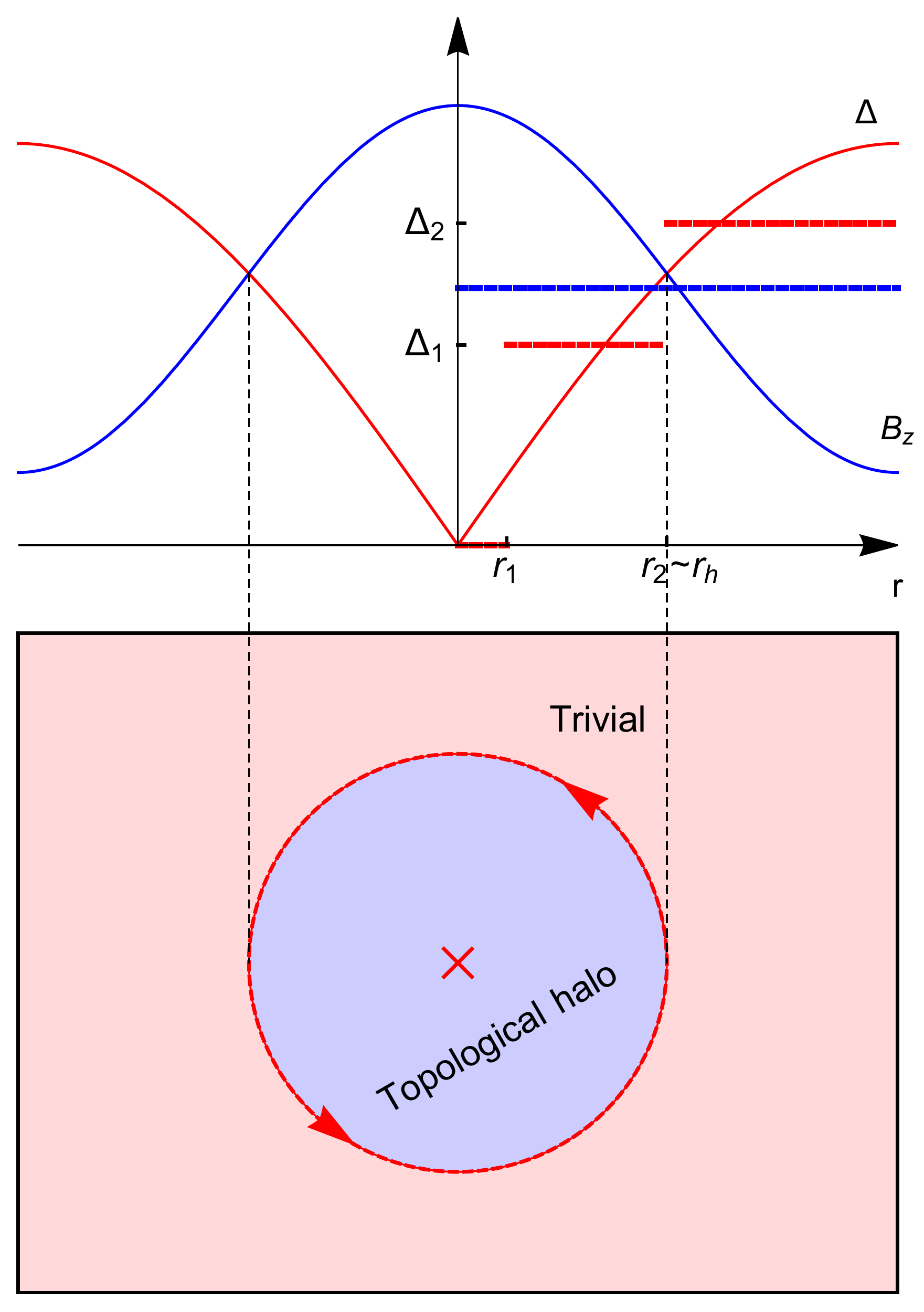}
    \caption{A schematic illustration of the single vortex problem. Upper panel - the radial profiles of the modulus of the order parameter and  the magnetic field (dashed semitransparent lines correspond to the considered toy model with the same color scheme as for the solid lines); lower panel - schematic illustration of the locally topological and trivial regions and MZMs in $xy$-plane in correspondence with the upper panel.}
    \label{fig: Vortexa}
\end{figure}

In the solution of the Ginzburg-Landau equations for a single vortex, the superconducting order parameter $\Delta(\b r)$ and magnetic field $B(\b r)$  both depend on the radial distance from the vortex core. Starting from the core and moving outwards, the order parameter is initially  zero, and adjusts back to its bulk value at a distance of the order of the coherence length ${\xi}$. The magnetic  field, on the other hand, is maximal at the core and gradually decays to zero at a distance given by the penetration depth $\lambda_L$ (we assume that $\lambda_L \gg \xi$). The dependence of these two fields is schematically plotted in \cref{fig: Vortexa}.  

In light of the discussion in \cref{sec: Model}, this implies that somewhere between the vortex core and $r\to \infty$ there is a ``halo'' radius $r_h$, where the critical threshold for creating Majorana-Weyl nodes $B(r_h) = \Delta(r_h)$ is satisfied (see \cref{fig: Vortexa}). Majorana-arc states then appear on a cylinder of radius $r_h$ and at the core of the vortex. Clearly, such states can only be observed if their localization length $l_M$ is significantly smaller than $r_h$. 

To obtain these states we solve the BdG equation explicitly (see \cref{app: BdG_vortex}). We consider two models. First we consider a toy model, which we solve analytically. 
In this model $B$ is taken to be constant and we mimic the spatial dependence of the gap near the vortex core by breaking it into two steps (see dashed lines in \cref{fig: Vortexa}). Namely,  the core region is defined to be in the region $r<r_1$, where the gap is zero. The second region is the topological ``halo'' defined in the region $r_1<r<r_2$ (where $r_2$ is the halo radius $r_h$ in this model). In this region the gap takes a non-zero value  $\Delta_1$, which is smaller than the field, such that the topological criterion $\Delta_1 < B$ is satisfied and there are Weyl nodes. The third region is $r>r_2$, where we assume $\Delta(r>r_2) = \Delta_2$ such that $\Delta_2 >B$ and therefore the superconducting state is trivial and fully gapped. 

The explicit solution shows there are two exponentially localized Majorana bound states, which  are slightly split in energy due to the finite spatial separation between the boundaries at $r_1$ and $r_2$. The key result we obtain from the toy model is an estimate of the localization length of these states
\begin{align}\label{eq: lM}
    l_M\sim 2\pi\xi_0 { \frac{\lambda}{v_F} }{\frac{\Delta_1 \Delta_2}{B^2 - \Delta_1^2}},
\end{align}
where $\xi_0=\frac{v_F}{\pi \Delta_2}$ is the Pippard coherence length estimated at the momentum $k_z$ located between the Weyl nodes.

As can be seen, the length scale Eq. \eqref{eq: lM} appears in units of $\xi_0$ and is proportional to the parameter $\lambda / v_F$.  Close to $H_{c2}$, the halo size becomes of the order of the correlation length. Therefore the Majorana arc states on the edge of the Halo can be resolved from the core Majorana states in the limit $\lambda \ll v_F$. 

Recent theoretical results estimate the electron coupling to the transverse optical phonon mode in STO~\cite{Gastiasoro2021}. Using the average displacement in the ferroelectric phase~\cite{salmani2020order,Salmani2021}, this coupling constant can be estimated to be $\lambda=254$ meV$\cdot \buildrel _{\circ} \over {\mathrm{A}}$. Using this value of $\lambda$, we find the concentration at which $v_F$ becomes greater than $\lambda$ (which happens when the Fermi surface crosses the Dirac point) is $n \approx 2.3 \cdot 10^{19}$ cm$^{-3}$. For higher densities, the ratio $\lambda/v_F$ diminishes. For reference, this parameter diminishes to $\lambda/v_F=1/5$ at $n \approx 1.3 \cdot 10^{21}$ cm$^{-3}$. It is worth noting however, that other estimates of $\lambda$ are smaller~\cite{ruhman2016superconductivity}.


\begin{figure}[!htbp]
    \centering
    \includegraphics[width=0.9\linewidth]{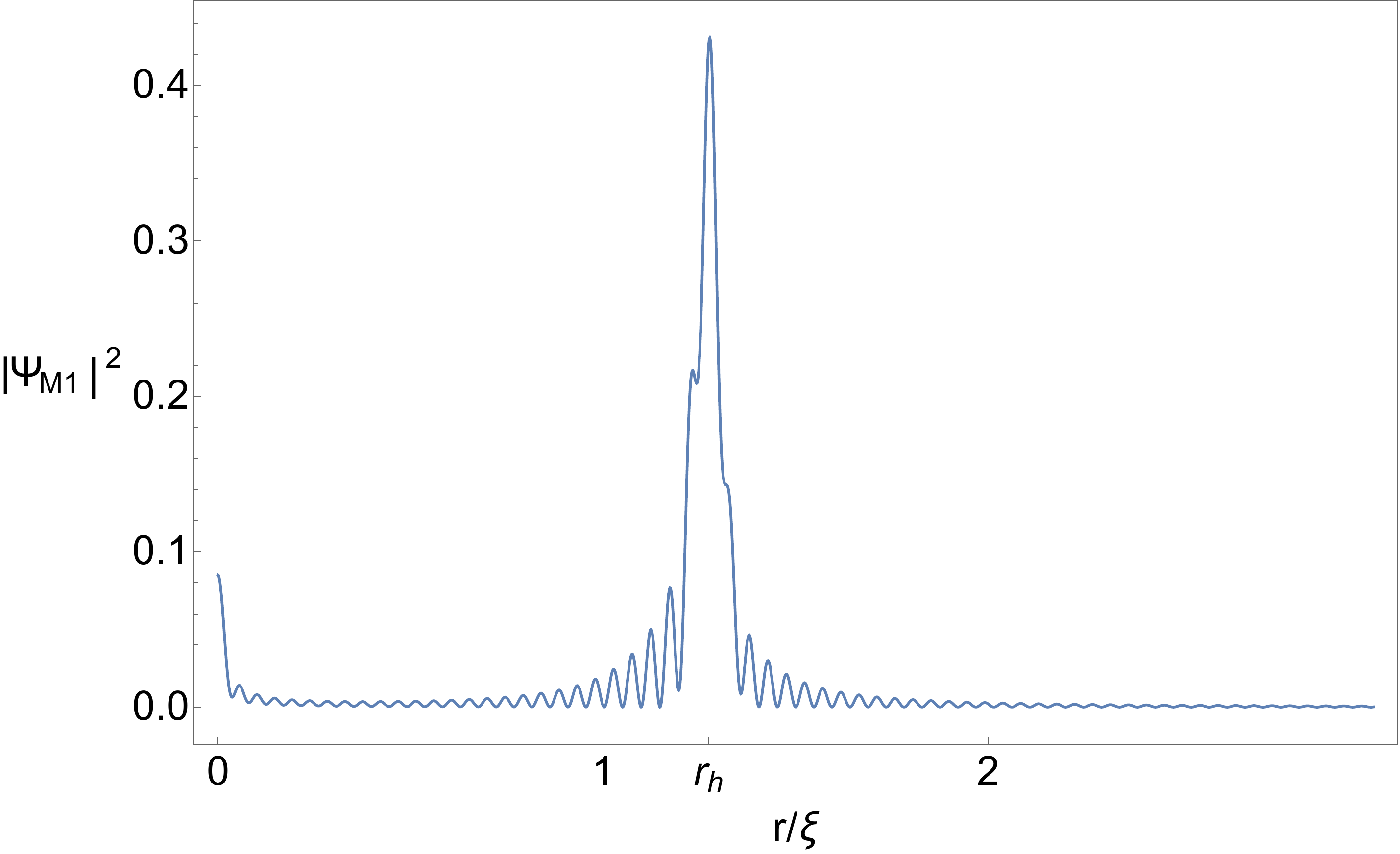}
    \caption{Squared absolute value of the wavefunction (not normalized) of the positive-energy state corresponding to a linear combination of two Majorana modes. The parameters used for the simulation are: $m=1,\; \Delta_0=2,\; B_z=1.71,\; \lambda=0.1,\; \mu=10,\; \xi=100,\; R=700$,
    $p_z$ is chosen such that $\epsilon_{p_z}=0$. 
    }
    \label{fig: Vortex Majorana wavefunction1}
\end{figure}

To confirm the results of the toy model we also solve the BdG problem numerically using a  more realistic profile of the gap, $\Delta(r) =\Delta_0 \tanh(r/\xi)$, where $\Delta_0 = \exp(i\phi)|\Delta_0|$ and $|\Delta_0| > B$. We solve the BdG problem inside the interior of a cylinder of radius $R$. 
As before, the topological criterion $\Delta(r)<B$ is only satisfied within a finite halo radius $r_h$ surrounding the core. The resulting amplitude of one of the two BdG wave functions with nearly zero energy is shown in \cref{fig: Vortex Majorana wavefunction1}. We observe two peaks, corresponding to location of the core and the critical radius $r_h$. 

An interesting aspect of the halo is that it realizes a local pseudo magnetic field~\cite{ilan2020pseudo}. The continuous variation of $|B(r)-\Delta(r) |$, which controls the distance between the Weyl nodes, therefore acts as  a pseudo gauge field in the z direction $\mathcal A_z (r)$. 
The resulting pseudo magnetic field looks like a vortex circulating the core of the halo. An important physical consequence of this field is the emergence of a whole spectrum of Landau levels, which in this case are labeled by angular momentum. These states are plotted in \cref{fig: Vortex energies} in \cref{app: BdG_vortex}.  
For more details regarding the analytic and numeric solutions of the BdG problem  we refer the reader to \cref{app: BdG_vortex}. 

\begin{figure}[!htbp]
    \begin{subfigure}[b]{0.46\textwidth}
    \centering
    \includegraphics[width=\textwidth]{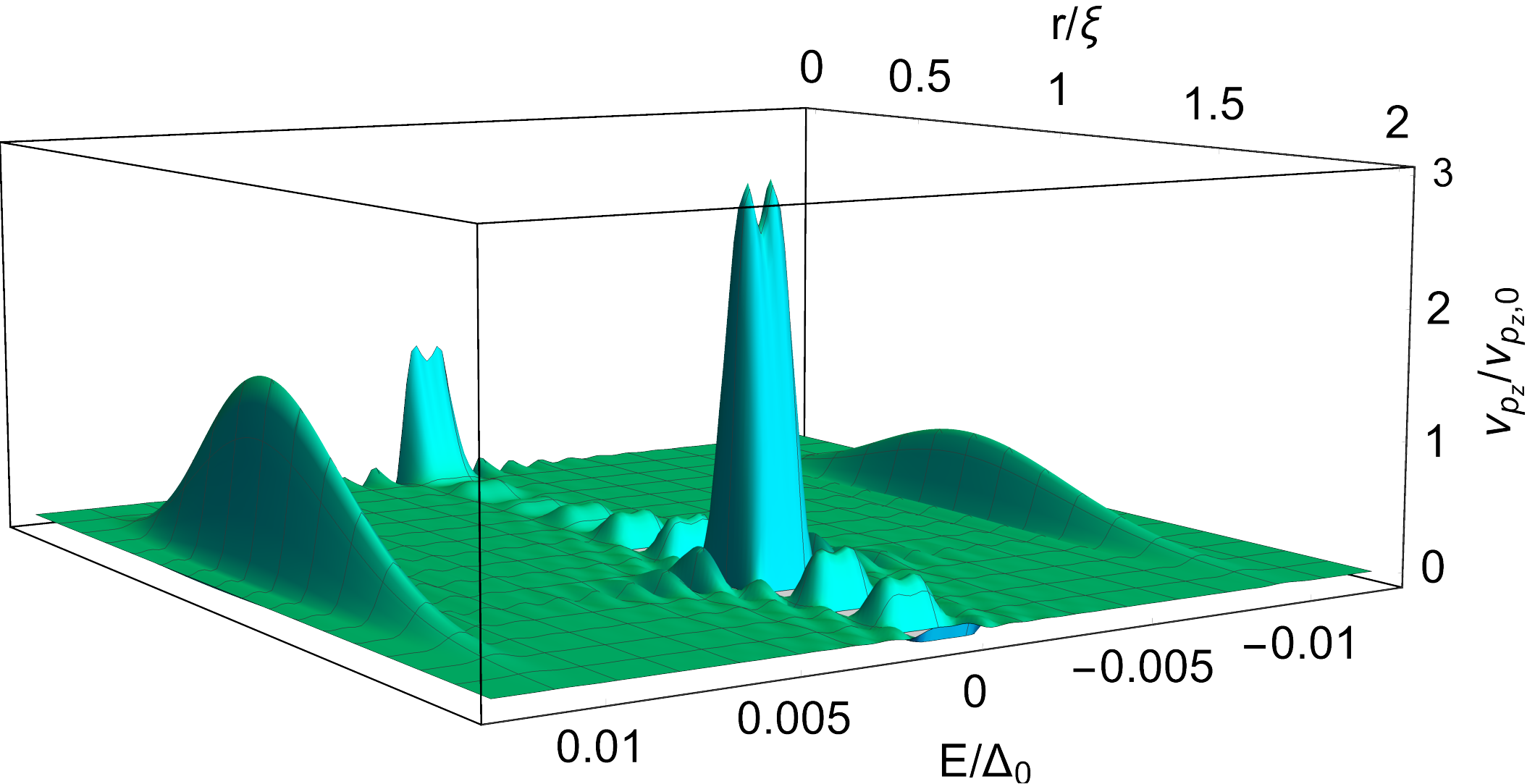}
    \caption{}
    \label{fig: Vortex dIdVa}
    \end{subfigure}
    \begin{subfigure}[b]{0.46\textwidth}
    \centering
    \includegraphics[width=\textwidth]{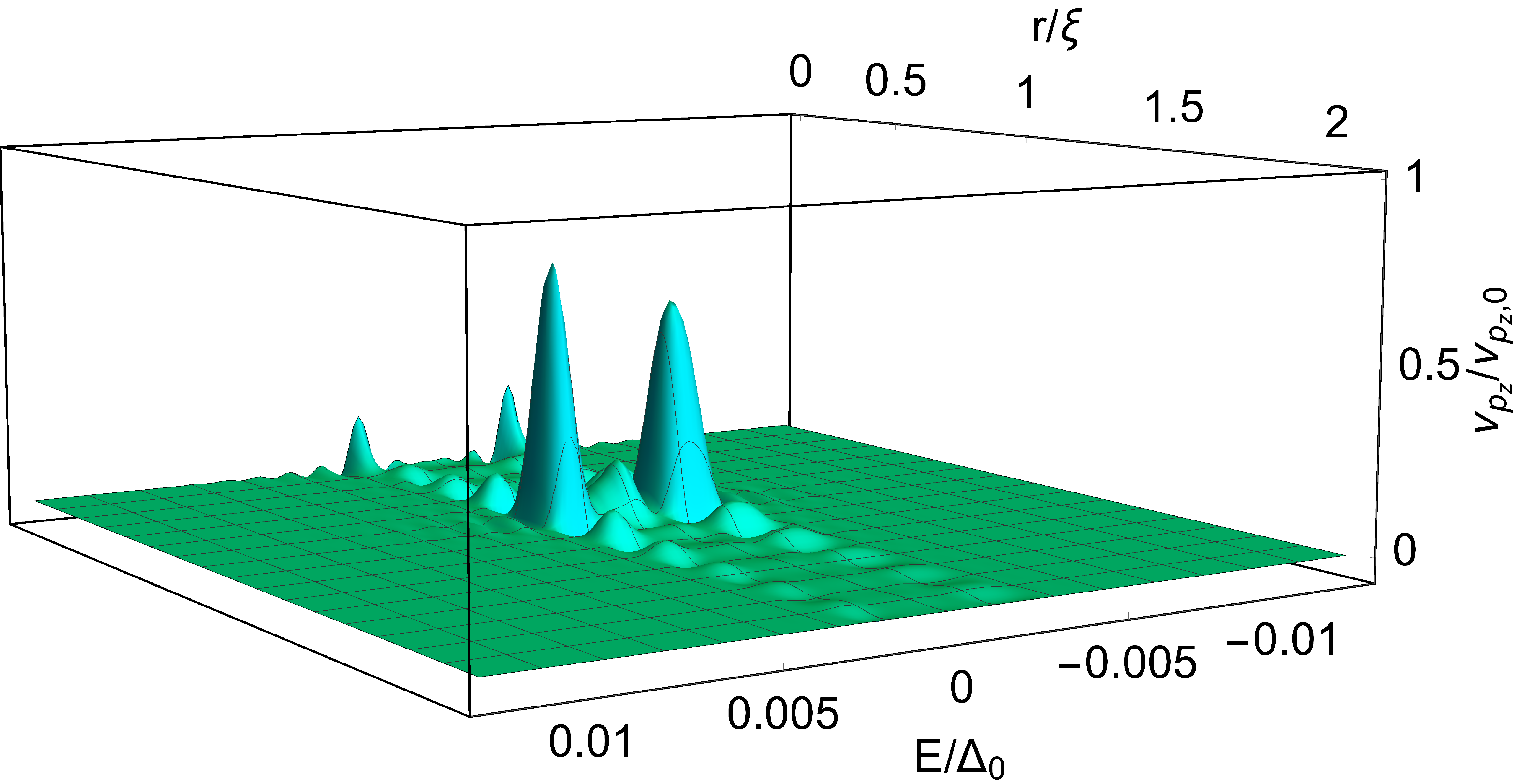}
    \caption{}
    \label{fig: Vortex dIdVb}
    \end{subfigure}
    \caption{a) $\nu_{p_z}(E,r)$ (normalized by $\nu_{p_z,0} = \nu_{p_z}(0,0)$) for $p_z=\sqrt{2m \mu}$ (such that $\epsilon_{p_z}=0$)  as a function of energy and distance from the vortex core; b) $\nu_{p_z}(E,r)$ (normalized by $\nu_{p_z,0}$, the peak value of $\nu_{p_z}(E,r)$ in the plotted region
    ) for $p_z=1.05\sqrt{2m \mu}$ as a function of energy and distance from the vortex core. The parameters used for the simulations are: $m=1,\; \Delta_0=2,\; B_z=1.71,\; \lambda=0.1,\; \mu=10,\; \xi=100,\; R=700$, and the temperature $T=5 \cdot 10^{-5} \Delta_0$.}
\end{figure}

\subsection{Local tunneling density of states in the vicinity of a single vortex }

Using our results from the previous subsection, we now compute the local tunneling density of states in the vicinity of a vortex.  The resolution of a typical scanning tunneling microscope is much smaller than the size of the vortex, and therefore it may be capable of distinguishing the core and edge states described above. 
The local density of states, which is often proportional to the differential conductance~\cite{blonder1982transition,gygi1991self}, is given by
\begin{align}
    \nu(E,r)  = &\int dp_z \nu_{p_z}(E,r) \\ \nn = & - \sum_{i,\sigma=1,2} \abs{\Psi_{\sigma, i}}^2(r)\nn n_F^{\prime} (E_i-E) \\ \nn
    &- \sum_{i,\sigma=3,4} \abs{\Psi_{\sigma, i}}^2(r) n_F^{\prime}(E_i+E).
\end{align}
Here $\Psi_{\sigma,i}(r)$ is the radial part of the $\sigma$-component of the Nambu wavefunction corresponding to the $i$-th eigenmode of energy $E_i$, $\nu_{p_z}(E,\b{r})$ stands for the contribution to the local density of states from eigenmodes corresponding to a particular $p_z$, and we substituted delta-functions with the negative derivatives of the Fermi-Dirac distribution at low temperature.


In~\cref{fig: Vortex dIdVa}, we plot $\nu_{p_z}(E,\b{r})$ for $p_z=\sqrt{2m \mu}$, as a function of energy and distance from the vortex core $r$. One can clearly see peaks at zero bias for $r=0$ and $r \approx r_h$. 
For $p_z$'s away from $\sqrt{2m \mu}$ (but for which MZMs still exist), the distance between the peaks of the MZMs deacreases, while the localization length of MZMs increases. This results in broadening of the peaks in $r$-direction and further separation in $E$-direction; see ~\cref{fig: Vortex dIdVb}~\footnote{We calculate values $\nu_{p_z}(E,r)$ on a relatively sparse grid of points in $(E,r)$-space, which does not include the point corresponding to the highest peak of $\nu_{p_z}(E,r)$, and then interpolate between the points. This results in that none of the peaks in the plot reach the value of 1.} in which $\nu_{p_z}(E,\b{r})$ is plotted for $p_z=1.05\sqrt{2m \mu}$. Consequently, the full density of states $\nu(E,r)$ (and the differential conductance) will have smeared zero-bias peaks.


\begin{figure*}[!htbp]
    \centering
    \begin{subfigure}[t]{0.35\linewidth}
    \centering
    \includegraphics[height=0.9\textwidth]{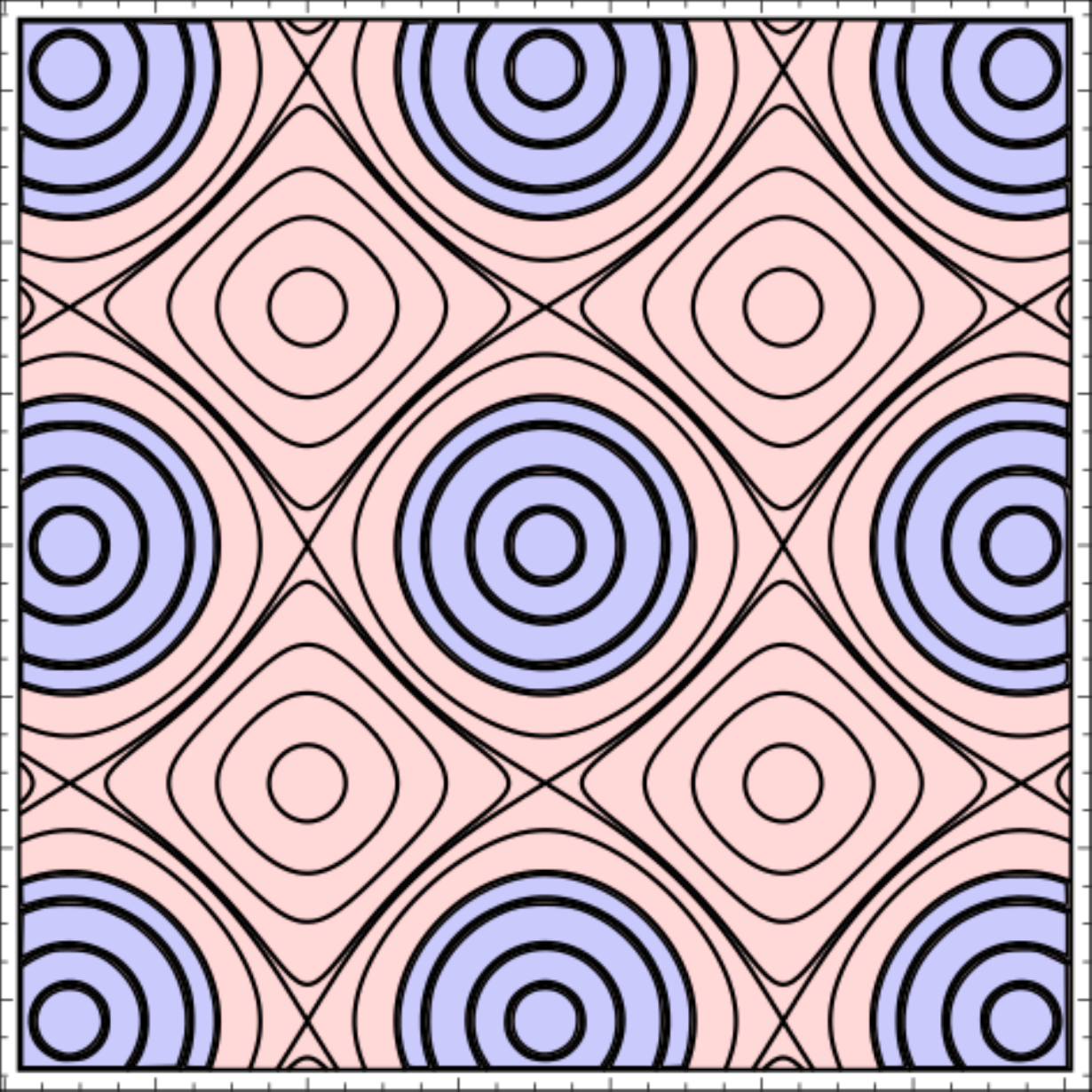}
    \caption{In a relatively small magnetic field, vortices in the Abrikosov lattice are far away from each other (in units of the coherence length), and topologically non-trivial regions locally surround each of them.}
    \label{fig: Percolation1}
    \end{subfigure}
    \begin{subfigure}[t]{0.35\linewidth}
    \centering
    \includegraphics[height=0.9\textwidth]{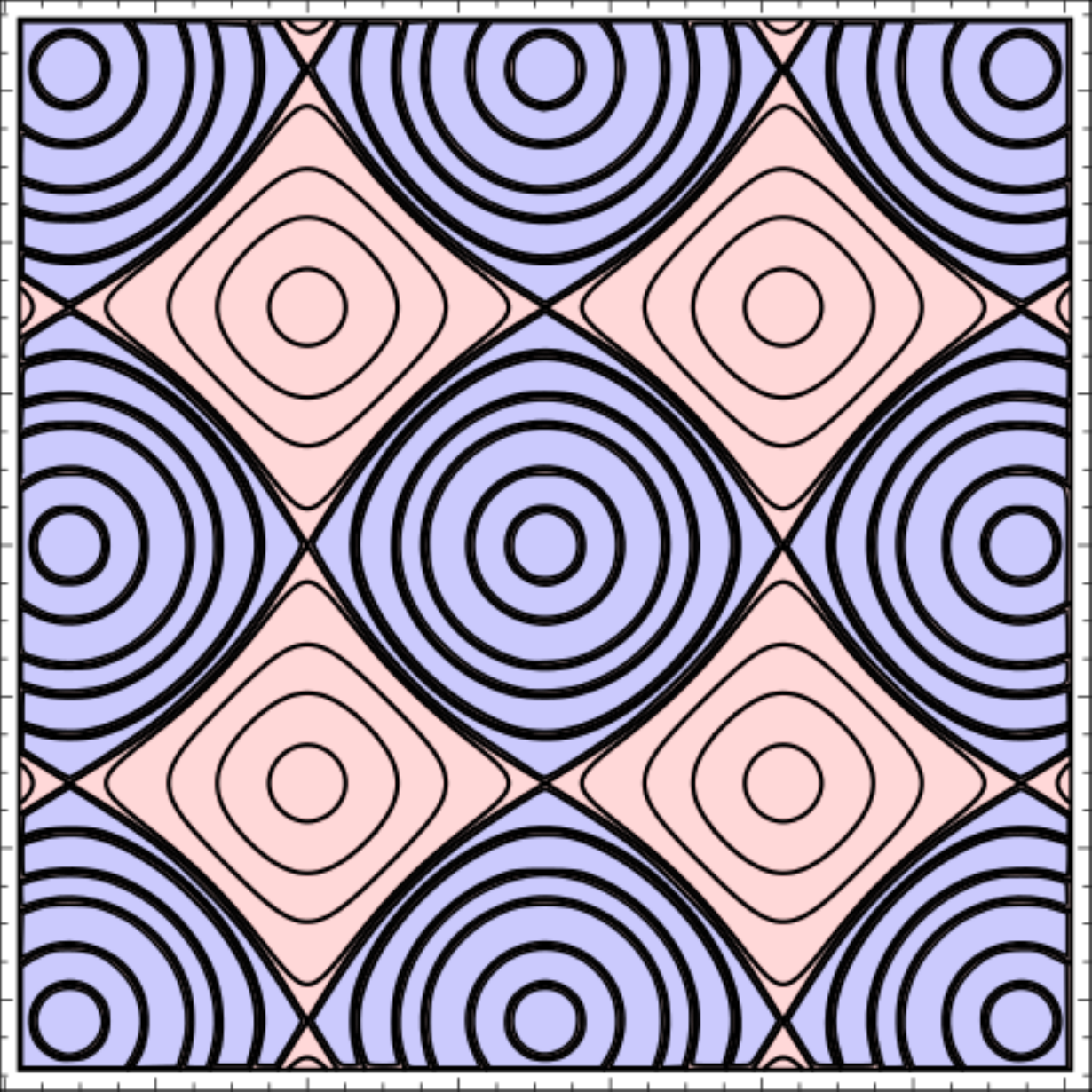}
    \caption{At the critical value $B_*$, previously bounded topologically non-trivial ``puddles" touch with the neighbouring "puddles" at one point.}
    \label{fig: Percolation2}
    \end{subfigure}
    \begin{subfigure}[t]{0.35\linewidth}
    \centering
    \includegraphics[height=0.9\textwidth]{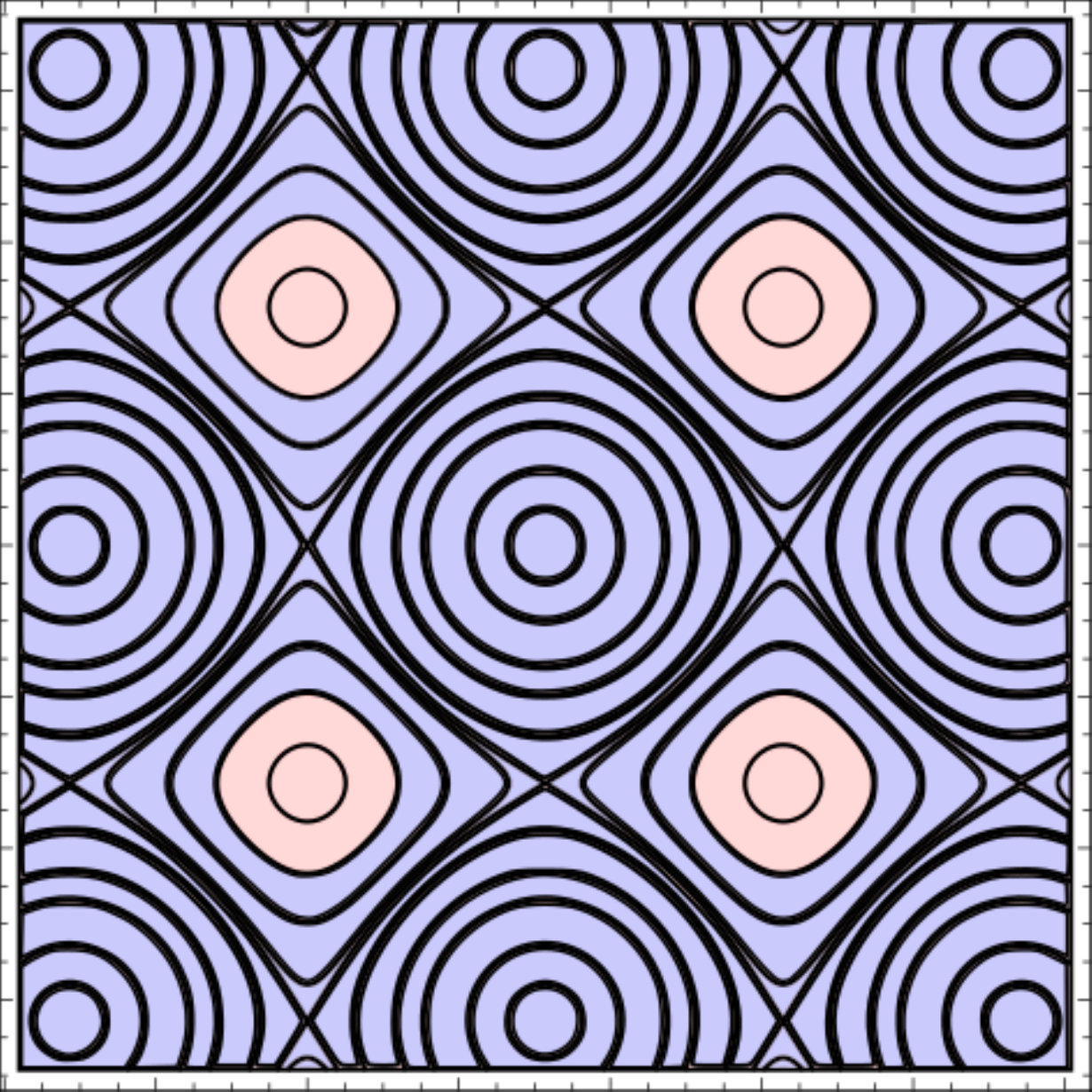}
    \caption{For large values of the magnetic field, topologically non-trivial ``puddles" around each vortex overlap, creating a topologically non-trivial ``sea".}
    \label{fig: Percolation3}
    \end{subfigure}
    \begin{subfigure}[t]{0.35\linewidth}
    \centering
    \includegraphics[height=0.9\textwidth]{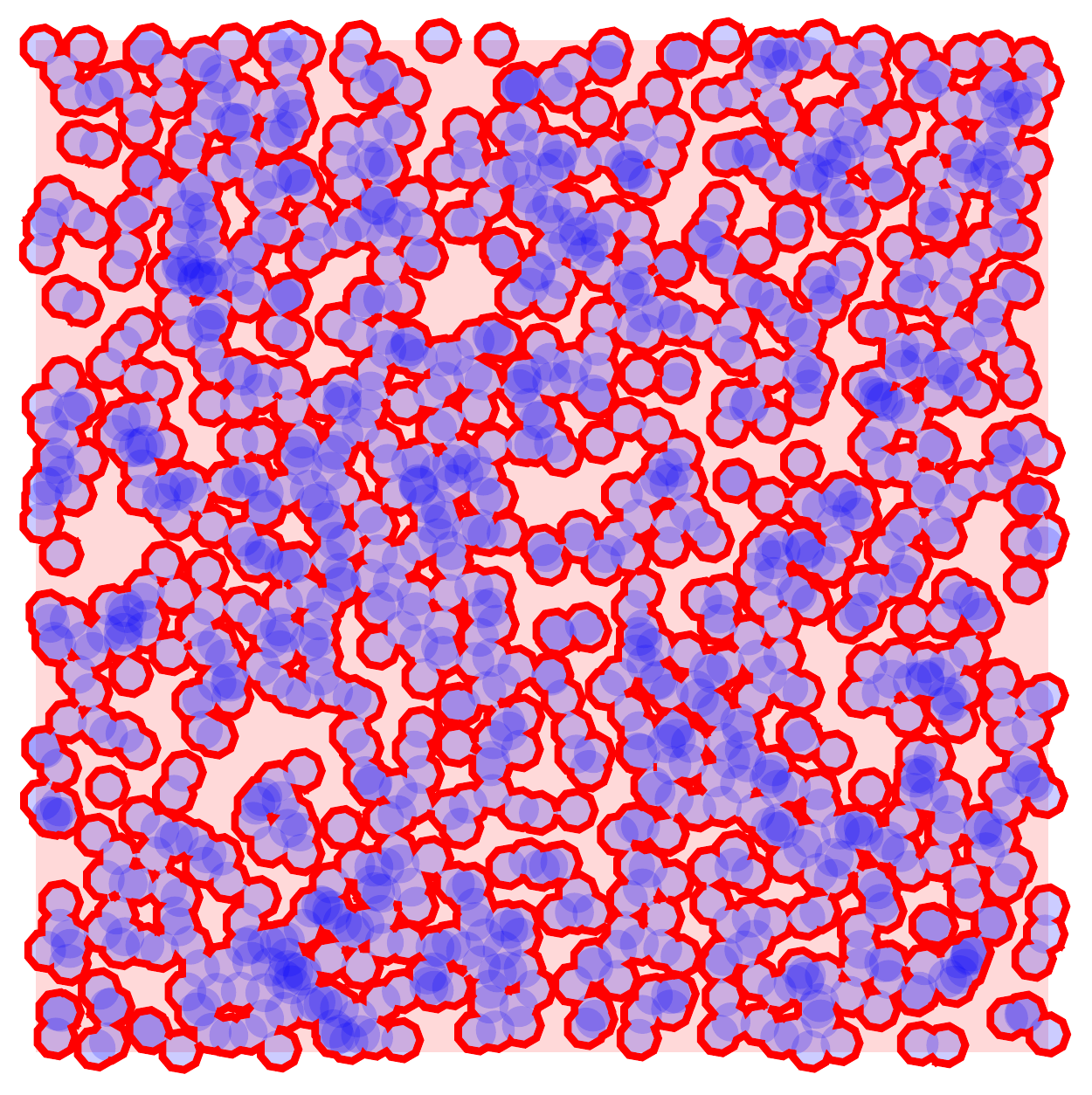}
    \caption{In the more realistic disordered network of vortices, the percolation will have a disordered character as well. 
    Red lines are Majorana ``halos" surrounding the topological phase.}
    \label{fig: VL3}
    \end{subfigure}
    \caption{Schematic illustration of the percolation of the topological phase. The black line structure is a contour plot for $\Delta(\b(r))$, orange color denotes topologically trivial regions, and blue color - topologically non-trivial regions. 
    }
    \label{fig: Percolation}
\end{figure*}

\begin{figure*}[!htbp]
    \centering
    \begin{subfigure}[t]{0.32\textwidth}
    \centering
    \includegraphics[width=\linewidth]{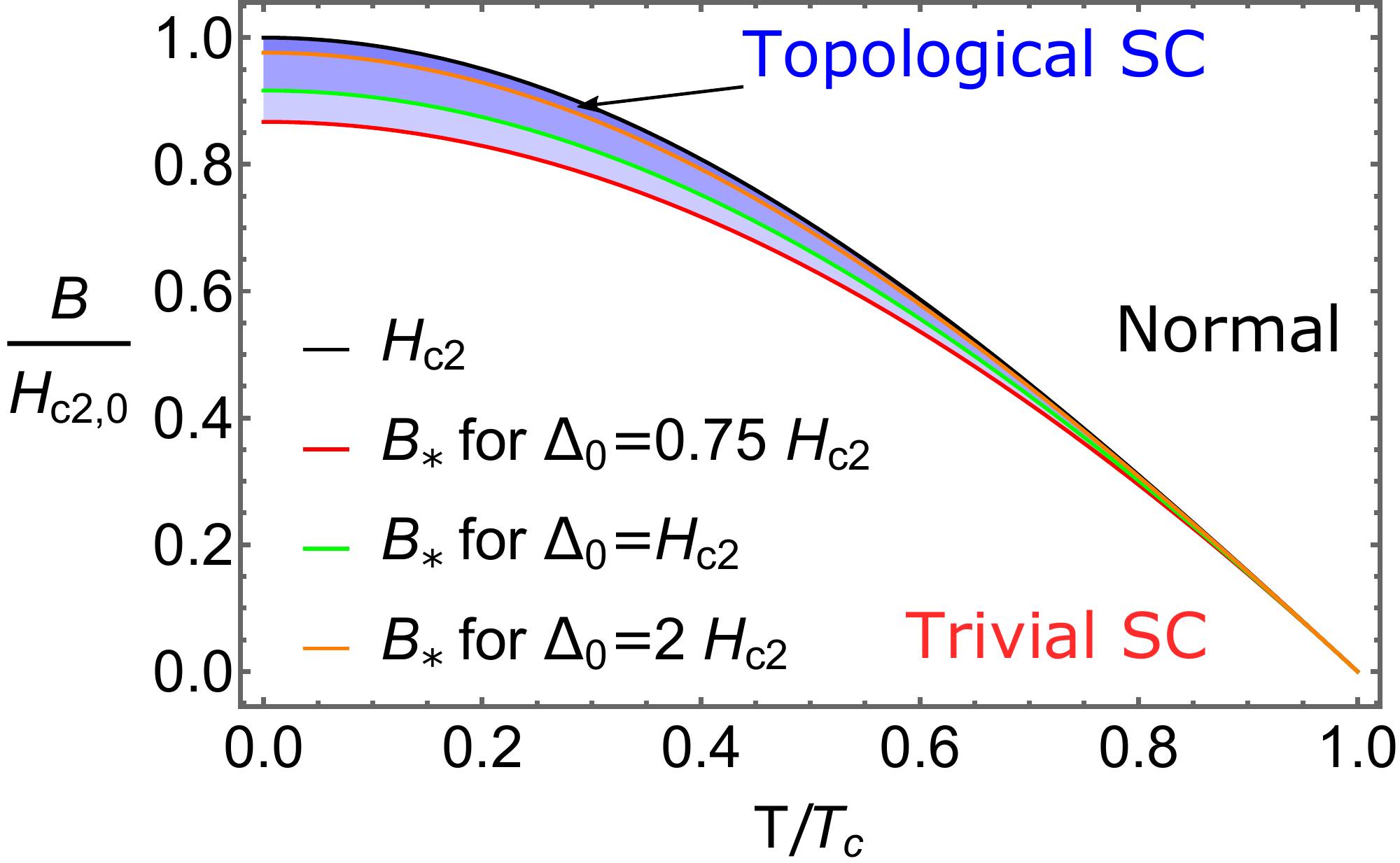}
    \caption{$K=0.1$}
    \end{subfigure}
    \centering
    \begin{subfigure}[t]{0.32\textwidth}
    \centering
    \includegraphics[width=\linewidth]{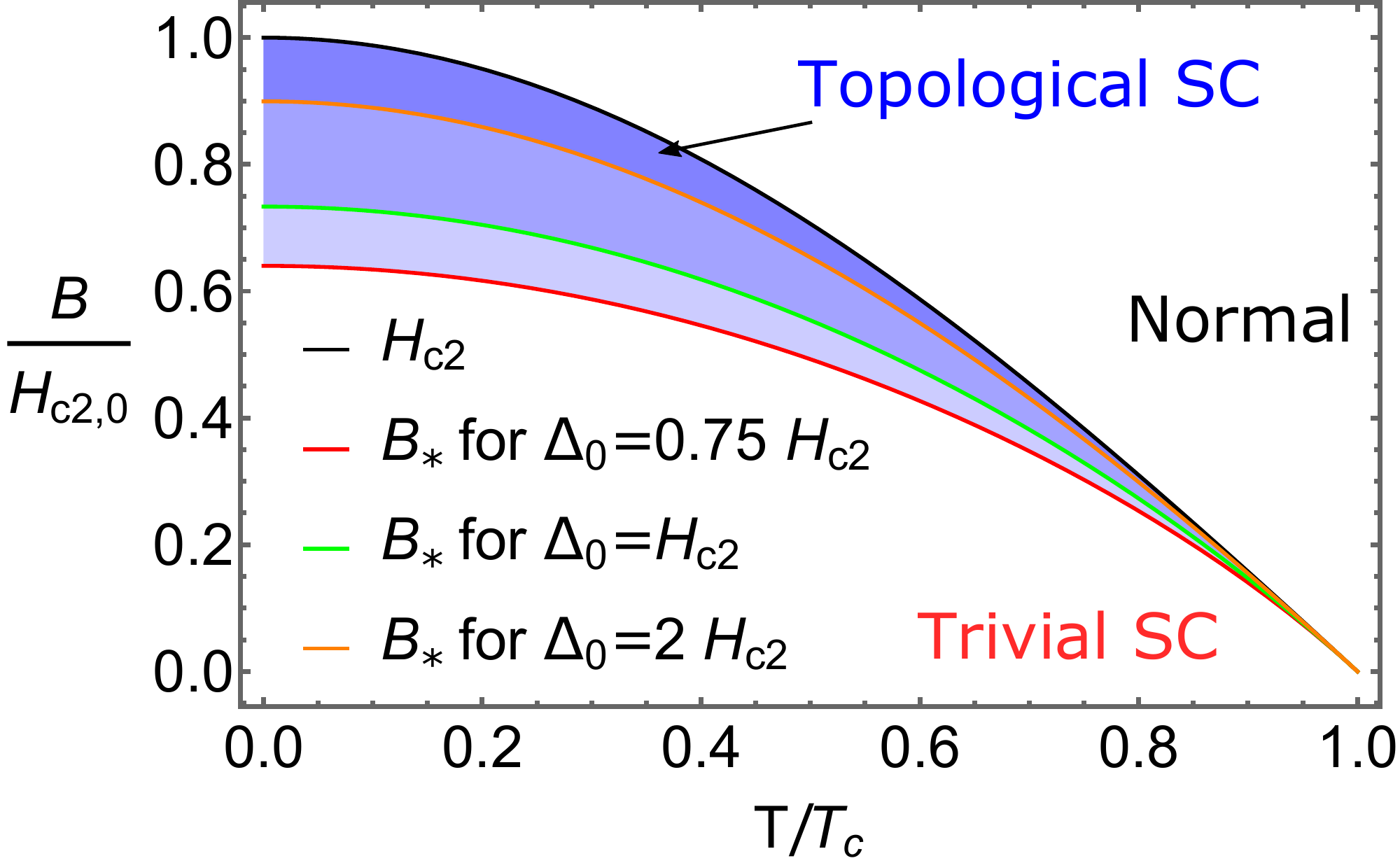}
    \caption{$K=0.5$}
    \end{subfigure}
    \centering
    \begin{subfigure}[t]{0.32\textwidth}
    \centering
    \includegraphics[width=\linewidth]{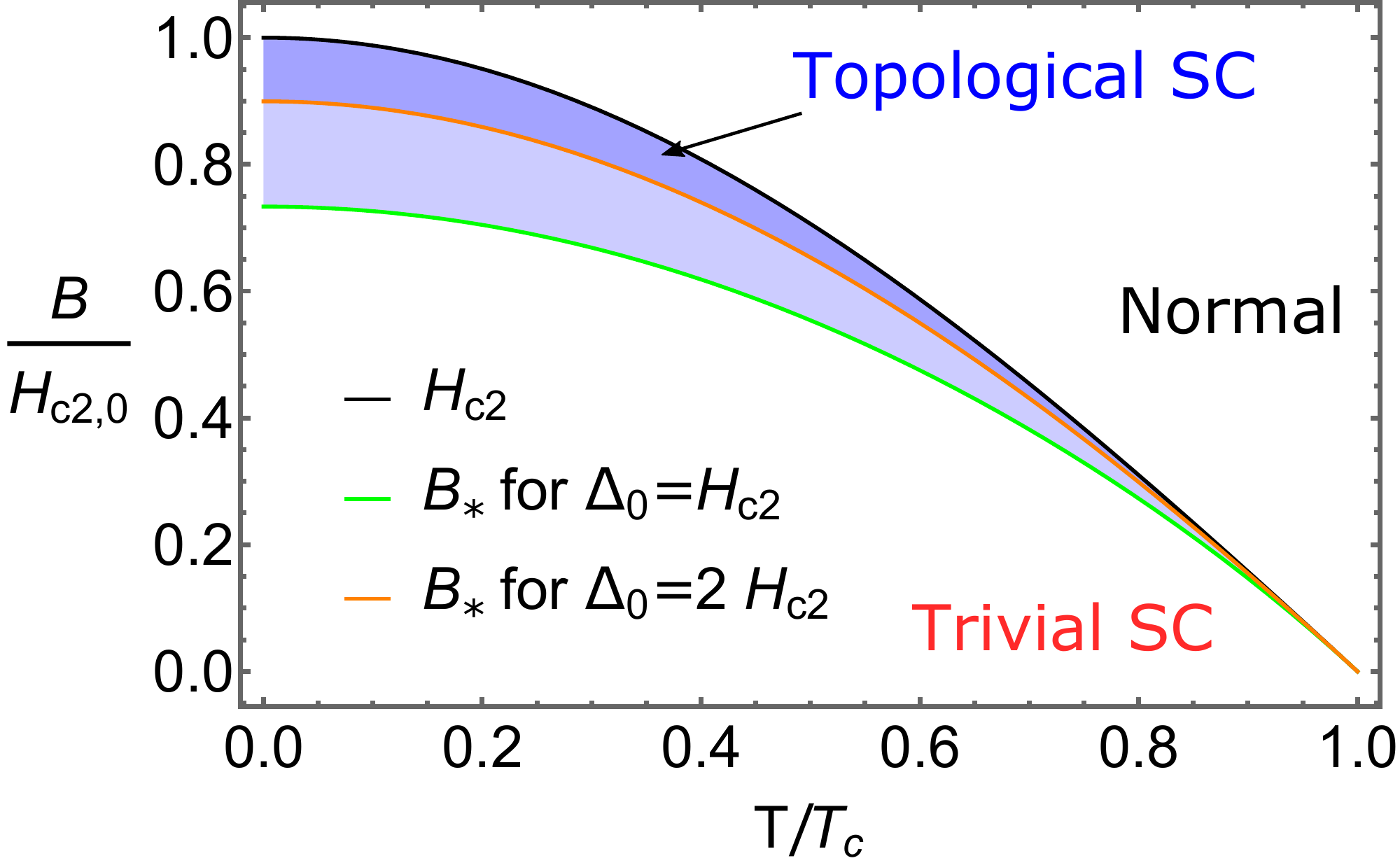}
    \caption{$K=1$}
    \end{subfigure}
    \caption{Schematic percolation-phase diagrams (based on~\cref{eq: H_percolation}) in the space of $B/H_{c2,0}$ vs $T/T_c$, where $H_{c2,0}$ is the upper critical field at $T=0$. Area between $B_*$ and $H_{c2}$ lines correspond to topological phase and denoted by shading. The increase of intensity of area shading corresponds to the overlapping of regions of topological phases corresponding to different values of input parameter $\Delta_0/H_{c2}$. 
    }
    \label{fig: PercolationPhaseDiagram}
\end{figure*}

\subsection{The many-vortex problem - Percolation of the topological phase}

The picture presented above, where each isolated vortex is surrounded by a topological halo, suggests the possibility of a percolation transition, where the halos overlap and the topological phase percolates through the system. 
At large field, $B \sim H_{c2}$, the ground state of the system is expected to be a vortex lattice. Therefore, let us assume the magnetic field is large enough such that the lattice state is formed, yet the halos are still separated and each vortex is encircled by chiral Majorana zero modes.  Upon increasing the field even further, the halos grow, and eventually touch, creating a connected sea of the topological phase. 
Below we develop a crude estimate for this percolation threshold $B_*$ and find that it is always smaller than $H_{c2}$. {Full microscopic calculations are required to verify this scenario more rigorously, especially considering that the proof for the existence of Majorana zero modes presented in this manuscript is strictly valid only at low magnetic fields.}

We use Abrikosov's theory~\cite{abrikosov1957magnetic}, applicable for magnetic fields close to the upper critical field $H_{c2}$. The harmonic approximation solution for the first Ginzburg-Landau equation can be written in the form

\begin{align}
\label{eq: Psi_l}
    \Delta(\b{r})&= \Delta_0 f(\b{r}),
\end{align}
where $\Delta_0$ is the gap function at zero magnetic field and
\begin{align}
\label{eq: f(r)}
    f(\b{r})=\sum_{n=-\infty}^{\infty} D_n e^{i n q y} e^{-\frac{(x-x_n^2)}{2 \xi^2(T)}},
\end{align}
where $\xi(T)$ is the coherence length at temperature $T$, $x_n$ is a position of $n$th vortex core on $x$-axis, ${2\pi}/{q}$ is the periodicity in the $y$-direction, and $D_n$ are dimensionless coefficients. 
Substituting~\cref{eq: Psi_l} into the second Ginzburg-Landau equation, one finds an expression for the magnetic field
\begin{align}
    B(\b{r}) = B_0 - \frac{H_{c_2}}{2 \kappa} f^2(\b{r}),
\end{align}
where $\kappa={\lambda_L}/{\xi}$ is the Ginzburg-Landau parameter. We note that Abrikosov's theory is valid for small ${f^2}$, therefore in what follows we will consider $B_0\gtrsim 0.9 H_{c_2}$.

To find the next order correction to $f$ in the small parameter $1-B_0/H_{c2}$ one requires that~\cite{abrikosov1957magnetic}
\begin{align}
    \overline{f^4} (1-\frac{1}{2 \kappa^2}) - \overline{f^2} (1-\frac{B_0}{H_{c_2}})  = 0,
\end{align}
where $\overline{O}$ stands for the averaging $O$ over one unit cell of the vortex lattice. Then, using a parameter 
\begin{align}
  \beta = \frac{\overline{f^4}}{\overline{f^2}^2},
\end{align}
which characterizes a lattice structure (for the square lattice $\beta=1.18$, and for the triangular one $\beta=1.16$), the non-zero solution for $\overline{f^2}$ is
\begin{align}
 \label{eq: f^2}
    \overline{f^2} = \frac{1-\frac{B_0}{H_{c_2}}}{1-\frac{1}{2 \kappa^2}} \beta^{-1}.
\end{align}

The spatial profile $f(\b{r})$ of the order parameter is defined by the coefficients $D_n$ in \cref{eq: Psi_l}. For simplicity, in the following we consider the case of the square lattice, for which $D_n$ are constants denoted by $D$, and $x_n = n q \xi^2$. 

Percolation of the topological phase will occur when the magnetic field at the half distance between the neighboring vortices' cores, $d/2$, reaches the critical value for the topological phase transition (see~\cref{fig: Percolation}), $B(d/2)=\Delta(d/2)$, i.e.,
\begin{align}
     B_0 - \frac{ H_{c_2} }{2 \kappa} D^2 f_0^2(d/2)  = \Delta(d/2),
\end{align}
where $f_0$ is given by~\cref{eq: f(r)} in which all $D_n$ set to one. 
An important remark here is that we use the topological criterion $B>\Delta$ derived in~\cref{sec: Model} for a uniform case and which remains valid for the vortex problem in a small magnetic field when the vector potential terms in BdG Hamiltonian can be neglected. In principle, in high magnetic fields, the contribution from these terms might modify the topological criterion. We leave the investigation of this question for future work.

Combining this equation with~\cref{eq: f^2}, we find the value of $B_*$ needed to be applied to reach the percolation point 
\begin{align} \label{eq: H_percolation}
   {B_*}=H_{c2}(1-\delta) ,
\end{align}
where 
\begin{align}
    \delta &= \frac{\beta f_0^2\left(\frac{d}{2}\right) \overline{f_0^2} \left(1-\frac{1}{2 \kappa^2}\right)}{\left( \beta \left(1-\frac{1}{2 \kappa^2}\right) \overline{f_0^2} + \frac{1}{2 \kappa^2} f_0^2(\frac{d}{2}) \right)^2} \bigg\{ -\frac{1}{\sqrt{K}} \frac{ \Delta_0}{2 H_{c2}} \frac{\xi(T)}{\xi_0}\nn\\&+ \sqrt{\frac{1}{K} \frac{{ \Delta_0^2}}{(2 H_{c2})^2} \left( \frac{\xi(T)}{\xi_0} \right)^2+ \beta\left(1-\frac{1}{2 \kappa^2}\right)\frac{\overline{f_0^2}}{f_0^2(\frac{d}{2})}+\frac{1}{2 \kappa^2}}   \; \bigg\}^2\,. \nn
\end{align}

{
The field $B_*$ at which the topological phase percolates is therefore controlled by two phenomenological parameters. The first is 
 $K = \rho_s / 2 \Delta_0^2 \mathcal N(0)\xi_0^2$, where $\rho_s$ is the superfluid stiffness and $\mathcal{N}(0)$ is the density of states of the underlying metal. Assuming a full volume fraction, a parabolic band dispersion $\rho_s={\hbar^2 n}/{4 m}$~\footnote{we restored $\hbar$ here} and  $\Delta_0 \approx 1.76 T_c$~\cite{Swartz2018}, we have $K\approx 0.1 (\mu/T_c)^2/(k_F \xi_0)^2$, which can be estimated directly from experiment (at $n = 10^{18}$ cm$^{-3}$\  $\mu = 2$ meV, $T_c = 200$ mK~\cite{Lin2014}) to be between 1 and 5 depending on the value of $\xi_0$ between 100 and 50 nm, respectively. It is interesting to  compare this result with the prediction of BCS theory $K=0.5$~\cite{gor1959microscopic}.
 The second parameter controlling $B_*$ is the ratio $\Delta_0 / H_{c2}$. Comparing with the experimental data of Ref.~\cite{schumann2020possible} we find that this parameter can be on the order of (and even larger than) 1.}

We plot the resulting schematic~\footnote{For the curve $H_{c2}(T)$ we use approximate Gor'kov's formula~\cite{Gorkov1960} $H_{c2}(T)/H_{c}(T)\approx \chi (1.77 - 0.43 (T/T_c)^2 + 0.07(T/T_c)^4)$, and we approximate $H_c(T)/H_c(0)\approx 1-(T/T_c)^2$. Taking $H_{c2}(0) \simeq \frac{\phi_0}{2 \pi \xi_0^2}$, where $\phi_0$ is the flux quantum, we find $\frac{\xi(T)}{\xi_0} = \sqrt{\frac{H_{c2}(0)}{H_{c2}(T)}}$. Also, we have neglected the dependence of the parameter $K$ on
temperature.} phase diagrams in the space of magnetic field $B$ and temperature  $T$ for different values of $\Delta_0/H_{c2}$ and $K$ in~\cref{fig: PercolationPhaseDiagram}, where we naively extended the use of~\cref{eq: H_percolation} beyond the region of validity of the Ginzburg-Landau theory. The value of $\delta$ in~\cref{eq: H_percolation} does not depend much on $\kappa$ for $\kappa>3$, and we fix it to equal to $10$. As can be seen, for all values of $K$ there is a topological  phase separating  between the trivial superconducting and normal state. This result is much more generic than our particular model. We predict that any noncentrosymmetric superconductor where inversion is broken by a vector~\cite{KoziiFu2015} will develop such topological halos above $H_{c1}$. 
Consequently, all such superconductors 
may
undergo a  percolation transition to a bulk topological phase before giving way to the normal state.

\section{ Conclusions and Discussion }
\label{sec: Conclusions}
We studied Majorana-Weyl superconductivity emerging in systems with intertwined superconducting and ferroelectric orders due to the application of a magnetic field. First, we considered the effect of a uniform Zeeman field. We confirmed that above the Clogston-Chandrasekhar threshold $g\mu_B B>2\Delta$, Weyl cones appear in the Bogoliubov quasiparticle spectrum along the axis of the polarization moment, regardless of the charge density. 
We also showed that rotating the magnetic field with respect to the polarization tilts the Weyl cones and eventually causes Bogoliubov Fermi surfaces shaped as bananas to appear. 

However, the magnetic field is not expected to be uniform in the superconducting state. Instead it threads through the sample in the form of vortices. Due to the vanishing of the gap at the core of each vortex, the critical threshold $g\mu_B B>2\Delta$ is always fulfilled in some area surrounding it, which we dub the ``halo''. 
Such halos are characterized by Majorana strings at their core and chiral Majorana arc states going around them. When the magnetic field is increased towards $H_{c2}$ the vortices become denser, the halos merge  and the system undergoes a percolation type phase transition to a bulk Majorana-Weyl superconductivity. This transition always precedes $H_{c2}$. 

Our predictions have a number of sharp experimental consequences. The first is the emergence of topological halos surrounding vortices at small magnetic fields above $H_{c1}$. These can be observed in the local tunneling density of states using an STM. However, we expect a clear separation of scales between the size of the halo and the arc state's localization length, only close to $H_{c2}$. This is because the magnetic field at the center of an isolated vortex is of order $H_{c1}$, which is much smaller than the critical threshold. Therefore, the halo radius is very small when the magnetic field is far from $H_{c2}$. 
In addition to the zero modes, the nodes also modify the tunneling density of states away from zero energy. Namely, due to the bulk nodes there will be a quadratic dependence on bias. 
The arc and nodal states can also be observed in the heat conductivity. For example, we anticipate that close to $H_{c2}$, in the topological phase, the system will become heat conducting albeit still superconducting. Finally, when tilting the magnetic field to be perpendicular to the polarization direction we expect Bogoliubov Fermi surfaces to emerge. Close to $H_{c2}$ these surfaces will contribute a $T$-linear term to the specific heat and a constant tunneling density of states. Finally, it is also possible that the existence of Majorana zero modes surrounding vortices will contribute a constant term to the specific heat close to $H_{c2}$, which will manifest itself as a Schottky anomaly at low temperatures. The size of the anomaly should diminish by a factor of $1/2$ when crossing to the topological phase.     

{Full quantum calculations are needed to verify the proposed scenario of the percolation of the topological phase.}
{In the presence of slowly varying disorder, naively, we may anticipate 
a scenario similar to
the transitions between integer quantum Hall states~\cite{chalker1988percolation}}, {where the topological nature of the phases is manifested as long as a network of edge-states percolates through the bulk}.
Furthermore, it is also interesting to consider the transition between the topological state considered here and the FFLO state, which is also a relevant ground state when the magnetic field is perpendicualr to the polarization~\cite{agterberg2003novel,Dimitrova2007,Michaeli2012}. To that end, one needs to solve self-consistently for the lowest energy ground state. We postpone the study of such questions to future work. 


\begin{acknowledgements}
We gratefully acknowledge helpful discussions with Susanne Stemmer, Maria Gastiasoro, Vlad Kozii and Rafael Fernandes. 
ES thanks  financial support by the
US-Israel Binational Science Foundation through
awards No. 2016130 and 2018726, and by the Israel Science Foundation (ISF) Grant No. 993/19. JR acknowledges funding by the Israeli Science Foundation under grant No. 3467/21.  RI is supported by the Israeli Science Foundation under grant No. 1790/18.  
\end{acknowledgements}


\appendix
\section{Bogoluibov Fermi surface in a tilted magnetic field}
\label{app: Appendix_Bogoluibov FS}

The energy dispersion of~\cref{eq: BDG hamiltonian} is determined from the equation
\begin{widetext}
\begin{align}\label{eq: Spectrum eq general}
& E_\b{k}^4 - 2(\epsilon_\b{k}^2+B^2 +\lambda^2 k_{\perp}^2+\Delta^2) E_\b{k}^2 + 8\lambda (k_y B_x - k_x B_y) \epsilon_\b{k} E_\b{k} \\ \nn & - 4 \lambda^2 (k_y B_x - k_x B_y)^2  + (\Delta^2+\epsilon_\b{k}^2 - B^2 -\lambda^2 k_{\perp}^2)^2 +4 \lambda^2 k_{\perp}^2 \Delta^2 =0. 
\end{align}
\end{widetext}
For $B_{x}=B_{y}=0$, its solutions are easily found and given in~\cref{eq: Spectrum eq}. Here we analyze its zero-energy solution for the case when the magnetic field is not parallel to the polarization.

We first show that the conditions for the gap closure is essentially the same as for the case of perpendicular magnetic field. For a generic quartic equation
\begin{align}
    x^4+bx^3+cx^2+d x+e=0,
\end{align}
a product of its roots $\prod_{i=1}^{4}x_i=e$. Considering~\cref{eq: Spectrum eq general} for $k_{\perp}=0$, $e=0$ is satisfied at $B^2=\Delta^2+\epsilon_\b{k}^2$, signifying that there is a zero root.
In addition, this root is double, as it can be readily seen form~\cref{eq: Spectrum eq general}: the free term is zero, and the linear term is zero at $k_\perp$ as well. Thus, the gap closes at $k_{\perp}=0$ for $B^2>\Delta^2$ at $k_z$ determined from the equation $B^2=\Delta^2+\epsilon_\b{k}^2$. Other non-degenerate zero solutions might be determined from equation $e=0$, where $e$ is the free term in~\cref{eq: Spectrum eq general}. Without loss of generality, choosing the direction of the magnetic field such that $B_y=0$, we recast this equation in a form

\begin{align} \label{eq: ky for BFS}
   4 \lambda^2 B_x^2 k_y^2 = (\Delta^2+\epsilon_\b{k}^2 - B^2 -\lambda^2 k_{\perp}^2)^2 +4 \lambda^2 k_{\perp}^2 \Delta^2, 
\end{align}
which determines the dependence of $k_y$ on $k_{\perp}$  and $k_z$ for the momenta satisfying the condition $e=0$. It is indeed a solution if $| k_y | \leq k_{\perp}$, which may happen only if $B_x^2>\Delta^2$. Also, note that these roots are non-degenerate, and there are roots of different sign amongst those four corresponding to the solution of~\cref{eq: Spectrum eq}. It is easy to see that the solution with $k_{\perp}=0, \Delta^2+\epsilon_\b{k}^2 - B^2 \neq 0$ (i.e. away from the Weyl nodes) is impossible. Thus, for $B_x^2>\Delta^2$, we infer that the momenta at which $E=0$ form closed surface(s) defining a 3D Bogoliubov Fermi surface. Numerical investigation shows that these surfaces connect the Weyl nodes at $\b{p}_{1,2}$ and $\b{p}_{3,4}$, respectively (see~\cref{fig: FS4}). In particular, for $B_x^2=\Delta^2, B_z=0$, zero solution can exist only for $k_x=0$, and from~\cref{eq: ky for BFS} we obtain

\begin{align}
    \epsilon_\b{k}^2 = \lambda^2 k_y^2,
\end{align}
which defines two intersecting circles
\begin{align}
k_z^2 + (k_y \pm m \lambda)^2 = 2 m \mu + m^2 \lambda^2.
\end{align}

We emphasize that this result is obtained under the assumption that the superconducting order parameter remains $s$-wave order.

\section{Bogoliubov FS in the presence of triplet component in the gap function}
\label{app: Appendix_pz-wave}
\begin{figure*}[!htbp]
    \centering
    \begin{subfigure}[t]{0.29\textwidth}
    \centering
    \includegraphics[width=\linewidth]{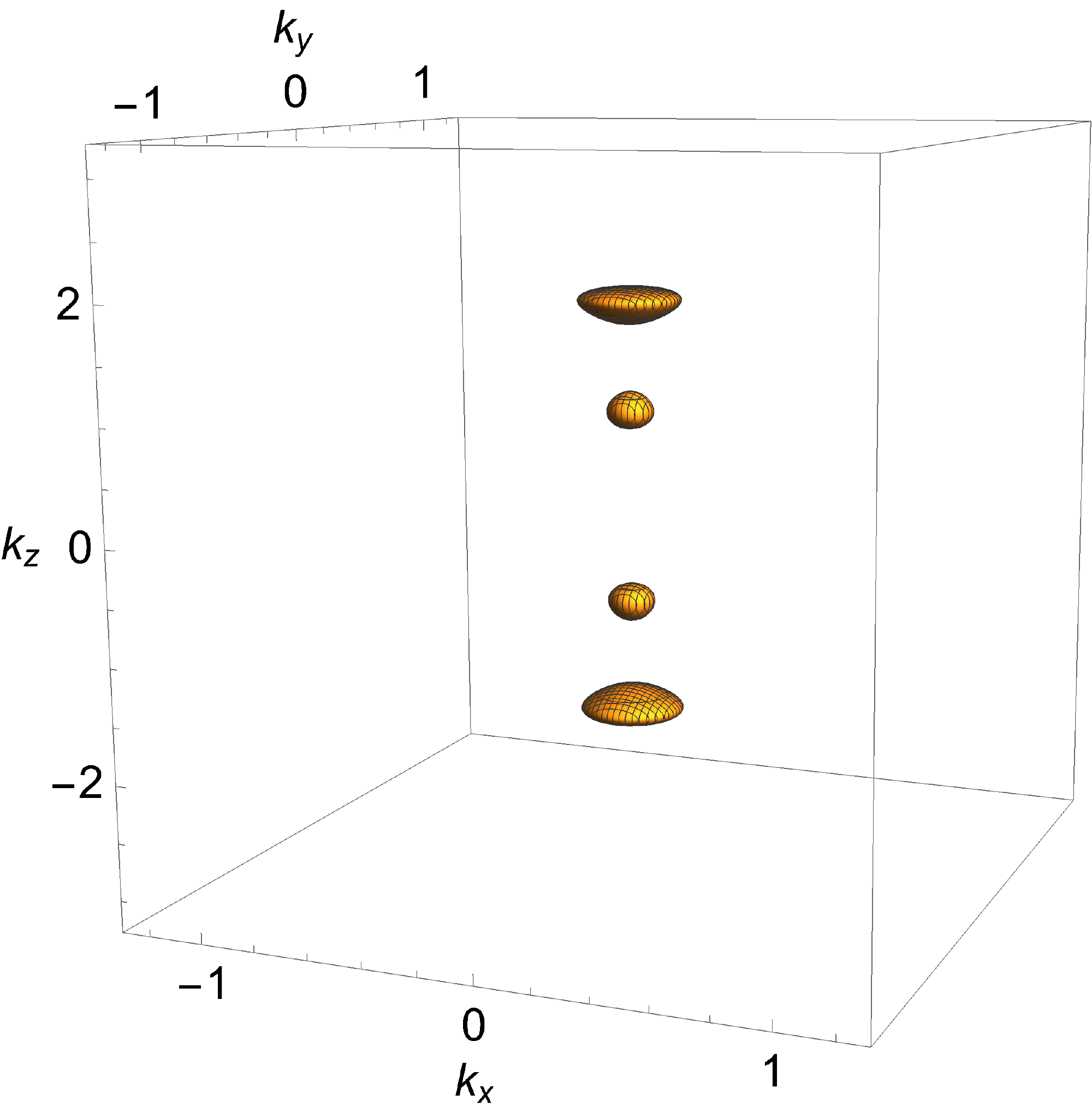}
    \caption{}
    \end{subfigure}
    \begin{subfigure}[t]{0.29\textwidth}
    \centering
    \includegraphics[width=\textwidth]{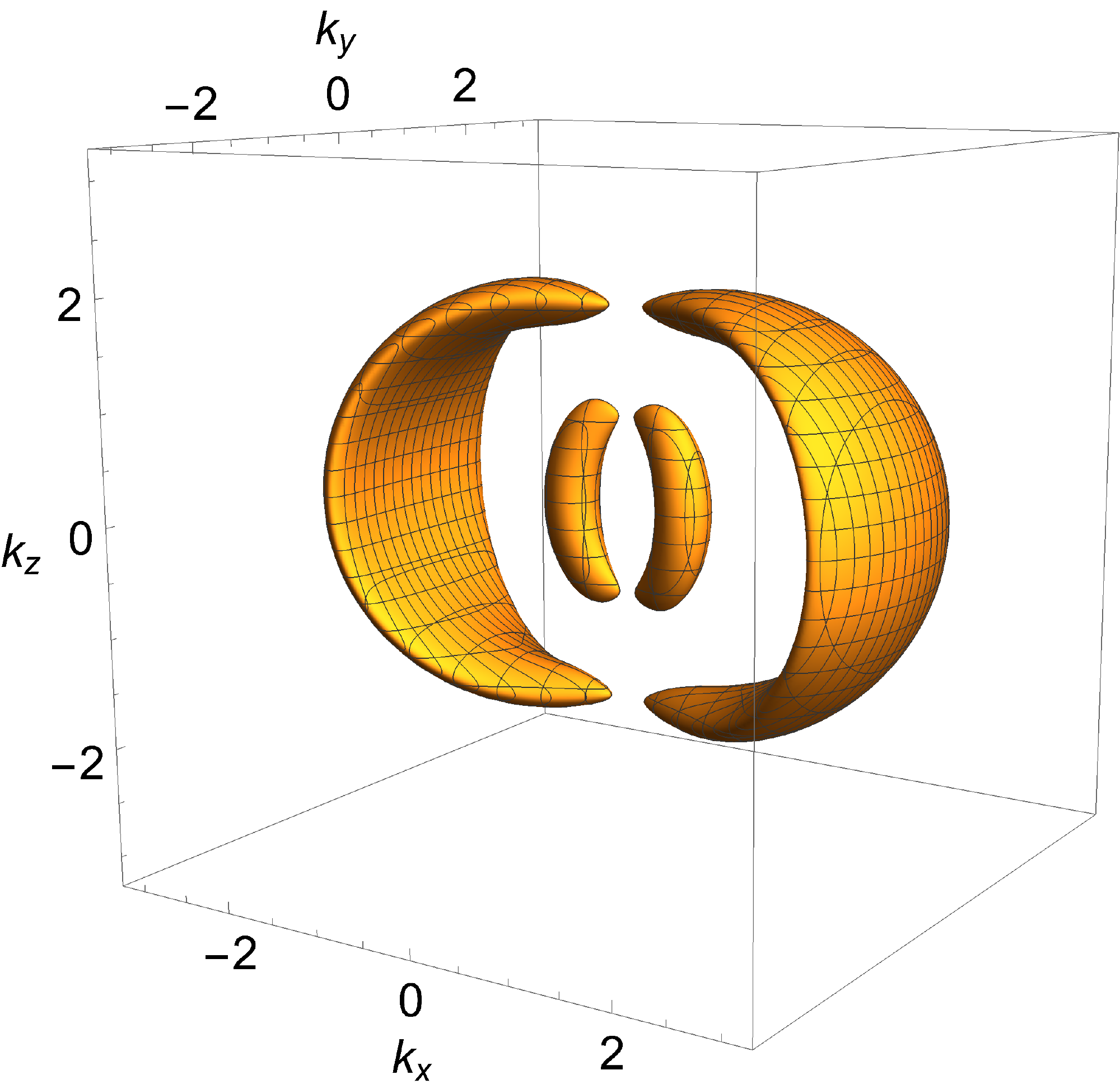}
    \caption{}
    \end{subfigure}
    \begin{subfigure}[t]{0.29\textwidth}
    \centering
    \includegraphics[width=\textwidth]{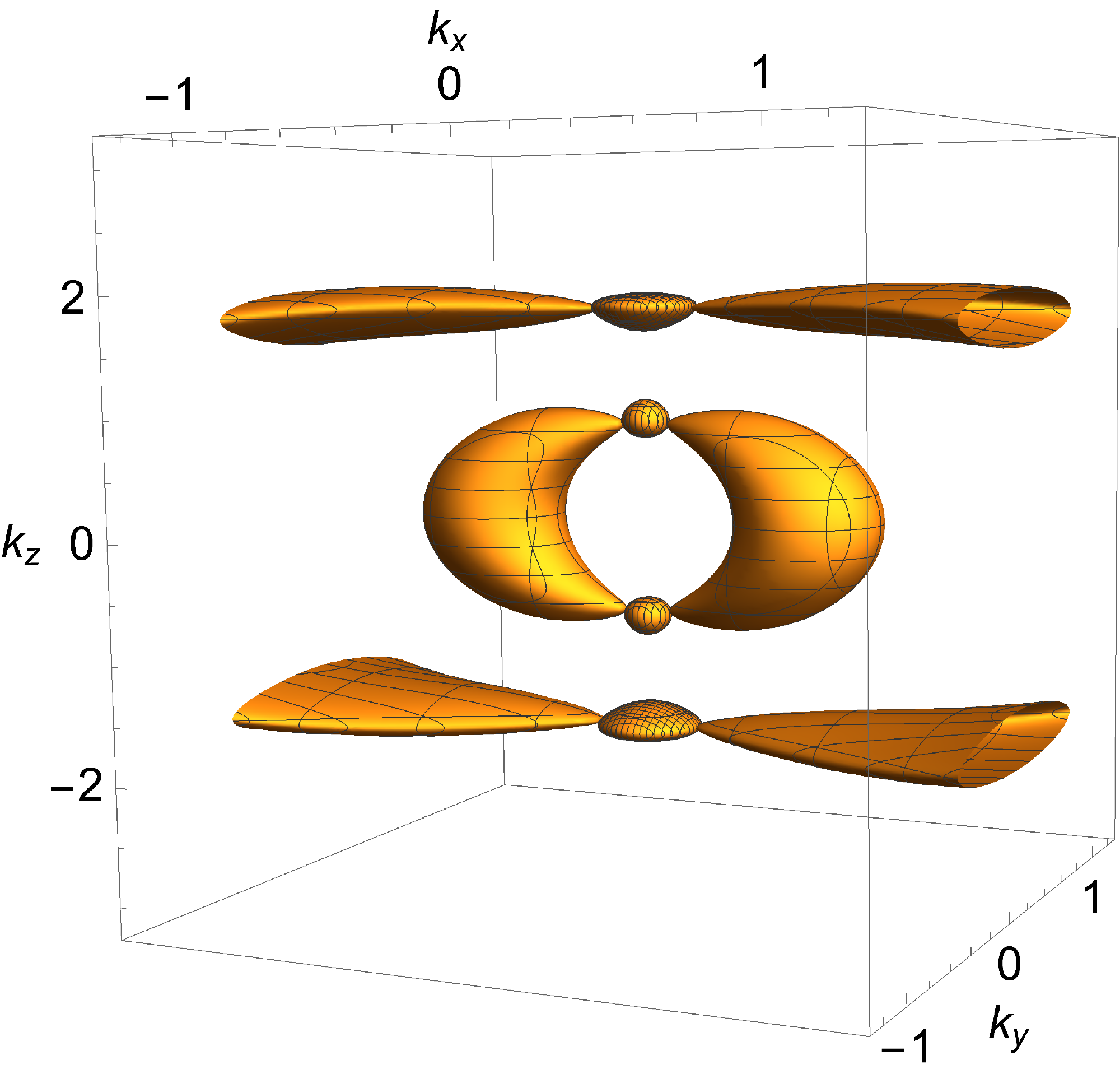}
    \caption{}
    \end{subfigure}
    \caption{(a) Bogoliubov Fermi surface for the mixed singlet-triplet gap function in the parallel magnetic field. (b) Bogoliubov Fermi surface for the mixed singlet-triplet gap function in the overtilted with respect to $\b{P}$ magnetic field. (c) The two plots, (a) and (b), are combined together.}
\label{fig: BFS_pz}
\end{figure*}

Here we illustrate that the presence of a triplet component ($k_z$ dependent) leads to the inflation of the Weyl nodes into Bogoliubov Fermi surfaces. We consider~\cref{eq: BDG hamiltonian} with $\b{\Delta} = i \sigma_y (\Delta + \Delta_1 k_z \sigma_z)$ treating $\Delta$ and $\Delta_1$ as parameters. In~\cref{fig: BFS_pz}, we plot zero-energy surfaces in $\b{k}$-space in the parallel (a) and not parallel (and overtilted) (b) to $\b{P}$ magnetic field for the case of $B>\Delta \gg \Delta_1$. We note that overtilting of the magnetic field produces large Bogoliubov Fermi surfaces.

\section{Additions to the ``Fermi arcs on surfaces and Domain-walls" section of the main text}
\label{app: Appendix_Domains}

The characteristic equation for $\alpha$ in the ansatz solution $\Psi_{\b{k}_{||}}(x) = \Psi_{0,\b{k}_{||}} e^{-\alpha x}$ of~\cref{eq: vacuum boundary problem 1} is
\begin{widetext}
\begin{align}
\label{eq: characteristic for alpha}
&\left[(\epsilon_{\b{k}_{||}}-\frac{\alpha^2}{2m} )^2+\lambda^2 \sin^2 \theta k_y^2-B^2+\Delta^2+\lambda^2 \cos^2\theta k_y^2 + \lambda^2 (\sin \theta k_z - i \cos \theta \alpha)^2\right]^2 - 4 \lambda^2 \bigg[(\epsilon_{\b{k}_{||}}-\frac{\alpha^2}{2m} ) \sin \theta k_y \\ \nn
&+i \big(B_y \cos \theta k_y -  B_x( \sin \theta k_z - i \cos \theta \alpha) \big) \bigg]^2 - 4\bigg( \big((\epsilon_{\b{k}_{||}}-\frac{\alpha^2}{2m} )^2 - B_z^2 \big) [\lambda^2 \cos^2\theta k_y^2 + \lambda^2 (\sin \theta k_z - i \cos \theta \alpha)^2] \\ \nn 
&+ 2 \lambda \sin \theta k_y B_x \big(i (\epsilon_{\b{k}_{||}}-\frac{\alpha^2}{2m} ) \lambda (\sin \theta k_z - i \cos \theta \alpha) -B_z \lambda \cos \theta k_y \big) - 2 i \lambda \sin \theta k_y B_y \big( (\epsilon_{\b{k}_{||}}-\frac{\alpha^2}{2m} ) \lambda \cos \theta k_y \\ \nn
&- i B_z \lambda (\sin \theta k_z - i \cos \theta \alpha) \big)  -\lambda^2 \sin^2 \theta k_y^2 (B_x+B_y)^2  \bigg) = 0.
\end{align}
\end{widetext}
One may view this equation as an equation with real coefficients with respect to $i \alpha$. Thus, its roots are symmetric with respect to the imaginary axis, i.e., if $\alpha$ is a root, then $-\alpha^*$ is also. Therefore,~\cref{eq: characteristic for alpha} can have four roots with positive real part.

In the main text, we showed the existence of the Majorna-Fermi arcs for the case of $B_{||}=0$. Here, we present a solution for an arbitrary $\b{B}$. To find a locus of Majorana zero modes in the $k_y k_z$-plane by substituting the general solution of~\cref{eq: vacuum boundary problem 1} into boundary conditions Eq.~(\ref{eq: vacuum boundary problem 2}) without any assumption is quite difficult. Instead, we check if $k_y=0$, $k_z \in (p_{z2},p_{z1})\bigcup (p_{z4},p_{z3})$ is the locus of zero-energy solutions.

For $k_y=0$ at artibitrary $\b{B}$, as for the case of $B_y=0$, the characteristic equation for $\alpha$, ~\cref{eq: characteristic for alpha}, splits into two simpler equations

\begin{align} \label{eq: characterictic for alpha at ky=0}
    (\epsilon_{k_z}&- \frac{\alpha^2}{2m})^2 - \lambda^2 (k_z \sin \theta - i \alpha \cos \theta)^2 \\&-B^2+\Delta^2 = 
     - 2 i \eta \lambda \sqrt{\Delta^2-B_y^2} (k_z \sin \theta - i \alpha \cos \theta),\nn
\end{align}
where $\eta=\pm1$, and we find 
\begin{align}
v_{k_z,s} = -\left(i \frac{B_y}{\Delta} + \eta\frac{\sqrt{\Delta^2-B_y^2}}{\Delta} \right) u_{k_z,s}.
\end{align}
Again, the problem separates into two sectors corresponding to $\eta=\pm1$, and the further analysis proceeds in analogy to the presented one in the main text.

In the main text, based on the low-energy theory, we pointed out that non-protected Fermi arcs still may exist in the case \textbf{II} of \textit{scenario (i)}, where the Weyl nodes of the same chiralities in two domains   project onto same points on the interface. Here we show that such solution exists in our continuous model.

We choose the coordinate system as described in the main text for the case of  a boundary between the single domain and vacuum with $x-$axis pointing into the first domain, $D_1$. The boundary problem to be solved is

\begin{subnumcases}{}
  H_1(-i \partial_x, \b{k}_{||}) \Psi_{1,\b{k}_{||}}(x)  = 0,  \label{eq: MZM eq D1} \\
  H_2(-i \partial_x, \b{k}_{||}) \Psi_{2,\b{k}_{||}}(x)  = 0,  \label{eq: MZM eq D2} \\
  \Psi_{1,\b{k}_{||}}(0) = \Psi_{2,\b{k}_{||}}(0),\  \partial_x \Psi_{1,\b{k}_{||}}(0) = \partial_x \Psi_{2,\b{k}_{||}}(0),  \label{eq: MZM bc 3}
\end{subnumcases}
where $H_{1(2)}$ and $\Psi_{1(2),\b{k}_{||}}(x)$ are the Hamiltonian and the wavefunction for the first (second) domain.

Again, we are looking the solutions for $k_y=0$ in the form $\Psi_{1(2),\b{k}_{||}}(x) = \Psi_{01(2),\b{k}_{||}} e^{-\alpha_{1(2)} x}$, where $\alpha_{1(2)}$ are determined from~\cref{eq: characterictic for alpha at ky=0}. We consider $\lambda>0$ in the first domain, and the flip of the Weyl node chiralities in the second domain correspond to the flip of the sign of $\lambda$, i.e., $\lambda<0$ in $D_2$. The decaying solutions in $D_{1(2)}$ imply $\alpha_{1(2)}>(<)0$. For $k_z \in (p_{z2},p_{z1})\bigcup (p_{z4},p_{z3})$, there are three $\alpha_1$ with $Re(\alpha_1)>0$ and three $\alpha_2$ with $Re(\alpha_2)<0$ in case of $\eta=1$. For $\eta=-1$, there are one $\alpha_1$ with $Re(\alpha_1)>0$ and one $\alpha_2$ with $Re(\alpha_2)<0$. For $k_z \not\in (p_{z2},p_{z1})\bigcup (p_{z4},p_{z3})$ there are two $\alpha_{1(2)}$ with $Re(\alpha_{1(2)})>(<0)$. Again, boundary condition Eq.~(\ref{eq: MZM bc 3}) imply that we can stitch solutions in $D_1$ and $D_2$ corresponding to the same $\eta=\pm 1$ only, and that the problem separates into two sectors $\eta=\pm 1$. This results in that that the boundary conditions effectively give us four constraints, and together with the normalization condition there are five constraints. For $k_z \in (p_{z2},p_{z1})\bigcup (p_{z4},p_{z3})$, the general solutions for Eqs.~(\ref{eq: MZM eq D1}) and~(\ref{eq: MZM eq D2}) are linear combinations of three functions, which gives us six unknown coefficients to be found. This is one more then the number of constraints we have, which implies that we get a family of solutions parametrized by one parameter, which might be thought of as an angle in two-dimensional vector space. Thus, this set of solutions can be thought as a linear combination of two orthogonal solutions.


\section{Majorana zero modes in the isolated vortex}
\label{app: BdG_vortex}

To obtain MZM in the presence of vortices in the low field regime, we solve the BdG problem in the vicinity of a single vortex. The corresponding BdG Hamiltonian in cylindrical coordinates assumes the form
\begin{align} \label{eq: BDG cylindrical}
   \mathcal{H}(p_z) = \begin{pmatrix}
    \hat{H}_{p_z}(\b{r}) & i \sigma_y \Delta(\b{r}) \\
   [i \sigma_y \Delta(\b{r})]^\dagger & -\hat{H}_{-p_z}(\b{r})^T  \\
    \end{pmatrix},
\end{align}
where the $z$-component of the momentum remains a good quantum number and
\begin{widetext}
\begin{align} 
    \hat{H}_{p_z} = \begin{pmatrix}
    -\frac{1}{2m} (\partial_r^2 + \frac{1}{r} \partial_r - \frac{1}{r^2} \partial_\theta^2) + \epsilon_{p_z} -B_z(r) & \lambda e^{-i \theta} (\partial_r - \frac{i}{r} \partial_\theta) \\
    -\lambda e^{i \theta} (\partial_r + \frac{i}{r} \partial_\theta) & -\frac{1}{2m} (\partial_r^2 + \frac{1}{r} \partial_r +\frac{1}{r^2} \partial_\theta^2) + \epsilon_{p_z} +B_z(r)  \\
    \end{pmatrix}\,.
\end{align}
\end{widetext}
Here $\epsilon_{p_z}=\frac{p_z^2}{2m}  - \mu$ and we have neglected the coupling to the vector potential~\cite{Caroli1964}.
The phase of the order parameter winds by $2 \pi$ around the vortex origin, $\Delta(\b{r}) = \Delta_0(r) e^{i \phi}$. The cylindrical form of~\cref{eq: BDG cylindrical} suggests to look for the energy eigenstates in the form

\begin{align} \label{eq: vortex wavefunction represantation}
    \Psi_{p_zE}(\b{r}) = \sum_{l=-\infty}^\infty a_{lE} e^{i l \theta}\begin{pmatrix} \Psi_{1El}(r) \\
    \Psi_{2El}(r) e^{i \theta} \\ \Psi_{3El}(r) \\ \Psi_{4El}(r) e^{-i \theta}
    \end{pmatrix} = \sum_{l-\infty}^\infty a_l \Psi_{El} (r,\theta).
\end{align}



Following Ref.~\cite{Sau2010}, 
in searching for the Majorana modes, we  focus on the $l=0$ channel, which is also justified by our numerical calculations. Substitution of~\cref{eq: vortex wavefunction represantation} into~\cref{eq: BDG cylindrical} leads to a system of ordinary differential equations (ODE) with real coefficients, and thus the functions $\Psi_{iEl}(r), i=1..4$ in~\cref{eq: vortex wavefunction represantation} are real. For $l=0$, the particle-hole  symmetry implies $\sigma_x \otimes \sigma_0 \Psi_{p_zE0}(\b{r})^* = \eta \Psi_{-p_z-E0}(\b{r})$, where $\eta$ is a phase-factor. Given that the Hamiltonian~\cref{eq: BDG cylindrical} is even in $p_z$, we obtain $\sigma_x \otimes \sigma_0 \Psi_{p_zE0}(\b{r})^* = \eta \Psi_{p_z-E0}(\b{r})$.  Combining this with the statement about the reality of $\Psi_i(r)$, we conclude $\eta=\pm 1$ and $\Psi_{3(4)E0}=\eta \Psi_{1(2)-E0}$. Although we anticipate the splitting in energy due to overlapping of Majorana states at $r=0$ and $r=r_h$, we start with seeking the zero-energy solution. In the following, we drop the subindices for the putative $E=0,\; l=0$ state and use $\Psi_i \equiv \Psi_{i00}$, and thus we have $\Psi_{3(4)}=\eta \Psi_{1(2)}$. Then the zero-energy eigenstate equation for BDG~\cref{eq: BDG cylindrical} reduces to the system of two ODEs
\begin{widetext}
\begin{align}\label{eq: ODEs}
   \begin{pmatrix}
    \epsilon_{p_z}-\frac{1}{2m} (\partial_r^2 + \frac{1}{r} \partial_r )  - B_z(r) & \lambda (\partial_r + \frac{1}{r}) + \eta \Delta_0(r) \\
    - \lambda \partial_r - \eta \Delta_0(r) & 
   \epsilon_{p_z}-\frac{1}{2m} (\partial_r^2 + \frac{1}{r} \partial_r - \frac{1}{r^2})  + B_z(r)
    \end{pmatrix}
    \times\Psi (r)=0,
\end{align}
\end{widetext}
where $\Psi(r)=(\Psi_1(r),\Psi_2(r))^T$.

In what follows, we first present an analytic analysis of Eq.~\eqref{eq: ODEs} for a simplified piece-wise constant model. Then in the next step we present numerical analysis for more realistic profiles of the gap and magnetic field, which continuously vary in space. 

The gap structure in the  simplified model constitutes of three regions
\begin{align} \label{eq: ProfileDelta_CrudeApproximation}
    \Delta_0(r)=\begin{cases} 0, \ \ \ \ \ 0 \leq r < r_1, \\
                            \Delta_1, \ \ \ \ \   r_1 \leq r < r_2, \\
                            \Delta_2, \ \ \ \ \  r \geq r_2,
    \end{cases}
\end{align}
We also assume the magnetic field is uniform (justified by the type II condition 
$\lambda_L\gg\xi$). We then focus on the limit $\Delta_1 < B_z$, in which case the intermediate region is ``topological".

In the region $0<r<r_1$, where $\Delta(r) \equiv 0$, we look for the solution in the form~\cite{Sau2010}

\begin{align}
    \begin{pmatrix}
    \Psi_1(r) \\
    \Psi_2(r)
    \end{pmatrix} = \begin{pmatrix}
    a J_0(\alpha r) \\
    b J_1(\alpha r)
    \end{pmatrix}
\end{align}
where $J_n(z)$ are Bessel functions.
Substituting this into~\cref{eq: ODEs}, we find a characteristic equation for $\alpha$

\begin{align}
    (\frac{\alpha^2}{2m}  + \epsilon_{p_z})^2 - B_z^2 - \lambda^2 \alpha^2 = 0,
\end{align}
which has four solutions: $\pm \alpha_1$ and  $\pm \alpha_2$. Thus, the general solution in this region is

\begin{align}
    \begin{pmatrix}
    \Psi_1(r) \\
    \Psi_2(r)
    \end{pmatrix} = C_1 \begin{pmatrix}
    a_1 J_0(\alpha_1 r) \\
    b_1 J_1(\alpha_1 r)
    \end{pmatrix} + C_2 \begin{pmatrix}
    a_2 J_0(\alpha_2 r) \\
    b_2 J_1(\alpha_2 r)
    \end{pmatrix}.
\end{align}
For the regions $r_1<r<r_2$ and $r>r_2$, where $\Delta_0(r) \neq 0$, we look for the solution in the form~\cite{Sau2010}
\begin{align}
    \begin{pmatrix}
    \Psi_1(r) \\
    \Psi_2(r)
    \end{pmatrix} = \frac{e^{i q r}}{\sqrt{r}} \sum_{n=0}^\infty \frac{1}{r^n} \begin{pmatrix}
    a_n \\
    b_n
    \end{pmatrix}
\end{align}
and get a set of algebraic equations for the coefficients $(a_n, b_n)$. For $n=0$, we obtain

\begin{align}
\begin{cases}
    (\frac{q^2}{2m} +\epsilon_{p_z} -B_z) a_0 +(i q \lambda + \eta \Delta) b_0 = 0,\\ \nn
    (-i q \lambda - \eta \Delta) a_0 + (\frac{q^2}{2m} + \epsilon_{p_z}+ B_z) b_0=0,
\end{cases}
\end{align}
where $\Delta$ stands for either $\Delta_1$ or $\Delta_2$ depending on the region under consideration. This gives the following equation for $\Tilde{q} = -i q$
\begin{align}
    \label{eq: Charcter eq for beta2}
    \frac{\Tilde{q}^4}{(2m)^2}  - 2\left( \frac{\epsilon_{p_z}}{2m} - \lambda^2\right) \Tilde{q}^2 - 2 \lambda \eta \Delta \Tilde{q} +\epsilon_{p_z}^2 - B_z^2 +\Delta^2=0.
\end{align}
The roots $\Tilde{q_i}$ of this equation satisfy the condition

\begin{align} \label{eq: number of coeffs criterion}
    \prod_{i=1}^4 \Tilde{q_i} = \epsilon_{p_z}^2 - B_z^2 +\Delta^2.
\end{align}

For the region $r>r_2$, the decaying solutions correspond to such $\Tilde{q}$  that $\Re(\Tilde{q})>0$. For $\prod_{i=1}^4 \Tilde{q_i}>0$, there are two such roots for either $\eta$; for $\prod_{i=1}^4 \Tilde{q_i}<0$, there are three such roots for $\eta=-1$, and one such root for $\eta=1$.


Two boundaries (at $r_1$ and $r_2$) with smooth continuity conditions for a two-component vector and one normalization condition bring nine conditions, in total. Now we count the number of yet unknown coefficients in the constructed solution to be obtained from these conditions focusing on the case $\prod_{i=1}^4\Tilde{q}_i > 0$ for the region $r>r_2$ for all $p_z$ (which correspond to $\Delta_2>B_z$), where the middle region (the halo) $r_1<r<r_2$ is in the topological phase, while the outer and inner regions are trivial. In the region $0 \leq r<r_1$, there are two coefficients; in the region $r_1 \leq r<r_2$, there are four coefficients; and in the region $r>r_2$, there are two coefficients. This brings in total eight coefficients, which is not enough to satisfy nine conditions. Thus, there is no zero-energy solutions. In fact, this is anticipated and corresponds to the overlapping of two Majorana states at $r_1$ and $r_2$. Specifically, removing the ``domain" wall at $r_2$ (or moving it to infinity), at the boundary $r=r_1$ we have to satisfy only five conditions at $r=r_1$. In this case, for $r>r_1$ we have to single out only decaying at infinity solutions, which gives three coefficients in this region. And it total we have five coefficients to satisfy five conditions. Analogously, moving $r_2$ to infinity, and requiring that physical solutions decay far away from $r_2$ in the topological phase, we look for such $\Tilde{q}$ in~\cref{eq: number of coeffs criterion} that $\Re(\Tilde{q})<0$. There is one such root for $\eta=-1$ sector, and three such roots for $\eta=1$. Then, in $\eta=1$ sector we again have equal number of constraints and coefficients. As we move $r_2$ from large distance closer to $r_1$, the overlapping of the two Majorana modes leads to splitting in energy of the states constructed out of the linear combinations of these Majoranas. 
Alternatively, we can think in the following way. For the case $B_z>\Delta_2$, we have one Majorana zero mode localized around the vortex core, which is essentially the case considered in Ref.~\cite{Sau2010}. But as we decrease $B_z$ to values just below $\Delta_2$, we lose the zero-energy solution. The only way it can happen is via pairing the Majorana zero mode at the vortex core with another one at $r=r_2$. We also note that while here we considered a crudely discretized model, the argument presented extends to the arbitrary fine discretization. Indeed, the introduction of a new segment brings in four new boundary conditions and, at the same time, four new constants to be found, thus leaving the balance between the number of conditions and the number of coefficients untouched.

However, because the halo has finite size it is essential to estimate the  splitting of the zero modes due to their overlap.
To this end, at zero temperature, we assume the separation between the two boundaries $r_2 - r_1 \sim \xi_0$ is on the order of the coherence length $\xi_0 = \frac{v_F}{\pi \Delta_2}$~\footnote{This is true only close to $H_{c2}$.}. This length should be compared with the localization length of the zero modes, ${l}_M$, which can be  estimated from the low-energy effective Hamiltonian~\cref{eq: 2D Weyl Hamiltonian} (for $\b{B} = B \hat{z}$)
\begin{align}
    H_{eff} = v k_y \sigma_x + v k_y \sigma_y + E_g \sigma_z.
\end{align}
where $E_g$ is the gap at a given $k_z$ away from the Weyl point. 
We then find that  $l_M\sim\frac{v}{E_g}$. Comparing with~\cref{eq: 2D Weyl Hamiltonian}, we find that $v=\frac{\lambda \Delta}{B}$, $E_g = \frac{p_z \epsilon_{\b{p}} k_z}{2mB}$ yielding
\begin{align}
    l_M\sim \frac{\lambda \Delta}{(p_z k_z/2m)\sqrt{B^2 - \Delta^2}}.
\end{align}
We then  evaluate $l_M$ for $k_z$ located at the middle point between the two Weyl points under assumption $\mu \gg \sqrt{B^2 - \Delta^2}$. Focusing on the region $r_1<r<r_2$, we find that $\frac{l_M}{\xi_0} \sim 2\pi \frac{\lambda}{v_F} \frac{\Delta_1 \Delta_2}{B^2-\Delta_1^2}$. Thus, the ratio of the length scales is controlled by the small parameter $\lambda/v_F$ and is therefore expected to be very small except for very close to the nodes or close to the transition point. 


To confirm our analytical considerations, we perform numerical calculations. For simplicity, we consider a cylinder of a radius $R$ with a single vortex located at the axis of the cylinder and impose zero boundary conditions at $r=R$. 
For $r<R$, we assume a radial dependence of the order parameter given by the function $\Delta (r) = \Delta_0 \tanh(r/\xi)$, which reflects a typical behavior in a vortex core center.
Also, we assume the magnetic field is uniform, and smaller than the bulk threshold  $B_z<\Delta_0$.

We represent the radial part of the spinor $\Psi_{p_zEl}(\b{r})$ in the Bessel-Fourier series form

\begin{align}
    \Psi_{p_zEl} (r,\theta) = e^{i l \theta}\sum_{i=1}^\infty \begin{pmatrix}
    a_i J_l (\mu_i^{l} \frac{r}{R}) \\
    b_i J_{l+1} (\mu_i^{l+1} \frac{r}{R}) e^{i \theta} \\
    c_i J_l (\mu_i^{l} \frac{r}{R}) \\
    d_i J_{l-1} (\mu_i^{l-1} \frac{r}{R}) e^{-i \theta},
    \end{pmatrix}
\end{align}
where ${\mu_i^{l}}$ is the set of roots of the equation $J_l (\mu_i^{l})=0$, which guarantees that the boundary conditions are  satisfied. Substituting this representation into~\cref{eq: BDG cylindrical} and projecting onto $J_\nu (\mu_i \frac{r}{R}), \nu={l-1,l,l+1}$, we obtain an infinite system of algebraic equations, which is solved approximately by truncation. In the calculations used for producing plots in this article, we cut the system of algebraic equations at size $600\cross600$.

\begin{figure}[!htbp]
    \centering
    \includegraphics[width=0.48\textwidth]{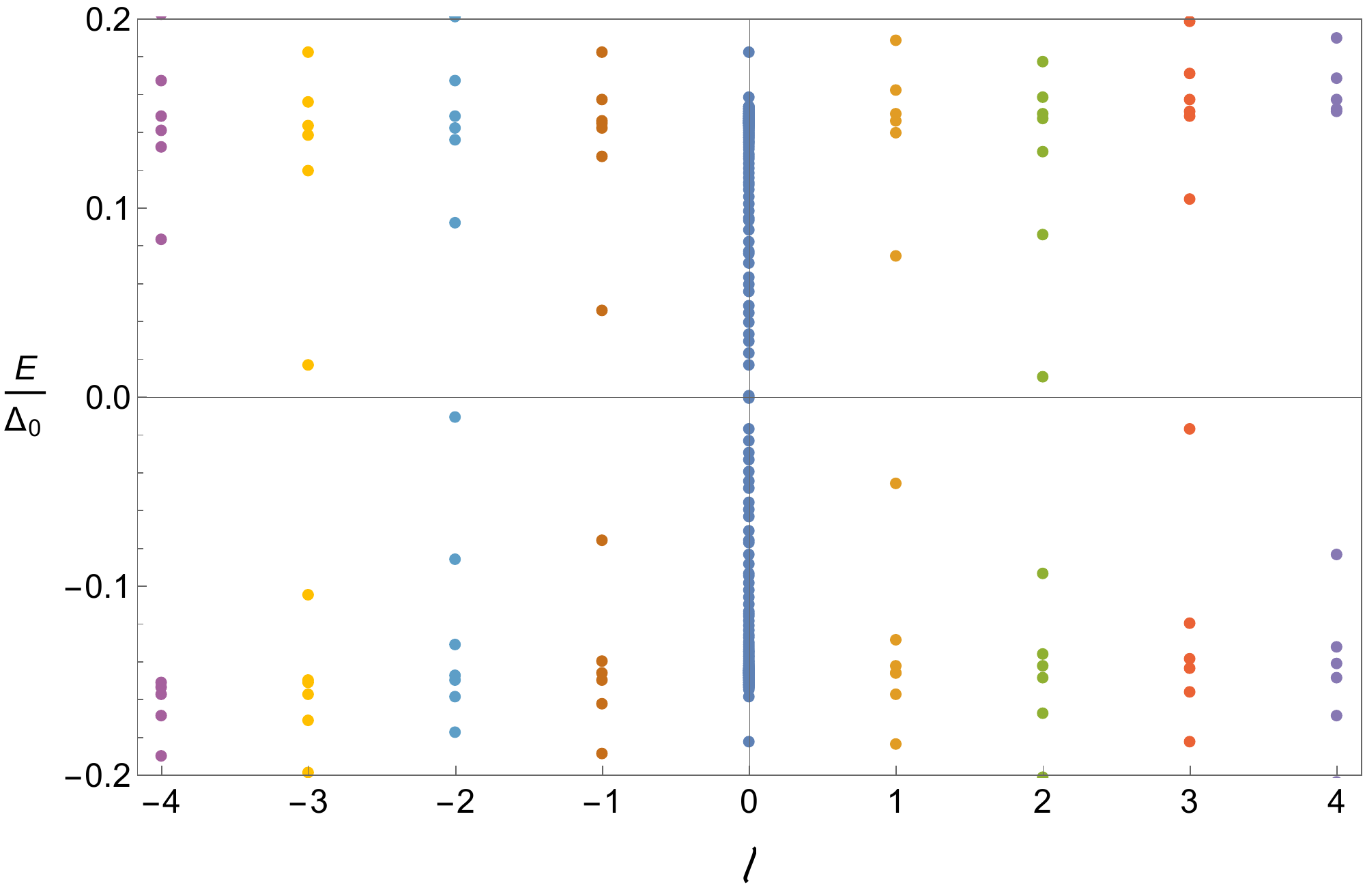}
    \caption{The energy spectrum of the single vortex problem. On the horizontal axis, $l$ is the angular momentum channel. The parameters used for the simulation are: $m=1,\; \Delta_0=2,\; B_z=1.71,\; \lambda=0.1,\; \mu=10,\; \xi=100,\; R=700$, and $p_z$ is chosen such that $\epsilon_{p_z}=0$. }
\label{fig: Vortex energies}
\end{figure}

We plot eigenenergies corresponding to the wavefunctions $\Psi_{El} (r,\theta)$ in~\cref{fig: Vortex energies}. 
Under PHS, $l \rightarrow -l$ and $E\rightarrow -E$. Thus, in fact, for $l=0$, there are two near-zero energy solutions (for the parameters considered, these energies are on the order of $10^{-4} \cdot \Delta_0$) that are indistinguishable in the plot and correspond to the states that are linear combinations of Majorana zero modes. 

\bibliography{library}

\begin{thebibliography}{86}%
\makeatletter
\providecommand \@ifxundefined [1]{%
 \@ifx{#1\undefined}
}%
\providecommand \@ifnum [1]{%
 \ifnum #1\expandafter \@firstoftwo
 \else \expandafter \@secondoftwo
 \fi
}%
\providecommand \@ifx [1]{%
 \ifx #1\expandafter \@firstoftwo
 \else \expandafter \@secondoftwo
 \fi
}%
\providecommand \natexlab [1]{#1}%
\providecommand \enquote  [1]{``#1''}%
\providecommand \bibnamefont  [1]{#1}%
\providecommand \bibfnamefont [1]{#1}%
\providecommand \citenamefont [1]{#1}%
\providecommand \href@noop [0]{\@secondoftwo}%
\providecommand \href [0]{\begingroup \@sanitize@url \@href}%
\providecommand \@href[1]{\@@startlink{#1}\@@href}%
\providecommand \@@href[1]{\endgroup#1\@@endlink}%
\providecommand \@sanitize@url [0]{\catcode `\\12\catcode `\$12\catcode
  `\&12\catcode `\#12\catcode `\^12\catcode `\_12\catcode `\%12\relax}%
\providecommand \@@startlink[1]{}%
\providecommand \@@endlink[0]{}%
\providecommand \url  [0]{\begingroup\@sanitize@url \@url }%
\providecommand \@url [1]{\endgroup\@href {#1}{\urlprefix }}%
\providecommand \urlprefix  [0]{URL }%
\providecommand \Eprint [0]{\href }%
\providecommand \doibase [0]{http://dx.doi.org/}%
\providecommand \selectlanguage [0]{\@gobble}%
\providecommand \bibinfo  [0]{\@secondoftwo}%
\providecommand \bibfield  [0]{\@secondoftwo}%
\providecommand \translation [1]{[#1]}%
\providecommand \BibitemOpen [0]{}%
\providecommand \bibitemStop [0]{}%
\providecommand \bibitemNoStop [0]{.\EOS\space}%
\providecommand \EOS [0]{\spacefactor3000\relax}%
\providecommand \BibitemShut  [1]{\csname bibitem#1\endcsname}%
\let\auto@bib@innerbib\@empty
\bibitem [{\citenamefont {Alicea}(2012)}]{alicea2012new}%
  \BibitemOpen
  \bibfield  {author} {\bibinfo {author} {\bibfnamefont {Jason}\ \bibnamefont
  {Alicea}},\ }\bibfield  {title} {\enquote {\bibinfo {title} {New directions
  in the pursuit of majorana fermions in solid state systems},}\ }\href
  {https://iopscience.iop.org/article/10.1088/0034-4885/75/7/076501/meta}
  {\bibfield  {journal} {\bibinfo  {journal} {Rep. Prog. Phys}\ }\textbf
  {\bibinfo {volume} {75}},\ \bibinfo {pages} {076501} (\bibinfo {year}
  {2012})}\BibitemShut {NoStop}%
\bibitem [{\citenamefont {Ando}\ and\ \citenamefont
  {Fu}(2015)}]{ando2015topological}%
  \BibitemOpen
  \bibfield  {author} {\bibinfo {author} {\bibfnamefont {Yoichi}\ \bibnamefont
  {Ando}}\ and\ \bibinfo {author} {\bibfnamefont {Liang}\ \bibnamefont {Fu}},\
  }\bibfield  {title} {\enquote {\bibinfo {title} {Topological crystalline
  insulators and topological superconductors: From concepts to materials},}\
  }\href {https://doi.org/10.1146/annurev-conmatphys-031214-014501} {\bibfield
  {journal} {\bibinfo  {journal} {Annual Review of Condensed Matter Physics}\
  }\textbf {\bibinfo {volume} {6}},\ \bibinfo {pages} {361--381} (\bibinfo
  {year} {2015})}\BibitemShut {NoStop}%
\bibitem [{\citenamefont {Lutchyn}\ \emph {et~al.}(2018)\citenamefont
  {Lutchyn}, \citenamefont {Bakkers}, \citenamefont {Kouwenhoven},
  \citenamefont {Krogstrup}, \citenamefont {Marcus},\ and\ \citenamefont
  {Oreg}}]{lutchyn2018majorana}%
  \BibitemOpen
  \bibfield  {author} {\bibinfo {author} {\bibfnamefont {Roman~M.}\
  \bibnamefont {Lutchyn}}, \bibinfo {author} {\bibfnamefont {Erik P. A.~M.}\
  \bibnamefont {Bakkers}}, \bibinfo {author} {\bibfnamefont {Leo~P.}\
  \bibnamefont {Kouwenhoven}}, \bibinfo {author} {\bibfnamefont {Peter}\
  \bibnamefont {Krogstrup}}, \bibinfo {author} {\bibfnamefont {Charles~M.}\
  \bibnamefont {Marcus}}, \ and\ \bibinfo {author} {\bibfnamefont {Yuval}\
  \bibnamefont {Oreg}},\ }\bibfield  {title} {\enquote {\bibinfo {title}
  {Majorana zero modes in superconductor--semiconductor heterostructures},}\
  }\href {https://doi.org/10.1038/s41578-018-0003-1} {\bibfield  {journal}
  {\bibinfo  {journal} {Nat. Rev. Mater.}\ }\textbf {\bibinfo {volume} {3}},\
  \bibinfo {pages} {52--68} (\bibinfo {year} {2018})}\BibitemShut {NoStop}%
\bibitem [{\citenamefont {Mackenzie}\ \emph {et~al.}(2017)\citenamefont
  {Mackenzie}, \citenamefont {Scaffidi}, \citenamefont {Hicks},\ and\
  \citenamefont {Maeno}}]{mackenzie2017even}%
  \BibitemOpen
  \bibfield  {author} {\bibinfo {author} {\bibfnamefont {Andrew~P}\
  \bibnamefont {Mackenzie}}, \bibinfo {author} {\bibfnamefont {Thomas}\
  \bibnamefont {Scaffidi}}, \bibinfo {author} {\bibfnamefont {Clifford~W}\
  \bibnamefont {Hicks}}, \ and\ \bibinfo {author} {\bibfnamefont {Yoshiteru}\
  \bibnamefont {Maeno}},\ }\bibfield  {title} {\enquote {\bibinfo {title} {Even
  odder after twenty-three years: The superconducting order parameter puzzle of
  {Sr}$_2${RuO}$_4$},}\ }\href {https://doi.org/10.1038/s41535-017-0045-4}
  {\bibfield  {journal} {\bibinfo  {journal} {npj Quant. Mater.}\ }\textbf
  {\bibinfo {volume} {2}},\ \bibinfo {pages} {1--9} (\bibinfo {year}
  {2017})}\BibitemShut {NoStop}%
\bibitem [{\citenamefont {Yu}\ \emph {et~al.}(2021)\citenamefont {Yu},
  \citenamefont {Chen}, \citenamefont {Gomanko}, \citenamefont {Badawy},
  \citenamefont {Bakkers}, \citenamefont {Zuo}, \citenamefont {Mourik},\ and\
  \citenamefont {Frolov}}]{Yu2021}%
  \BibitemOpen
  \bibfield  {author} {\bibinfo {author} {\bibfnamefont {P.}~\bibnamefont
  {Yu}}, \bibinfo {author} {\bibfnamefont {J.}~\bibnamefont {Chen}}, \bibinfo
  {author} {\bibfnamefont {M.}~\bibnamefont {Gomanko}}, \bibinfo {author}
  {\bibfnamefont {G.}~\bibnamefont {Badawy}}, \bibinfo {author} {\bibfnamefont
  {E.~P. A.~M.}\ \bibnamefont {Bakkers}}, \bibinfo {author} {\bibfnamefont
  {K.}~\bibnamefont {Zuo}}, \bibinfo {author} {\bibfnamefont {V.}~\bibnamefont
  {Mourik}}, \ and\ \bibinfo {author} {\bibfnamefont {S.~M.}\ \bibnamefont
  {Frolov}},\ }\bibfield  {title} {\enquote {\bibinfo {title} {Non-majorana
  states yield nearly quantized conductance in proximatized nanowires},}\
  }\href {\doibase 10.1038/s41567-020-01107-w} {\bibfield  {journal} {\bibinfo
  {journal} {Nat. Phys.}\ }\textbf {\bibinfo {volume} {17}},\ \bibinfo {pages}
  {482–488} (\bibinfo {year} {2021})}\BibitemShut {NoStop}%
\bibitem [{\citenamefont {Frolov}\ and\ \citenamefont
  {Mourik}(2022)}]{frolov2022believe}%
  \BibitemOpen
  \bibfield  {author} {\bibinfo {author} {\bibfnamefont {Sergey}\ \bibnamefont
  {Frolov}}\ and\ \bibinfo {author} {\bibfnamefont {Vincent}\ \bibnamefont
  {Mourik}},\ }\href@noop {} {\enquote {\bibinfo {title} {We cannot believe we
  overlooked these {Majorana} discoveries},}\ } (\bibinfo {year} {2022}),\
  \Eprint {http://arxiv.org/abs/2203.17060} {arXiv:2203.17060
  [cond-mat.mes-hall]} \BibitemShut {NoStop}%
\bibitem [{\citenamefont {Fu}\ and\ \citenamefont
  {Kane}(2008)}]{fu2008superconducting}%
  \BibitemOpen
  \bibfield  {author} {\bibinfo {author} {\bibfnamefont {Liang}\ \bibnamefont
  {Fu}}\ and\ \bibinfo {author} {\bibfnamefont {Charles~L.}\ \bibnamefont
  {Kane}},\ }\bibfield  {title} {\enquote {\bibinfo {title} {Superconducting
  proximity effect and majorana fermions at the surface of a topological
  insulator},}\ }\href
  {https://link.aps.org/doi/10.1103/PhysRevLett.100.096407} {\bibfield
  {journal} {\bibinfo  {journal} {Phys. Rev. Lett.}\ }\textbf {\bibinfo
  {volume} {100}},\ \bibinfo {pages} {096407} (\bibinfo {year}
  {2008})}\BibitemShut {NoStop}%
\bibitem [{\citenamefont {Sau}\ \emph {et~al.}(2010)\citenamefont {Sau},
  \citenamefont {Lutchyn}, \citenamefont {Tewari},\ and\ \citenamefont
  {Sarma}}]{Sau2010}%
  \BibitemOpen
  \bibfield  {author} {\bibinfo {author} {\bibfnamefont {Jay~D.}\ \bibnamefont
  {Sau}}, \bibinfo {author} {\bibfnamefont {Roman~M.}\ \bibnamefont {Lutchyn}},
  \bibinfo {author} {\bibfnamefont {Sumanta}\ \bibnamefont {Tewari}}, \ and\
  \bibinfo {author} {\bibfnamefont {S.~Das}\ \bibnamefont {Sarma}},\ }\bibfield
   {title} {\enquote {\bibinfo {title} {Generic new platform for topological
  quantum computation using semiconductor heterostructures},}\ }\href {\doibase
  10.1103/PhysRevLett.104.040502} {\bibfield  {journal} {\bibinfo  {journal}
  {Phys. Rev. Lett.}\ }\textbf {\bibinfo {volume} {104}},\ \bibinfo {pages}
  {040502} (\bibinfo {year} {2010})}\BibitemShut {NoStop}%
\bibitem [{\citenamefont {Lutchyn}\ \emph {et~al.}(2010)\citenamefont
  {Lutchyn}, \citenamefont {Sau},\ and\ \citenamefont
  {Das~Sarma}}]{Lutchyn2011}%
  \BibitemOpen
  \bibfield  {author} {\bibinfo {author} {\bibfnamefont {Roman~M.}\
  \bibnamefont {Lutchyn}}, \bibinfo {author} {\bibfnamefont {Jay~D.}\
  \bibnamefont {Sau}}, \ and\ \bibinfo {author} {\bibfnamefont
  {S.}~\bibnamefont {Das~Sarma}},\ }\bibfield  {title} {\enquote {\bibinfo
  {title} {Majorana fermions and a topological phase transition in
  semiconductor-superconductor heterostructures},}\ }\href {\doibase
  10.1103/PhysRevLett.105.077001} {\bibfield  {journal} {\bibinfo  {journal}
  {Phys. Rev. Lett.}\ }\textbf {\bibinfo {volume} {105}},\ \bibinfo {pages}
  {077001} (\bibinfo {year} {2010})}\BibitemShut {NoStop}%
\bibitem [{\citenamefont {Oreg}\ \emph {et~al.}(2010)\citenamefont {Oreg},
  \citenamefont {Refael},\ and\ \citenamefont {von Oppen}}]{Oreg2011}%
  \BibitemOpen
  \bibfield  {author} {\bibinfo {author} {\bibfnamefont {Yuval}\ \bibnamefont
  {Oreg}}, \bibinfo {author} {\bibfnamefont {Gil}\ \bibnamefont {Refael}}, \
  and\ \bibinfo {author} {\bibfnamefont {Felix}\ \bibnamefont {von Oppen}},\
  }\bibfield  {title} {\enquote {\bibinfo {title} {Helical liquids and
  {Majorana} bound states in quantum wires},}\ }\href {\doibase
  10.1103/PhysRevLett.105.177002} {\bibfield  {journal} {\bibinfo  {journal}
  {Phys. Rev. Lett.}\ }\textbf {\bibinfo {volume} {105}},\ \bibinfo {pages}
  {177002} (\bibinfo {year} {2010})}\BibitemShut {NoStop}%
\bibitem [{\citenamefont {Potter}\ and\ \citenamefont
  {Lee}(2012)}]{Potter2012}%
  \BibitemOpen
  \bibfield  {author} {\bibinfo {author} {\bibfnamefont {Andrew~C.}\
  \bibnamefont {Potter}}\ and\ \bibinfo {author} {\bibfnamefont {Patrick~A.}\
  \bibnamefont {Lee}},\ }\bibfield  {title} {\enquote {\bibinfo {title}
  {Topological superconductivity and {Majorana} fermions in metallic surface
  states},}\ }\href {\doibase 10.1103/PhysRevB.85.094516} {\bibfield  {journal}
  {\bibinfo  {journal} {Phys. Rev. B}\ }\textbf {\bibinfo {volume} {85}},\
  \bibinfo {pages} {094516} (\bibinfo {year} {2012})}\BibitemShut {NoStop}%
\bibitem [{\citenamefont {Agterberg}(2003)}]{agterberg2003novel}%
  \BibitemOpen
  \bibfield  {author} {\bibinfo {author} {\bibfnamefont {D.~F.}\ \bibnamefont
  {Agterberg}},\ }\bibfield  {title} {\enquote {\bibinfo {title} {Novel
  magnetic field effects in unconventional superconductors},}\ }\href {\doibase
  https://doi.org/10.1016/S0921-4534(03)00634-8} {\bibfield  {journal}
  {\bibinfo  {journal} {Physica C: Superconductivity}\ }\textbf {\bibinfo
  {volume} {387}},\ \bibinfo {pages} {13--16} (\bibinfo {year}
  {2003})}\BibitemShut {NoStop}%
\bibitem [{\citenamefont {Dimitrova}\ and\ \citenamefont
  {Feigel'man}(2007)}]{Dimitrova2007}%
  \BibitemOpen
  \bibfield  {author} {\bibinfo {author} {\bibfnamefont {Ol'ga}\ \bibnamefont
  {Dimitrova}}\ and\ \bibinfo {author} {\bibfnamefont {M.~V.}\ \bibnamefont
  {Feigel'man}},\ }\bibfield  {title} {\enquote {\bibinfo {title} {Theory of a
  two-dimensional superconductor with broken inversion symmetry},}\ }\href
  {\doibase 10.1103/PhysRevB.76.014522} {\bibfield  {journal} {\bibinfo
  {journal} {Phys. Rev. B}\ }\textbf {\bibinfo {volume} {76}},\ \bibinfo
  {pages} {014522} (\bibinfo {year} {2007})}\BibitemShut {NoStop}%
\bibitem [{\citenamefont {Michaeli}\ \emph {et~al.}(2012)\citenamefont
  {Michaeli}, \citenamefont {Potter},\ and\ \citenamefont
  {Lee}}]{Michaeli2012}%
  \BibitemOpen
  \bibfield  {author} {\bibinfo {author} {\bibfnamefont {Karen}\ \bibnamefont
  {Michaeli}}, \bibinfo {author} {\bibfnamefont {Andrew~C.}\ \bibnamefont
  {Potter}}, \ and\ \bibinfo {author} {\bibfnamefont {Patrick~A.}\ \bibnamefont
  {Lee}},\ }\bibfield  {title} {\enquote {\bibinfo {title} {Superconducting and
  ferromagnetic phases in $\mathrm{SrTiO}_{3}/\mathrm{LaAlO}_{3}$ oxide
  interface structures: Possibility of finite momentum pairing},}\ }\href
  {\doibase 10.1103/PhysRevLett.108.117003} {\bibfield  {journal} {\bibinfo
  {journal} {Phys. Rev. Lett.}\ }\textbf {\bibinfo {volume} {108}},\ \bibinfo
  {pages} {117003} (\bibinfo {year} {2012})}\BibitemShut {NoStop}%
\bibitem [{\citenamefont {Loder}\ \emph {et~al.}(2015)\citenamefont {Loder},
  \citenamefont {Kampf},\ and\ \citenamefont {Kopp}}]{loder2015route}%
  \BibitemOpen
  \bibfield  {author} {\bibinfo {author} {\bibfnamefont {Florian}\ \bibnamefont
  {Loder}}, \bibinfo {author} {\bibfnamefont {Arno~P.}\ \bibnamefont {Kampf}},
  \ and\ \bibinfo {author} {\bibfnamefont {Thilo}\ \bibnamefont {Kopp}},\
  }\bibfield  {title} {\enquote {\bibinfo {title} {Route to topological
  superconductivity via magnetic field rotation},}\ }\href
  {https://doi.org/10.1038/srep15302} {\bibfield  {journal} {\bibinfo
  {journal} {Sci. Rep.}\ }\textbf {\bibinfo {volume} {5}},\ \bibinfo {pages}
  {1--10} (\bibinfo {year} {2015})}\BibitemShut {NoStop}%
\bibitem [{\citenamefont {Sato}\ \emph {et~al.}(2009)\citenamefont {Sato},
  \citenamefont {Takahashi},\ and\ \citenamefont {Fujimoto}}]{Sato2009}%
  \BibitemOpen
  \bibfield  {author} {\bibinfo {author} {\bibfnamefont {Masatoshi}\
  \bibnamefont {Sato}}, \bibinfo {author} {\bibfnamefont {Yoshiro}\
  \bibnamefont {Takahashi}}, \ and\ \bibinfo {author} {\bibfnamefont {Satoshi}\
  \bibnamefont {Fujimoto}},\ }\bibfield  {title} {\enquote {\bibinfo {title}
  {Non-abelian topological order in $s$-wave superfluids of ultracold fermionic
  atoms},}\ }\href {\doibase 10.1103/PhysRevLett.103.020401} {\bibfield
  {journal} {\bibinfo  {journal} {Phys. Rev. Lett.}\ }\textbf {\bibinfo
  {volume} {103}},\ \bibinfo {pages} {020401} (\bibinfo {year}
  {2009})}\BibitemShut {NoStop}%
\bibitem [{\citenamefont {Sato}\ \emph {et~al.}(2010)\citenamefont {Sato},
  \citenamefont {Takahashi},\ and\ \citenamefont {Fujimoto}}]{Sato2010}%
  \BibitemOpen
  \bibfield  {author} {\bibinfo {author} {\bibfnamefont {Masatoshi}\
  \bibnamefont {Sato}}, \bibinfo {author} {\bibfnamefont {Yoshiro}\
  \bibnamefont {Takahashi}}, \ and\ \bibinfo {author} {\bibfnamefont {Satoshi}\
  \bibnamefont {Fujimoto}},\ }\bibfield  {title} {\enquote {\bibinfo {title}
  {Non-abelian topological orders and {Majorana} fermions in spin-singlet
  superconductors},}\ }\href {\doibase 10.1103/PhysRevB.82.134521} {\bibfield
  {journal} {\bibinfo  {journal} {Phys. Rev. B}\ }\textbf {\bibinfo {volume}
  {82}},\ \bibinfo {pages} {134521} (\bibinfo {year} {2010})}\BibitemShut
  {NoStop}%
\bibitem [{\citenamefont {Gong}\ \emph {et~al.}(2011)\citenamefont {Gong},
  \citenamefont {Tewari},\ and\ \citenamefont {Zhang}}]{Gong2011}%
  \BibitemOpen
  \bibfield  {author} {\bibinfo {author} {\bibfnamefont {Ming}\ \bibnamefont
  {Gong}}, \bibinfo {author} {\bibfnamefont {Sumanta}\ \bibnamefont {Tewari}},
  \ and\ \bibinfo {author} {\bibfnamefont {Chuanwei}\ \bibnamefont {Zhang}},\
  }\bibfield  {title} {\enquote {\bibinfo {title} {{BCS-BEC} crossover and
  topological phase transition in {3D} spin-orbit coupled degenerate {Fermi}
  gases},}\ }\href {\doibase 10.1103/PhysRevLett.107.195303} {\bibfield
  {journal} {\bibinfo  {journal} {Phys. Rev. Lett.}\ }\textbf {\bibinfo
  {volume} {107}},\ \bibinfo {pages} {195303} (\bibinfo {year}
  {2011})}\BibitemShut {NoStop}%
\bibitem [{\citenamefont {Jiang}\ \emph {et~al.}(2011)\citenamefont {Jiang},
  \citenamefont {Liu}, \citenamefont {Hu},\ and\ \citenamefont
  {Pu}}]{Jiang2011}%
  \BibitemOpen
  \bibfield  {author} {\bibinfo {author} {\bibfnamefont {Lei}\ \bibnamefont
  {Jiang}}, \bibinfo {author} {\bibfnamefont {Xia-Ji}\ \bibnamefont {Liu}},
  \bibinfo {author} {\bibfnamefont {Hui}\ \bibnamefont {Hu}}, \ and\ \bibinfo
  {author} {\bibfnamefont {Han}\ \bibnamefont {Pu}},\ }\bibfield  {title}
  {\enquote {\bibinfo {title} {Rashba spin-orbit-coupled atomic {Fermi}
  gases},}\ }\href {\doibase 10.1103/PhysRevA.84.063618} {\bibfield  {journal}
  {\bibinfo  {journal} {Phys. Rev. A}\ }\textbf {\bibinfo {volume} {84}},\
  \bibinfo {pages} {063618} (\bibinfo {year} {2011})}\BibitemShut {NoStop}%
\bibitem [{\citenamefont {Seo}\ \emph {et~al.}(2012)\citenamefont {Seo},
  \citenamefont {Han},\ and\ \citenamefont {de~Melo}}]{Seo2012}%
  \BibitemOpen
  \bibfield  {author} {\bibinfo {author} {\bibfnamefont {Kangjun}\ \bibnamefont
  {Seo}}, \bibinfo {author} {\bibfnamefont {Li}~\bibnamefont {Han}}, \ and\
  \bibinfo {author} {\bibfnamefont {C.~A. R.~S{\'{a}}}\ \bibnamefont
  {de~Melo}},\ }\bibfield  {title} {\enquote {\bibinfo {title} {Topological
  phase transitions in ultracold {Fermi} superfluids: The evolution from
  {Bardeen-Cooper-Schrieffer} to {Bose-Einstein-condensate} superfluids under
  artificial spin-orbit fields},}\ }\href {\doibase 10.1103/PhysRevA.85.033601}
  {\bibfield  {journal} {\bibinfo  {journal} {Phys. Rev. A}\ }\textbf {\bibinfo
  {volume} {85}},\ \bibinfo {pages} {033601} (\bibinfo {year}
  {2012})}\BibitemShut {NoStop}%
\bibitem [{\citenamefont {Seo}\ \emph {et~al.}(2013)\citenamefont {Seo},
  \citenamefont {Zhang},\ and\ \citenamefont {Tewari}}]{Seo2013}%
  \BibitemOpen
  \bibfield  {author} {\bibinfo {author} {\bibfnamefont {Kangjun}\ \bibnamefont
  {Seo}}, \bibinfo {author} {\bibfnamefont {Chuanwei}\ \bibnamefont {Zhang}}, \
  and\ \bibinfo {author} {\bibfnamefont {Sumanta}\ \bibnamefont {Tewari}},\
  }\bibfield  {title} {\enquote {\bibinfo {title} {Thermodynamic signatures for
  topological phase transitions to {Majorana} and {Weyl} superfluids in
  ultracold {Fermi} gases},}\ }\href {\doibase 10.1103/PhysRevA.87.063618}
  {\bibfield  {journal} {\bibinfo  {journal} {Phys. Rev. A}\ }\textbf {\bibinfo
  {volume} {87}},\ \bibinfo {pages} {063618} (\bibinfo {year}
  {2013})}\BibitemShut {NoStop}%
\bibitem [{\citenamefont {Xi}\ \emph {et~al.}(2016)\citenamefont {Xi},
  \citenamefont {Wang}, \citenamefont {Zhao}, \citenamefont {Park},
  \citenamefont {Law}, \citenamefont {Berger}, \citenamefont {Forr{\'o}},
  \citenamefont {Shan},\ and\ \citenamefont {Mak}}]{xi2016ising}%
  \BibitemOpen
  \bibfield  {author} {\bibinfo {author} {\bibfnamefont {Xiaoxiang}\
  \bibnamefont {Xi}}, \bibinfo {author} {\bibfnamefont {Zefang}\ \bibnamefont
  {Wang}}, \bibinfo {author} {\bibfnamefont {Weiwei}\ \bibnamefont {Zhao}},
  \bibinfo {author} {\bibfnamefont {Ju-Hyun}\ \bibnamefont {Park}}, \bibinfo
  {author} {\bibfnamefont {Kam~Tuen}\ \bibnamefont {Law}}, \bibinfo {author}
  {\bibfnamefont {Helmuth}\ \bibnamefont {Berger}}, \bibinfo {author}
  {\bibfnamefont {L{\'a}szl{\'o}}\ \bibnamefont {Forr{\'o}}}, \bibinfo {author}
  {\bibfnamefont {Jie}\ \bibnamefont {Shan}}, \ and\ \bibinfo {author}
  {\bibfnamefont {Kin~Fai}\ \bibnamefont {Mak}},\ }\bibfield  {title} {\enquote
  {\bibinfo {title} {Ising pairing in superconducting {NbSe}$_2$ atomic
  layers},}\ }\href {https://doi.org/10.1038/nphys3538} {\bibfield  {journal}
  {\bibinfo  {journal} {Nat. Phys.}\ }\textbf {\bibinfo {volume} {12}},\
  \bibinfo {pages} {139--143} (\bibinfo {year} {2016})}\BibitemShut {NoStop}%
\bibitem [{\citenamefont {Hsu}\ \emph {et~al.}(2017)\citenamefont {Hsu},
  \citenamefont {Vaezi}, \citenamefont {Fischer},\ and\ \citenamefont
  {Kim}}]{hsu2017topological}%
  \BibitemOpen
  \bibfield  {author} {\bibinfo {author} {\bibfnamefont {Yi-Ting}\ \bibnamefont
  {Hsu}}, \bibinfo {author} {\bibfnamefont {Abolhassan}\ \bibnamefont {Vaezi}},
  \bibinfo {author} {\bibfnamefont {Mark~H}\ \bibnamefont {Fischer}}, \ and\
  \bibinfo {author} {\bibfnamefont {Eun-Ah}\ \bibnamefont {Kim}},\ }\bibfield
  {title} {\enquote {\bibinfo {title} {Topological superconductivity in
  monolayer transition metal dichalcogenides},}\ }\href
  {https://doi.org/10.1038/ncomms14985} {\bibfield  {journal} {\bibinfo
  {journal} {Nat. Commun.}\ }\textbf {\bibinfo {volume} {8}},\ \bibinfo {pages}
  {1--6} (\bibinfo {year} {2017})}\BibitemShut {NoStop}%
\bibitem [{\citenamefont {M{\"o}ckli}\ and\ \citenamefont
  {Khodas}(2018)}]{mockli2018robust}%
  \BibitemOpen
  \bibfield  {author} {\bibinfo {author} {\bibfnamefont {David}\ \bibnamefont
  {M{\"o}ckli}}\ and\ \bibinfo {author} {\bibfnamefont {Maxim}\ \bibnamefont
  {Khodas}},\ }\bibfield  {title} {\enquote {\bibinfo {title} {Robust
  parity-mixed superconductivity in disordered monolayer transition metal
  dichalcogenides},}\ }\href {\doibase 10.1103/PhysRevB.98.144518} {\bibfield
  {journal} {\bibinfo  {journal} {Phys. Rev. B}\ }\textbf {\bibinfo {volume}
  {98}},\ \bibinfo {pages} {144518} (\bibinfo {year} {2018})}\BibitemShut
  {NoStop}%
\bibitem [{\citenamefont {Wickramaratne}\ \emph {et~al.}(2020)\citenamefont
  {Wickramaratne}, \citenamefont {Khmelevskyi}, \citenamefont {Agterberg},\
  and\ \citenamefont {Mazin}}]{wickramaratne2020ising}%
  \BibitemOpen
  \bibfield  {author} {\bibinfo {author} {\bibfnamefont {Darshana}\
  \bibnamefont {Wickramaratne}}, \bibinfo {author} {\bibfnamefont {Sergii}\
  \bibnamefont {Khmelevskyi}}, \bibinfo {author} {\bibfnamefont {Daniel~F.}\
  \bibnamefont {Agterberg}}, \ and\ \bibinfo {author} {\bibfnamefont {I.I.}\
  \bibnamefont {Mazin}},\ }\bibfield  {title} {\enquote {\bibinfo {title}
  {Ising superconductivity and magnetism in {NbSe}$_2$},}\ }\href {\doibase
  10.1103/PhysRevX.10.041003} {\bibfield  {journal} {\bibinfo  {journal} {Phys.
  Rev. X}\ }\textbf {\bibinfo {volume} {10}},\ \bibinfo {pages} {041003}
  (\bibinfo {year} {2020})}\BibitemShut {NoStop}%
\bibitem [{\citenamefont {Rischau}\ \emph {et~al.}(2017)\citenamefont
  {Rischau}, \citenamefont {Lin}, \citenamefont {Grams}, \citenamefont {Finck},
  \citenamefont {Harms}, \citenamefont {Engelmayer}, \citenamefont {Lorenz},
  \citenamefont {Gallais}, \citenamefont {Fauqué}, \citenamefont {Hemberger},\
  and\ \citenamefont {Behnia}}]{Rischau2017}%
  \BibitemOpen
  \bibfield  {author} {\bibinfo {author} {\bibfnamefont {Carl~Willem}\
  \bibnamefont {Rischau}}, \bibinfo {author} {\bibfnamefont {Xiao}\
  \bibnamefont {Lin}}, \bibinfo {author} {\bibfnamefont {Christoph~P.}\
  \bibnamefont {Grams}}, \bibinfo {author} {\bibfnamefont {Dennis}\
  \bibnamefont {Finck}}, \bibinfo {author} {\bibfnamefont {Steffen}\
  \bibnamefont {Harms}}, \bibinfo {author} {\bibfnamefont {Johannes}\
  \bibnamefont {Engelmayer}}, \bibinfo {author} {\bibfnamefont {Thomas}\
  \bibnamefont {Lorenz}}, \bibinfo {author} {\bibfnamefont {Yann}\ \bibnamefont
  {Gallais}}, \bibinfo {author} {\bibfnamefont {Benoît}\ \bibnamefont
  {Fauqué}}, \bibinfo {author} {\bibfnamefont {Joachim}\ \bibnamefont
  {Hemberger}}, \ and\ \bibinfo {author} {\bibfnamefont {Kamran}\ \bibnamefont
  {Behnia}},\ }\bibfield  {title} {\enquote {\bibinfo {title} {A ferroelectric
  quantum phase transition inside the superconducting dome of
  {Sr}$_{1-x}${Ca}$_x${Ti}{O}$_{3-\delta}$},}\ }\href {\doibase
  10.1038/nphys4085} {\bibfield  {journal} {\bibinfo  {journal} {Nat. Phys.}\
  }\textbf {\bibinfo {volume} {13}},\ \bibinfo {pages} {643--648} (\bibinfo
  {year} {2017})}\BibitemShut {NoStop}%
\bibitem [{\citenamefont {Fei}\ \emph {et~al.}(2018)\citenamefont {Fei},
  \citenamefont {Zhao}, \citenamefont {Palomaki}, \citenamefont {Sun},
  \citenamefont {Miller}, \citenamefont {Zhao}, \citenamefont {Yan},
  \citenamefont {Xu},\ and\ \citenamefont {Cobden}}]{fei2018ferroelectric}%
  \BibitemOpen
  \bibfield  {author} {\bibinfo {author} {\bibfnamefont {Zaiyao}\ \bibnamefont
  {Fei}}, \bibinfo {author} {\bibfnamefont {Wenjin}\ \bibnamefont {Zhao}},
  \bibinfo {author} {\bibfnamefont {Tauno~A.}\ \bibnamefont {Palomaki}},
  \bibinfo {author} {\bibfnamefont {Bosong}\ \bibnamefont {Sun}}, \bibinfo
  {author} {\bibfnamefont {Moira~K.}\ \bibnamefont {Miller}}, \bibinfo {author}
  {\bibfnamefont {Zhiying}\ \bibnamefont {Zhao}}, \bibinfo {author}
  {\bibfnamefont {Jiaqiang}\ \bibnamefont {Yan}}, \bibinfo {author}
  {\bibfnamefont {Xiaodong}\ \bibnamefont {Xu}}, \ and\ \bibinfo {author}
  {\bibfnamefont {David~H.}\ \bibnamefont {Cobden}},\ }\bibfield  {title}
  {\enquote {\bibinfo {title} {Ferroelectric switching of a two-dimensional
  metal},}\ }\href {https://doi.org/10.1038/s41586-018-0336-3} {\bibfield
  {journal} {\bibinfo  {journal} {Nature}\ }\textbf {\bibinfo {volume} {560}},\
  \bibinfo {pages} {336--339} (\bibinfo {year} {2018})}\BibitemShut {NoStop}%
\bibitem [{\citenamefont {Russell}\ \emph {et~al.}(2019)\citenamefont
  {Russell}, \citenamefont {Ratcliff}, \citenamefont {Ahadi}, \citenamefont
  {Dong}, \citenamefont {Stemmer},\ and\ \citenamefont
  {Harter}}]{russell2019ferroelectric}%
  \BibitemOpen
  \bibfield  {author} {\bibinfo {author} {\bibfnamefont {Ryan}\ \bibnamefont
  {Russell}}, \bibinfo {author} {\bibfnamefont {Noah}\ \bibnamefont
  {Ratcliff}}, \bibinfo {author} {\bibfnamefont {Kaveh}\ \bibnamefont {Ahadi}},
  \bibinfo {author} {\bibfnamefont {Lianyang}\ \bibnamefont {Dong}}, \bibinfo
  {author} {\bibfnamefont {Susanne}\ \bibnamefont {Stemmer}}, \ and\ \bibinfo
  {author} {\bibfnamefont {John~W.}\ \bibnamefont {Harter}},\ }\bibfield
  {title} {\enquote {\bibinfo {title} {Ferroelectric enhancement of
  superconductivity in compressively strained {SrTiO}$_3$ films},}\ }\href
  {\doibase 10.1103/PhysRevMaterials.3.091401} {\bibfield  {journal} {\bibinfo
  {journal} {Phys. Rev. Mater.}\ }\textbf {\bibinfo {volume} {3}},\ \bibinfo
  {pages} {091401(R)} (\bibinfo {year} {2019})}\BibitemShut {NoStop}%
\bibitem [{\citenamefont {Tomioka}\ \emph {et~al.}(2022)\citenamefont
  {Tomioka}, \citenamefont {Shirakawa},\ and\ \citenamefont
  {Inoue}}]{tomioka2022superconductivity}%
  \BibitemOpen
  \bibfield  {author} {\bibinfo {author} {\bibfnamefont {Yasuhide}\
  \bibnamefont {Tomioka}}, \bibinfo {author} {\bibfnamefont {Naoki}\
  \bibnamefont {Shirakawa}}, \ and\ \bibinfo {author} {\bibfnamefont {Isao~H.}\
  \bibnamefont {Inoue}},\ }\href@noop {} {\enquote {\bibinfo {title}
  {Superconductivity enhanced in the polar metal region of
  {Sr$_{0.95}$Ba$_{0.05}$TiO$_3$} and {Sr$_{0.985}$Ca$_{0.015}$TiO$_3$}
  revealed by the systematic {Nb} doping},}\ } (\bibinfo {year} {2022}),\
  \Eprint {http://arxiv.org/abs/2203.16208} {arXiv:2203.16208
  [cond-mat.supr-con]} \BibitemShut {NoStop}%
\bibitem [{\citenamefont {Scheerer}\ \emph {et~al.}(2020)\citenamefont
  {Scheerer}, \citenamefont {Boselli}, \citenamefont {Pulmannova},
  \citenamefont {Rischau}, \citenamefont {Waelchli}, \citenamefont {Gariglio},
  \citenamefont {Giannini}, \citenamefont {van~der Marel},\ and\ \citenamefont
  {Triscone}}]{Scheerer2020}%
  \BibitemOpen
  \bibfield  {author} {\bibinfo {author} {\bibfnamefont {Gernot}\ \bibnamefont
  {Scheerer}}, \bibinfo {author} {\bibfnamefont {Margherita}\ \bibnamefont
  {Boselli}}, \bibinfo {author} {\bibfnamefont {Dorota}\ \bibnamefont
  {Pulmannova}}, \bibinfo {author} {\bibfnamefont {Carl~Willem}\ \bibnamefont
  {Rischau}}, \bibinfo {author} {\bibfnamefont {Adrien}\ \bibnamefont
  {Waelchli}}, \bibinfo {author} {\bibfnamefont {Stefano}\ \bibnamefont
  {Gariglio}}, \bibinfo {author} {\bibfnamefont {Enrico}\ \bibnamefont
  {Giannini}}, \bibinfo {author} {\bibfnamefont {Dirk}\ \bibnamefont {van~der
  Marel}}, \ and\ \bibinfo {author} {\bibfnamefont {Jean-Marc}\ \bibnamefont
  {Triscone}},\ }\bibfield  {title} {\enquote {\bibinfo {title}
  {Ferroelectricity, superconductivity, and {SrTiO}$_3$—{Passions} of {K.A.
  Müller}},}\ }\href {https://www.mdpi.com/2410-3896/5/4/60} {\bibfield
  {journal} {\bibinfo  {journal} {Condensed Matter}\ }\textbf {\bibinfo
  {volume} {5}} (\bibinfo {year} {2020})}\BibitemShut {NoStop}%
\bibitem [{\citenamefont {Tuvia}\ \emph {et~al.}(2020)\citenamefont {Tuvia},
  \citenamefont {Frenkel}, \citenamefont {Rout}, \citenamefont {Silber},
  \citenamefont {Kalisky},\ and\ \citenamefont {Dagan}}]{Tuvia2020}%
  \BibitemOpen
  \bibfield  {author} {\bibinfo {author} {\bibfnamefont {Gal}\ \bibnamefont
  {Tuvia}}, \bibinfo {author} {\bibfnamefont {Yiftach}\ \bibnamefont
  {Frenkel}}, \bibinfo {author} {\bibfnamefont {Prasanna~K.}\ \bibnamefont
  {Rout}}, \bibinfo {author} {\bibfnamefont {Itai}\ \bibnamefont {Silber}},
  \bibinfo {author} {\bibfnamefont {Beena}\ \bibnamefont {Kalisky}}, \ and\
  \bibinfo {author} {\bibfnamefont {Yoram}\ \bibnamefont {Dagan}},\ }\bibfield
  {title} {\enquote {\bibinfo {title} {Ferroelectric exchange bias affects
  interfacial electronic states},}\ }\href {\doibase
  https://doi.org/10.1002/adma.202000216} {\bibfield  {journal} {\bibinfo
  {journal} {Advanced Materials}\ }\textbf {\bibinfo {volume} {32}},\ \bibinfo
  {pages} {2000216} (\bibinfo {year} {2020})}\BibitemShut {NoStop}%
\bibitem [{\citenamefont {Weaver}(1959)}]{WEAVER1959274}%
  \BibitemOpen
  \bibfield  {author} {\bibinfo {author} {\bibfnamefont {H.~E.}\ \bibnamefont
  {Weaver}},\ }\bibfield  {title} {\enquote {\bibinfo {title} {Dielectric
  properties of single crystals of {SrTiO}$_3$ at low temperatures},}\ }\href
  {\doibase https://doi.org/10.1016/0022-3697(59)90226-4} {\bibfield  {journal}
  {\bibinfo  {journal} {J. Phys. Chem. Solids}\ }\textbf {\bibinfo {volume}
  {11}},\ \bibinfo {pages} {274--277} (\bibinfo {year} {1959})}\BibitemShut
  {NoStop}%
\bibitem [{\citenamefont {M\"uller}\ and\ \citenamefont
  {Burkard}(1979)}]{Muller1979}%
  \BibitemOpen
  \bibfield  {author} {\bibinfo {author} {\bibfnamefont {K.~A.}\ \bibnamefont
  {M\"uller}}\ and\ \bibinfo {author} {\bibfnamefont {H.}~\bibnamefont
  {Burkard}},\ }\bibfield  {title} {\enquote {\bibinfo {title}
  {{SrTi}{${\mathrm{O}}_{3}$}: An intrinsic quantum paraelectric below 4
  {K}},}\ }\href {\doibase 10.1103/PhysRevB.19.3593} {\bibfield  {journal}
  {\bibinfo  {journal} {Phys. Rev. B}\ }\textbf {\bibinfo {volume} {19}},\
  \bibinfo {pages} {3593--3602} (\bibinfo {year} {1979})}\BibitemShut {NoStop}%
\bibitem [{\citenamefont {Ambwani}\ \emph {et~al.}(2016)\citenamefont
  {Ambwani}, \citenamefont {Xu}, \citenamefont {Haugstad}, \citenamefont
  {Jeong}, \citenamefont {Deng}, \citenamefont {Mkhoyan}, \citenamefont
  {Jalan},\ and\ \citenamefont {Leighton}}]{ambwani2016defects}%
  \BibitemOpen
  \bibfield  {author} {\bibinfo {author} {\bibfnamefont {P.}~\bibnamefont
  {Ambwani}}, \bibinfo {author} {\bibfnamefont {P.}~\bibnamefont {Xu}},
  \bibinfo {author} {\bibfnamefont {G.}~\bibnamefont {Haugstad}}, \bibinfo
  {author} {\bibfnamefont {J.~S.}\ \bibnamefont {Jeong}}, \bibinfo {author}
  {\bibfnamefont {R.}~\bibnamefont {Deng}}, \bibinfo {author} {\bibfnamefont
  {K.~A.}\ \bibnamefont {Mkhoyan}}, \bibinfo {author} {\bibfnamefont
  {B.}~\bibnamefont {Jalan}}, \ and\ \bibinfo {author} {\bibfnamefont
  {C.}~\bibnamefont {Leighton}},\ }\bibfield  {title} {\enquote {\bibinfo
  {title} {Defects, stoichiometry, and electronic transport in
  {SrTiO}$_{3-\delta}$ epilayers: A high pressure oxygen sputter deposition
  study},}\ }\href {https://aip.scitation.org/doi/abs/10.1063/1.4960343}
  {\bibfield  {journal} {\bibinfo  {journal} {J. Appl. Phys.}\ }\textbf
  {\bibinfo {volume} {120}},\ \bibinfo {pages} {055704} (\bibinfo {year}
  {2016})}\BibitemShut {NoStop}%
\bibitem [{\citenamefont {Collignon}\ \emph {et~al.}(2019)\citenamefont
  {Collignon}, \citenamefont {Lin}, \citenamefont {Rischau}, \citenamefont
  {Fauqué},\ and\ \citenamefont {Behnia}}]{Collignon2019}%
  \BibitemOpen
  \bibfield  {author} {\bibinfo {author} {\bibfnamefont {Clément}\
  \bibnamefont {Collignon}}, \bibinfo {author} {\bibfnamefont {Xiao}\
  \bibnamefont {Lin}}, \bibinfo {author} {\bibfnamefont {Carl~Willem}\
  \bibnamefont {Rischau}}, \bibinfo {author} {\bibfnamefont {Benoît}\
  \bibnamefont {Fauqué}}, \ and\ \bibinfo {author} {\bibfnamefont {Kamran}\
  \bibnamefont {Behnia}},\ }\bibfield  {title} {\enquote {\bibinfo {title}
  {Metallicity and superconductivity in doped strontium titanate},}\ }\href
  {\doibase 10.1146/annurev-conmatphys-031218-013144} {\bibfield  {journal}
  {\bibinfo  {journal} {Annual Review of Condensed Matter Physics}\ }\textbf
  {\bibinfo {volume} {10}},\ \bibinfo {pages} {25--44} (\bibinfo {year}
  {2019})}\BibitemShut {NoStop}%
\bibitem [{\citenamefont {Gastiasoro}\ \emph
  {et~al.}(2020{\natexlab{a}})\citenamefont {Gastiasoro}, \citenamefont
  {Ruhman},\ and\ \citenamefont {Fernandes}}]{Gastiasoro2020b}%
  \BibitemOpen
  \bibfield  {author} {\bibinfo {author} {\bibfnamefont {Maria~N.}\
  \bibnamefont {Gastiasoro}}, \bibinfo {author} {\bibfnamefont {Jonathan}\
  \bibnamefont {Ruhman}}, \ and\ \bibinfo {author} {\bibfnamefont {Rafael~M.}\
  \bibnamefont {Fernandes}},\ }\bibfield  {title} {\enquote {\bibinfo {title}
  {Superconductivity in dilute {SrTiO$_3$}: A review},}\ }\href {\doibase
  10.1016/j.aop.2020.168107} {\bibfield  {journal} {\bibinfo  {journal} {Ann.
  Phys.}\ }\textbf {\bibinfo {volume} {417}},\ \bibinfo {pages} {168107}
  (\bibinfo {year} {2020}{\natexlab{a}})}\BibitemShut {NoStop}%
\bibitem [{\citenamefont {Rowley}\ \emph {et~al.}(2014)\citenamefont {Rowley},
  \citenamefont {Spalek}, \citenamefont {Smith}, \citenamefont {Dean},
  \citenamefont {Itoh}, \citenamefont {Scott}, \citenamefont {Lonzarich},\ and\
  \citenamefont {Saxena}}]{rowley2014ferroelectric}%
  \BibitemOpen
  \bibfield  {author} {\bibinfo {author} {\bibfnamefont {S.~E.}\ \bibnamefont
  {Rowley}}, \bibinfo {author} {\bibfnamefont {L.~J.}\ \bibnamefont {Spalek}},
  \bibinfo {author} {\bibfnamefont {R.~P.}\ \bibnamefont {Smith}}, \bibinfo
  {author} {\bibfnamefont {M.~P.~M.}\ \bibnamefont {Dean}}, \bibinfo {author}
  {\bibfnamefont {M.}~\bibnamefont {Itoh}}, \bibinfo {author} {\bibfnamefont
  {J.~F.}\ \bibnamefont {Scott}}, \bibinfo {author} {\bibfnamefont {G.~G.}\
  \bibnamefont {Lonzarich}}, \ and\ \bibinfo {author} {\bibfnamefont {S.~S.}\
  \bibnamefont {Saxena}},\ }\bibfield  {title} {\enquote {\bibinfo {title}
  {Ferroelectric quantum criticality},}\ }\href {\doibase 10.1038/nphys2924}
  {\bibfield  {journal} {\bibinfo  {journal} {Nat. Phys.}\ }\textbf {\bibinfo
  {volume} {10}},\ \bibinfo {pages} {367--372} (\bibinfo {year}
  {2014})}\BibitemShut {NoStop}%
\bibitem [{\citenamefont {Stucky}\ \emph {et~al.}(2016)\citenamefont {Stucky},
  \citenamefont {Scheerer}, \citenamefont {Ren}, \citenamefont {Jaccard},
  \citenamefont {Poumirol}, \citenamefont {Barreteau}, \citenamefont
  {Giannini},\ and\ \citenamefont {van~der Marel}}]{Stucky2016}%
  \BibitemOpen
  \bibfield  {author} {\bibinfo {author} {\bibfnamefont {A.}~\bibnamefont
  {Stucky}}, \bibinfo {author} {\bibfnamefont {G.~W.}\ \bibnamefont
  {Scheerer}}, \bibinfo {author} {\bibfnamefont {Z.}~\bibnamefont {Ren}},
  \bibinfo {author} {\bibfnamefont {D.}~\bibnamefont {Jaccard}}, \bibinfo
  {author} {\bibfnamefont {J.-M.}\ \bibnamefont {Poumirol}}, \bibinfo {author}
  {\bibfnamefont {C.}~\bibnamefont {Barreteau}}, \bibinfo {author}
  {\bibfnamefont {E.}~\bibnamefont {Giannini}}, \ and\ \bibinfo {author}
  {\bibfnamefont {D.}~\bibnamefont {van~der Marel}},\ }\bibfield  {title}
  {\enquote {\bibinfo {title} {Isotope effect in superconducting n-doped
  {SrTiO}$_3$},}\ }\href {\doibase 10.1038/srep37582} {\bibfield  {journal}
  {\bibinfo  {journal} {Sci. Rep.}\ }\textbf {\bibinfo {volume} {6}},\ \bibinfo
  {pages} {37582} (\bibinfo {year} {2016})}\BibitemShut {NoStop}%
\bibitem [{\citenamefont {Salmani-Rezaie}\ \emph {et~al.}(2020)\citenamefont
  {Salmani-Rezaie}, \citenamefont {Ahadi}, \citenamefont {Strickland},\ and\
  \citenamefont {Stemmer}}]{salmani2020order}%
  \BibitemOpen
  \bibfield  {author} {\bibinfo {author} {\bibfnamefont {Salva}\ \bibnamefont
  {Salmani-Rezaie}}, \bibinfo {author} {\bibfnamefont {Kaveh}\ \bibnamefont
  {Ahadi}}, \bibinfo {author} {\bibfnamefont {William~M}\ \bibnamefont
  {Strickland}}, \ and\ \bibinfo {author} {\bibfnamefont {Susanne}\
  \bibnamefont {Stemmer}},\ }\bibfield  {title} {\enquote {\bibinfo {title}
  {Order-disorder ferroelectric transition of strained {SrTiO}$_3$},}\ }\href
  {\doibase 10.1103/PhysRevLett.125.087601} {\bibfield  {journal} {\bibinfo
  {journal} {Phys. Rev. Lett.}\ }\textbf {\bibinfo {volume} {125}},\ \bibinfo
  {pages} {087601} (\bibinfo {year} {2020})}\BibitemShut {NoStop}%
\bibitem [{\citenamefont {Sakai}\ \emph {et~al.}(2016)\citenamefont {Sakai},
  \citenamefont {Ikeura}, \citenamefont {Bahramy}, \citenamefont {Ogawa},
  \citenamefont {Hashizume}, \citenamefont {Fujioka}, \citenamefont {Tokura},\
  and\ \citenamefont {Ishiwata}}]{sakai2016critical}%
  \BibitemOpen
  \bibfield  {author} {\bibinfo {author} {\bibfnamefont {Hideaki}\ \bibnamefont
  {Sakai}}, \bibinfo {author} {\bibfnamefont {Koji}\ \bibnamefont {Ikeura}},
  \bibinfo {author} {\bibfnamefont {Mohammad~Saeed}\ \bibnamefont {Bahramy}},
  \bibinfo {author} {\bibfnamefont {Naoki}\ \bibnamefont {Ogawa}}, \bibinfo
  {author} {\bibfnamefont {Daisuke}\ \bibnamefont {Hashizume}}, \bibinfo
  {author} {\bibfnamefont {Jun}\ \bibnamefont {Fujioka}}, \bibinfo {author}
  {\bibfnamefont {Yoshinori}\ \bibnamefont {Tokura}}, \ and\ \bibinfo {author}
  {\bibfnamefont {Shintaro}\ \bibnamefont {Ishiwata}},\ }\bibfield  {title}
  {\enquote {\bibinfo {title} {Critical enhancement of thermopower in a
  chemically tuned polar semimetal $\text{MoTe}_2$},}\ }\href
  {https://www.science.org/doi/10.1126/sciadv.1601378} {\bibfield  {journal}
  {\bibinfo  {journal} {Sci. Adv.}\ }\textbf {\bibinfo {volume} {2}},\ \bibinfo
  {pages} {e1601378} (\bibinfo {year} {2016})}\BibitemShut {NoStop}%
\bibitem [{\citenamefont {Herrera}\ \emph {et~al.}(2019)\citenamefont
  {Herrera}, \citenamefont {Cerbin}, \citenamefont {Jayakody}, \citenamefont
  {Dunnett}, \citenamefont {Balatsky},\ and\ \citenamefont
  {Sochnikov}}]{Herrera_2019}%
  \BibitemOpen
  \bibfield  {author} {\bibinfo {author} {\bibfnamefont {Chloe}\ \bibnamefont
  {Herrera}}, \bibinfo {author} {\bibfnamefont {Jonah}\ \bibnamefont {Cerbin}},
  \bibinfo {author} {\bibfnamefont {Amani}\ \bibnamefont {Jayakody}}, \bibinfo
  {author} {\bibfnamefont {Kirsty}\ \bibnamefont {Dunnett}}, \bibinfo {author}
  {\bibfnamefont {Alexander~V.}\ \bibnamefont {Balatsky}}, \ and\ \bibinfo
  {author} {\bibfnamefont {Ilya}\ \bibnamefont {Sochnikov}},\ }\bibfield
  {title} {\enquote {\bibinfo {title} {Strain-engineered interaction of quantum
  polar and superconducting phases},}\ }\href {\doibase
  10.1103/PhysRevMaterials.3.124801} {\bibfield  {journal} {\bibinfo  {journal}
  {Phys. Rev. Mater.}\ }\textbf {\bibinfo {volume} {3}},\ \bibinfo {pages}
  {124801} (\bibinfo {year} {2019})}\BibitemShut {NoStop}%
\bibitem [{\citenamefont {Engelmayer}\ \emph {et~al.}(2019)\citenamefont
  {Engelmayer}, \citenamefont {Lin}, \citenamefont
  {Ko\ifmmode~\mbox{\c{c}}\else \c{c}\fi{}}, \citenamefont {Grams},
  \citenamefont {Hemberger}, \citenamefont {Behnia},\ and\ \citenamefont
  {Lorenz}}]{engelmayer2019ferroelectric}%
  \BibitemOpen
  \bibfield  {author} {\bibinfo {author} {\bibfnamefont {Johannes}\
  \bibnamefont {Engelmayer}}, \bibinfo {author} {\bibfnamefont {Xiao}\
  \bibnamefont {Lin}}, \bibinfo {author} {\bibfnamefont {Fulya}\ \bibnamefont
  {Ko\ifmmode~\mbox{\c{c}}\else \c{c}\fi{}}}, \bibinfo {author} {\bibfnamefont
  {Christoph~P.}\ \bibnamefont {Grams}}, \bibinfo {author} {\bibfnamefont
  {Joachim}\ \bibnamefont {Hemberger}}, \bibinfo {author} {\bibfnamefont
  {Kamran}\ \bibnamefont {Behnia}}, \ and\ \bibinfo {author} {\bibfnamefont
  {Thomas}\ \bibnamefont {Lorenz}},\ }\bibfield  {title} {\enquote {\bibinfo
  {title} {Ferroelectric order versus metallicity in
  $\text{Sr}_{1\ensuremath{-}x}\text{Ca}_{x}\text{TiO}_{3\ensuremath{-}\ensuremath{\delta}}$
  ($x=0.009$)},}\ }\href {\doibase 10.1103/PhysRevB.100.195121} {\bibfield
  {journal} {\bibinfo  {journal} {Phys. Rev. B}\ }\textbf {\bibinfo {volume}
  {100}},\ \bibinfo {pages} {195121} (\bibinfo {year} {2019})}\BibitemShut
  {NoStop}%
\bibitem [{\citenamefont {Wang}\ \emph {et~al.}(2019)\citenamefont {Wang},
  \citenamefont {Yang}, \citenamefont {Rischau}, \citenamefont {Xu},
  \citenamefont {Ren}, \citenamefont {Lorenz}, \citenamefont {Hemberger},
  \citenamefont {Lin},\ and\ \citenamefont {Behnia}}]{wang2019charge}%
  \BibitemOpen
  \bibfield  {author} {\bibinfo {author} {\bibfnamefont {Jialu}\ \bibnamefont
  {Wang}}, \bibinfo {author} {\bibfnamefont {Liangwei}\ \bibnamefont {Yang}},
  \bibinfo {author} {\bibfnamefont {Carl~Willem}\ \bibnamefont {Rischau}},
  \bibinfo {author} {\bibfnamefont {Zhuokai}\ \bibnamefont {Xu}}, \bibinfo
  {author} {\bibfnamefont {Zhi}\ \bibnamefont {Ren}}, \bibinfo {author}
  {\bibfnamefont {Thomas}\ \bibnamefont {Lorenz}}, \bibinfo {author}
  {\bibfnamefont {Joachim}\ \bibnamefont {Hemberger}}, \bibinfo {author}
  {\bibfnamefont {Xiao}\ \bibnamefont {Lin}}, \ and\ \bibinfo {author}
  {\bibfnamefont {Kamran}\ \bibnamefont {Behnia}},\ }\bibfield  {title}
  {\enquote {\bibinfo {title} {Charge transport in a polar metal},}\ }\href
  {\doibase 10.1038/s41535-019-0200-1} {\bibfield  {journal} {\bibinfo
  {journal} {npj Quantum Mater.}\ }\textbf {\bibinfo {volume} {4}},\ \bibinfo
  {pages} {1--8} (\bibinfo {year} {2019})}\BibitemShut {NoStop}%
\bibitem [{\citenamefont {Enderlein}\ \emph {et~al.}(2020)\citenamefont
  {Enderlein}, \citenamefont {de~Oliveira}, \citenamefont {Tompsett},
  \citenamefont {Saitovitch}, \citenamefont {Saxena}, \citenamefont
  {Lonzarich},\ and\ \citenamefont {Rowley}}]{enderlein2020superconductivity}%
  \BibitemOpen
  \bibfield  {author} {\bibinfo {author} {\bibfnamefont {C.}~\bibnamefont
  {Enderlein}}, \bibinfo {author} {\bibfnamefont {J.~Ferreira}\ \bibnamefont
  {de~Oliveira}}, \bibinfo {author} {\bibfnamefont {D.~A.}\ \bibnamefont
  {Tompsett}}, \bibinfo {author} {\bibfnamefont {E.~Baggio}\ \bibnamefont
  {Saitovitch}}, \bibinfo {author} {\bibfnamefont {S.~S.}\ \bibnamefont
  {Saxena}}, \bibinfo {author} {\bibfnamefont {G.~G.}\ \bibnamefont
  {Lonzarich}}, \ and\ \bibinfo {author} {\bibfnamefont {S.~E.}\ \bibnamefont
  {Rowley}},\ }\bibfield  {title} {\enquote {\bibinfo {title}
  {Superconductivity mediated by polar modes in ferroelectric metals},}\ }\href
  {https://doi.org/10.1038/s41467-020-18438-0} {\bibfield  {journal} {\bibinfo
  {journal} {Nat. Commun.}\ }\textbf {\bibinfo {volume} {11}},\ \bibinfo
  {pages} {4852} (\bibinfo {year} {2020})}\BibitemShut {NoStop}%
\bibitem [{\citenamefont {Salmani-Rezaie}\ \emph {et~al.}(2021)\citenamefont
  {Salmani-Rezaie}, \citenamefont {Jeong}, \citenamefont {Russell},
  \citenamefont {Harter},\ and\ \citenamefont {Stemmer}}]{Salmani2021}%
  \BibitemOpen
  \bibfield  {author} {\bibinfo {author} {\bibfnamefont {Salva}\ \bibnamefont
  {Salmani-Rezaie}}, \bibinfo {author} {\bibfnamefont {Hanbyeol}\ \bibnamefont
  {Jeong}}, \bibinfo {author} {\bibfnamefont {Ryan}\ \bibnamefont {Russell}},
  \bibinfo {author} {\bibfnamefont {John~W.}\ \bibnamefont {Harter}}, \ and\
  \bibinfo {author} {\bibfnamefont {Susanne}\ \bibnamefont {Stemmer}},\
  }\bibfield  {title} {\enquote {\bibinfo {title} {Role of locally polar
  regions in the superconductivity of $\mathrm{SrTi}{{\mathrm{O}}}_{3}$},}\
  }\href {\doibase 10.1103/PhysRevMaterials.5.104801} {\bibfield  {journal}
  {\bibinfo  {journal} {Phys. Rev. Mater.}\ }\textbf {\bibinfo {volume} {5}},\
  \bibinfo {pages} {104801} (\bibinfo {year} {2021})}\BibitemShut {NoStop}%
\bibitem [{\citenamefont {Ahadi}\ \emph {et~al.}(2019)\citenamefont {Ahadi},
  \citenamefont {Galletti}, \citenamefont {Li}, \citenamefont {Salmani-Rezaie},
  \citenamefont {Wu},\ and\ \citenamefont {Stemmer}}]{ahadi2019enhancing}%
  \BibitemOpen
  \bibfield  {author} {\bibinfo {author} {\bibfnamefont {Kaveh}\ \bibnamefont
  {Ahadi}}, \bibinfo {author} {\bibfnamefont {Luca}\ \bibnamefont {Galletti}},
  \bibinfo {author} {\bibfnamefont {Yuntian}\ \bibnamefont {Li}}, \bibinfo
  {author} {\bibfnamefont {Salva}\ \bibnamefont {Salmani-Rezaie}}, \bibinfo
  {author} {\bibfnamefont {Wangzhou}\ \bibnamefont {Wu}}, \ and\ \bibinfo
  {author} {\bibfnamefont {Susanne}\ \bibnamefont {Stemmer}},\ }\bibfield
  {title} {\enquote {\bibinfo {title} {Enhancing superconductivity in
  {SrTiO}$_3$ films with strain},}\ }\href {\doibase 10.1126/sciadv.aaw0120}
  {\bibfield  {journal} {\bibinfo  {journal} {Sci. Adv.}\ }\textbf {\bibinfo
  {volume} {5}},\ \bibinfo {pages} {eaaw0120} (\bibinfo {year}
  {2019})}\BibitemShut {NoStop}%
\bibitem [{\citenamefont {Kanasugi}\ and\ \citenamefont
  {Yanase}(2018)}]{Kanasugi2018}%
  \BibitemOpen
  \bibfield  {author} {\bibinfo {author} {\bibfnamefont {Shota}\ \bibnamefont
  {Kanasugi}}\ and\ \bibinfo {author} {\bibfnamefont {Youichi}\ \bibnamefont
  {Yanase}},\ }\bibfield  {title} {\enquote {\bibinfo {title}
  {Spin-orbit-coupled ferroelectric superconductivity},}\ }\href {\doibase
  10.1103/PhysRevB.98.024521} {\bibfield  {journal} {\bibinfo  {journal} {Phys.
  Rev. B}\ }\textbf {\bibinfo {volume} {98}},\ \bibinfo {pages} {024521}
  (\bibinfo {year} {2018})}\BibitemShut {NoStop}%
\bibitem [{\citenamefont {Kanasugi}\ and\ \citenamefont
  {Yanase}(2019)}]{Kanasugi2019}%
  \BibitemOpen
  \bibfield  {author} {\bibinfo {author} {\bibfnamefont {Shota}\ \bibnamefont
  {Kanasugi}}\ and\ \bibinfo {author} {\bibfnamefont {Youichi}\ \bibnamefont
  {Yanase}},\ }\bibfield  {title} {\enquote {\bibinfo {title} {Multiorbital
  ferroelectric superconductivity in doped {SrTiO}$_{3}$},}\ }\href {\doibase
  10.1103/PhysRevB.100.094504} {\bibfield  {journal} {\bibinfo  {journal}
  {Phys. Rev. B}\ }\textbf {\bibinfo {volume} {100}},\ \bibinfo {pages}
  {094504} (\bibinfo {year} {2019})}\BibitemShut {NoStop}%
\bibitem [{\citenamefont {Petersen}\ and\ \citenamefont
  {Hedegård}(2000)}]{PETERSEN200049}%
  \BibitemOpen
  \bibfield  {author} {\bibinfo {author} {\bibfnamefont {L.}~\bibnamefont
  {Petersen}}\ and\ \bibinfo {author} {\bibfnamefont {P.}~\bibnamefont
  {Hedegård}},\ }\bibfield  {title} {\enquote {\bibinfo {title} {A simple
  tight-binding model of spin–orbit splitting of sp-derived surface
  states},}\ }\href {\doibase https://doi.org/10.1016/S0039-6028(00)00441-6}
  {\bibfield  {journal} {\bibinfo  {journal} {Surf. Sci.}\ }\textbf {\bibinfo
  {volume} {459}},\ \bibinfo {pages} {49--56} (\bibinfo {year}
  {2000})}\BibitemShut {NoStop}%
\bibitem [{\citenamefont {Khalsa}\ and\ \citenamefont
  {MacDonald}(2012)}]{khalsa2012theory}%
  \BibitemOpen
  \bibfield  {author} {\bibinfo {author} {\bibfnamefont {Guru}\ \bibnamefont
  {Khalsa}}\ and\ \bibinfo {author} {\bibfnamefont {A.~H.}\ \bibnamefont
  {MacDonald}},\ }\bibfield  {title} {\enquote {\bibinfo {title} {Theory of the
  {SrTiO}$_3$ surface state two-dimensional electron gas},}\ }\href {\doibase
  10.1103/PhysRevB.86.125121} {\bibfield  {journal} {\bibinfo  {journal} {Phys.
  Rev. B}\ }\textbf {\bibinfo {volume} {86}},\ \bibinfo {pages} {125121}
  (\bibinfo {year} {2012})}\BibitemShut {NoStop}%
\bibitem [{\citenamefont {Agterberg}\ \emph {et~al.}(2017)\citenamefont
  {Agterberg}, \citenamefont {Brydon},\ and\ \citenamefont
  {Timm}}]{agterberg2017bogoliubov}%
  \BibitemOpen
  \bibfield  {author} {\bibinfo {author} {\bibfnamefont {D.~F.}\ \bibnamefont
  {Agterberg}}, \bibinfo {author} {\bibfnamefont {P.~M.~R.}\ \bibnamefont
  {Brydon}}, \ and\ \bibinfo {author} {\bibfnamefont {C.}~\bibnamefont
  {Timm}},\ }\bibfield  {title} {\enquote {\bibinfo {title} {Bogoliubov {Fermi}
  surfaces in superconductors with broken time-reversal symmetry},}\ }\href
  {\doibase 10.1103/PhysRevLett.118.127001} {\bibfield  {journal} {\bibinfo
  {journal} {Phys. Rev. Lett.}\ }\textbf {\bibinfo {volume} {118}},\ \bibinfo
  {pages} {127001} (\bibinfo {year} {2017})}\BibitemShut {NoStop}%
\bibitem [{\citenamefont {Venderbos}\ \emph {et~al.}(2018)\citenamefont
  {Venderbos}, \citenamefont {Savary}, \citenamefont {Ruhman}, \citenamefont
  {Lee},\ and\ \citenamefont {Fu}}]{venderbos2018pairing}%
  \BibitemOpen
  \bibfield  {author} {\bibinfo {author} {\bibfnamefont {J{\"o}rn~WF}\
  \bibnamefont {Venderbos}}, \bibinfo {author} {\bibfnamefont {Lucile}\
  \bibnamefont {Savary}}, \bibinfo {author} {\bibfnamefont {Jonathan}\
  \bibnamefont {Ruhman}}, \bibinfo {author} {\bibfnamefont {Patrick~A}\
  \bibnamefont {Lee}}, \ and\ \bibinfo {author} {\bibfnamefont {Liang}\
  \bibnamefont {Fu}},\ }\bibfield  {title} {\enquote {\bibinfo {title} {Pairing
  states of spin-$\frac{3}{2}$ fermions: Symmetry-enforced topological gap
  functions},}\ }\href {\doibase 10.1103/PhysRevX.8.011029} {\bibfield
  {journal} {\bibinfo  {journal} {Phys. Rev. X}\ }\textbf {\bibinfo {volume}
  {8}},\ \bibinfo {pages} {011029} (\bibinfo {year} {2018})}\BibitemShut
  {NoStop}%
\bibitem [{\citenamefont {Kozii}\ and\ \citenamefont {Fu}(2015)}]{KoziiFu2015}%
  \BibitemOpen
  \bibfield  {author} {\bibinfo {author} {\bibfnamefont {Vladyslav}\
  \bibnamefont {Kozii}}\ and\ \bibinfo {author} {\bibfnamefont {Liang}\
  \bibnamefont {Fu}},\ }\bibfield  {title} {\enquote {\bibinfo {title}
  {Odd-parity superconductivity in the vicinity of inversion symmetry breaking
  in spin-orbit-coupled systems},}\ }\href {\doibase
  10.1103/PhysRevLett.115.207002} {\bibfield  {journal} {\bibinfo  {journal}
  {Phys. Rev. Lett.}\ }\textbf {\bibinfo {volume} {115}},\ \bibinfo {pages}
  {207002} (\bibinfo {year} {2015})}\BibitemShut {NoStop}%
\bibitem [{\citenamefont {Ruhman}\ and\ \citenamefont
  {Lee}(2016)}]{ruhman2016superconductivity}%
  \BibitemOpen
  \bibfield  {author} {\bibinfo {author} {\bibfnamefont {Jonathan}\
  \bibnamefont {Ruhman}}\ and\ \bibinfo {author} {\bibfnamefont {Patrick~A}\
  \bibnamefont {Lee}},\ }\bibfield  {title} {\enquote {\bibinfo {title}
  {Superconductivity at very low density: The case of strontium titanate},}\
  }\href {\doibase 10.1103/PhysRevB.94.224515} {\bibfield  {journal} {\bibinfo
  {journal} {Phys. Rev. B}\ }\textbf {\bibinfo {volume} {94}},\ \bibinfo
  {pages} {224515} (\bibinfo {year} {2016})}\BibitemShut {NoStop}%
\bibitem [{\citenamefont {Gastiasoro}\ \emph
  {et~al.}(2020{\natexlab{b}})\citenamefont {Gastiasoro}, \citenamefont
  {Trevisan},\ and\ \citenamefont {Fernandes}}]{Gastiasoro2020}%
  \BibitemOpen
  \bibfield  {author} {\bibinfo {author} {\bibfnamefont {Maria~N.}\
  \bibnamefont {Gastiasoro}}, \bibinfo {author} {\bibfnamefont {Tha\'{\i}s~V.}\
  \bibnamefont {Trevisan}}, \ and\ \bibinfo {author} {\bibfnamefont
  {Rafael~M.}\ \bibnamefont {Fernandes}},\ }\bibfield  {title} {\enquote
  {\bibinfo {title} {Anisotropic superconductivity mediated by ferroelectric
  fluctuations in cubic systems with spin-orbit coupling},}\ }\href {\doibase
  10.1103/PhysRevB.101.174501} {\bibfield  {journal} {\bibinfo  {journal}
  {Phys. Rev. B}\ }\textbf {\bibinfo {volume} {101}},\ \bibinfo {pages}
  {174501} (\bibinfo {year} {2020}{\natexlab{b}})}\BibitemShut {NoStop}%
\bibitem [{\citenamefont {Kumar}\ \emph {et~al.}(2022)\citenamefont {Kumar},
  \citenamefont {Chandra},\ and\ \citenamefont {Volkov}}]{kumar2021spinphonon}%
  \BibitemOpen
  \bibfield  {author} {\bibinfo {author} {\bibfnamefont {Abhishek}\
  \bibnamefont {Kumar}}, \bibinfo {author} {\bibfnamefont {Premala}\
  \bibnamefont {Chandra}}, \ and\ \bibinfo {author} {\bibfnamefont {Pavel~A.}\
  \bibnamefont {Volkov}},\ }\bibfield  {title} {\enquote {\bibinfo {title}
  {Spin-phonon resonances in nearly polar metals with spin-orbit coupling},}\
  }\href {\doibase 10.1103/PhysRevB.105.125142} {\bibfield  {journal} {\bibinfo
   {journal} {Phys. Rev. B}\ }\textbf {\bibinfo {volume} {105}},\ \bibinfo
  {pages} {125142} (\bibinfo {year} {2022})}\BibitemShut {NoStop}%
\bibitem [{\citenamefont {Gastiasoro}\ \emph {et~al.}(2021)\citenamefont
  {Gastiasoro}, \citenamefont {Temperini}, \citenamefont {Barone},\ and\
  \citenamefont {Lorenzana}}]{Gastiasoro2021}%
  \BibitemOpen
  \bibfield  {author} {\bibinfo {author} {\bibfnamefont {Maria~N.}\
  \bibnamefont {Gastiasoro}}, \bibinfo {author} {\bibfnamefont
  {Maria~Eleonora}\ \bibnamefont {Temperini}}, \bibinfo {author} {\bibfnamefont
  {Paolo}\ \bibnamefont {Barone}}, \ and\ \bibinfo {author} {\bibfnamefont
  {Jose}\ \bibnamefont {Lorenzana}},\ }\bibfield  {title} {\enquote {\bibinfo
  {title} {Theory of {Rashba} coupling mediated superconductivity in incipient
  ferroelectrics},}\ }\href {https://arxiv.org/abs/2109.13207} {\bibfield
  {journal} {\bibinfo  {journal} {arXiv:2109.13207}\ } (\bibinfo {year}
  {2021})}\BibitemShut {NoStop}%
\bibitem [{Note1()}]{Note1}%
  \BibitemOpen
  \bibinfo {note} {This is justified in the regime $\protect \frac {g \mu _B
  B}{\mu } \ll 1$ (where $\mu $ is the chemical potential, $\mu _B$ is the Bohr
  magneton and $g$ is the Lande' g-factor).}\BibitemShut {Stop}%
\bibitem [{Note2()}]{Note2}%
  \BibitemOpen
  \bibinfo {note} {In general, inversion breaking leads to a state of mixed
  singlet-triplet superconducting states~\cite {Gorkov2001}.}\BibitemShut
  {Stop}%
\bibitem [{\citenamefont {Soluyanov}\ \emph {et~al.}(2015)\citenamefont
  {Soluyanov}, \citenamefont {Gresch}, \citenamefont {Wang}, \citenamefont
  {Wu}, \citenamefont {Troyer}, \citenamefont {Dai},\ and\ \citenamefont
  {Bernevig}}]{Soluyanov2015}%
  \BibitemOpen
  \bibfield  {author} {\bibinfo {author} {\bibfnamefont {Alexey~A.}\
  \bibnamefont {Soluyanov}}, \bibinfo {author} {\bibfnamefont {Dominik}\
  \bibnamefont {Gresch}}, \bibinfo {author} {\bibfnamefont {Zhijun}\
  \bibnamefont {Wang}}, \bibinfo {author} {\bibfnamefont {Quansheng}\
  \bibnamefont {Wu}}, \bibinfo {author} {\bibfnamefont {Matthias}\ \bibnamefont
  {Troyer}}, \bibinfo {author} {\bibfnamefont {Xi}~\bibnamefont {Dai}}, \ and\
  \bibinfo {author} {\bibfnamefont {B.~Andrei}\ \bibnamefont {Bernevig}},\
  }\bibfield  {title} {\enquote {\bibinfo {title} {Type-{II} {Weyl}
  semimetals},}\ }\href {\doibase 10.1038/nature15768} {\bibfield  {journal}
  {\bibinfo  {journal} {Nature}\ }\textbf {\bibinfo {volume} {527}},\ \bibinfo
  {pages} {495--498} (\bibinfo {year} {2015})}\BibitemShut {NoStop}%
\bibitem [{\citenamefont {Volovik}(2018)}]{volovik2018}%
  \BibitemOpen
  \bibfield  {author} {\bibinfo {author} {\bibfnamefont {G.~E.}\ \bibnamefont
  {Volovik}},\ }\bibfield  {title} {\enquote {\bibinfo {title} {Exotic
  {Lifshitz} transitions in topological materials},}\ }\href {\doibase
  10.3367/ufne.2017.01.038218} {\bibfield  {journal} {\bibinfo  {journal}
  {Phys.-Usp.}\ }\textbf {\bibinfo {volume} {61}},\ \bibinfo {pages} {89--98}
  (\bibinfo {year} {2018})}\BibitemShut {NoStop}%
\bibitem [{\citenamefont {Yuan}\ and\ \citenamefont {Fu}(2018)}]{Yuan2018}%
  \BibitemOpen
  \bibfield  {author} {\bibinfo {author} {\bibfnamefont {Noah F.~Q.}\
  \bibnamefont {Yuan}}\ and\ \bibinfo {author} {\bibfnamefont {Liang}\
  \bibnamefont {Fu}},\ }\bibfield  {title} {\enquote {\bibinfo {title}
  {Zeeman-induced gapless superconductivity with a partial {Fermi} surface},}\
  }\href {\doibase 10.1103/PhysRevB.97.115139} {\bibfield  {journal} {\bibinfo
  {journal} {Phys. Rev. B}\ }\textbf {\bibinfo {volume} {97}},\ \bibinfo
  {pages} {115139} (\bibinfo {year} {2018})}\BibitemShut {NoStop}%
\bibitem [{\citenamefont {Cowley}(1964)}]{cowley1964lattice}%
  \BibitemOpen
  \bibfield  {author} {\bibinfo {author} {\bibfnamefont {RA}~\bibnamefont
  {Cowley}},\ }\bibfield  {title} {\enquote {\bibinfo {title} {Lattice dynamics
  and phase transitions of strontium titanate},}\ }\href {\doibase
  10.1103/PhysRev.134.A981} {\bibfield  {journal} {\bibinfo  {journal} {Phys.
  Rev.}\ }\textbf {\bibinfo {volume} {134}},\ \bibinfo {pages} {A981} (\bibinfo
  {year} {1964})}\BibitemShut {NoStop}%
\bibitem [{\citenamefont {Bednorz}\ and\ \citenamefont
  {M\"uller}(1984)}]{Bednorz1984}%
  \BibitemOpen
  \bibfield  {author} {\bibinfo {author} {\bibfnamefont {J.~G.}\ \bibnamefont
  {Bednorz}}\ and\ \bibinfo {author} {\bibfnamefont {K.~A.}\ \bibnamefont
  {M\"uller}},\ }\bibfield  {title} {\enquote {\bibinfo {title}
  {{${\mathrm{Sr}}_{1\ensuremath{-}x}{\mathrm{Ca}}_{x}\mathrm{Ti}{\mathrm{O}}_{3}$}:
  An $\mathrm{XY}$ quantum ferroelectric with transition to randomness},}\
  }\href {\doibase 10.1103/PhysRevLett.52.2289} {\bibfield  {journal} {\bibinfo
   {journal} {Phys. Rev. Lett.}\ }\textbf {\bibinfo {volume} {52}},\ \bibinfo
  {pages} {2289--2292} (\bibinfo {year} {1984})}\BibitemShut {NoStop}%
\bibitem [{\citenamefont {Kleemann}\ \emph {et~al.}(1997)\citenamefont
  {Kleemann}, \citenamefont {Albertini}, \citenamefont {Kuss},\ and\
  \citenamefont {Lindner}}]{Kleemann1997}%
  \BibitemOpen
  \bibfield  {author} {\bibinfo {author} {\bibfnamefont {W.}~\bibnamefont
  {Kleemann}}, \bibinfo {author} {\bibfnamefont {A.}~\bibnamefont {Albertini}},
  \bibinfo {author} {\bibfnamefont {M.}~\bibnamefont {Kuss}}, \ and\ \bibinfo
  {author} {\bibfnamefont {R.}~\bibnamefont {Lindner}},\ }\bibfield  {title}
  {\enquote {\bibinfo {title} {Optical detection of symmetry breaking on a
  nanoscale in {SrTiO}$_3$:{Ca}},}\ }\href {\doibase 10.1080/00150199708012832}
  {\bibfield  {journal} {\bibinfo  {journal} {Ferroelectrics}\ }\textbf
  {\bibinfo {volume} {203}},\ \bibinfo {pages} {57--74} (\bibinfo {year}
  {1997})}\BibitemShut {NoStop}%
\bibitem [{\citenamefont {Kalisky}\ \emph {et~al.}(2013)\citenamefont
  {Kalisky}, \citenamefont {Spanton}, \citenamefont {Noad}, \citenamefont
  {Kirtley}, \citenamefont {Nowack}, \citenamefont {Bell}, \citenamefont
  {Sato}, \citenamefont {Hosoda}, \citenamefont {Xie}, \citenamefont {Hikita}
  \emph {et~al.}}]{kalisky2013locally}%
  \BibitemOpen
  \bibfield  {author} {\bibinfo {author} {\bibfnamefont {Beena}\ \bibnamefont
  {Kalisky}}, \bibinfo {author} {\bibfnamefont {Eric~M}\ \bibnamefont
  {Spanton}}, \bibinfo {author} {\bibfnamefont {Hilary}\ \bibnamefont {Noad}},
  \bibinfo {author} {\bibfnamefont {John~R}\ \bibnamefont {Kirtley}}, \bibinfo
  {author} {\bibfnamefont {Katja~C}\ \bibnamefont {Nowack}}, \bibinfo {author}
  {\bibfnamefont {Christopher}\ \bibnamefont {Bell}}, \bibinfo {author}
  {\bibfnamefont {Hiroki~K}\ \bibnamefont {Sato}}, \bibinfo {author}
  {\bibfnamefont {Masayuki}\ \bibnamefont {Hosoda}}, \bibinfo {author}
  {\bibfnamefont {Yanwu}\ \bibnamefont {Xie}}, \bibinfo {author} {\bibfnamefont
  {Yasuyuki}\ \bibnamefont {Hikita}},  \emph {et~al.},\ }\bibfield  {title}
  {\enquote {\bibinfo {title} {Locally enhanced conductivity due to the
  tetragonal domain structure in {LaAlO}$_3$/{SrTiO}$_3$ heterointerfaces},}\
  }\href {https://doi.org/10.1038/nmat3753} {\bibfield  {journal} {\bibinfo
  {journal} {Nat. Mater.}\ }\textbf {\bibinfo {volume} {12}},\ \bibinfo {pages}
  {1091--1095} (\bibinfo {year} {2013})}\BibitemShut {NoStop}%
\bibitem [{\citenamefont {Honig}\ \emph {et~al.}(2013)\citenamefont {Honig},
  \citenamefont {Sulpizio}, \citenamefont {Drori}, \citenamefont {Joshua},
  \citenamefont {Zeldov},\ and\ \citenamefont {Ilani}}]{honig2013local}%
  \BibitemOpen
  \bibfield  {author} {\bibinfo {author} {\bibfnamefont {Maayan}\ \bibnamefont
  {Honig}}, \bibinfo {author} {\bibfnamefont {Joseph~A}\ \bibnamefont
  {Sulpizio}}, \bibinfo {author} {\bibfnamefont {Jonathan}\ \bibnamefont
  {Drori}}, \bibinfo {author} {\bibfnamefont {Arjun}\ \bibnamefont {Joshua}},
  \bibinfo {author} {\bibfnamefont {Eli}\ \bibnamefont {Zeldov}}, \ and\
  \bibinfo {author} {\bibfnamefont {Shahal}\ \bibnamefont {Ilani}},\ }\bibfield
   {title} {\enquote {\bibinfo {title} {Local electrostatic imaging of striped
  domain order in {LaAlO}$_3$/{SrTiO}$_3$},}\ }\href
  {https://doi.org/10.1038/nmat3810} {\bibfield  {journal} {\bibinfo  {journal}
  {Nat. Mater.}\ }\textbf {\bibinfo {volume} {12}},\ \bibinfo {pages}
  {1112--1118} (\bibinfo {year} {2013})}\BibitemShut {NoStop}%
\bibitem [{\citenamefont {Hellberg}(2019)}]{Hellberg2019}%
  \BibitemOpen
  \bibfield  {author} {\bibinfo {author} {\bibfnamefont {C.~Stephen}\
  \bibnamefont {Hellberg}},\ }\bibfield  {title} {\enquote {\bibinfo {title}
  {Domain walls in strontium titanate},}\ }\href {\doibase
  10.1088/1742-6596/1252/1/012006} {\bibfield  {journal} {\bibinfo  {journal}
  {J. Phys. Conf. Ser.}\ }\textbf {\bibinfo {volume} {1252}},\ \bibinfo {pages}
  {012006} (\bibinfo {year} {2019})}\BibitemShut {NoStop}%
\bibitem [{\citenamefont {Dwivedi}(2018)}]{Dwivedi2018}%
  \BibitemOpen
  \bibfield  {author} {\bibinfo {author} {\bibfnamefont {Vatsal}\ \bibnamefont
  {Dwivedi}},\ }\bibfield  {title} {\enquote {\bibinfo {title} {Fermi arc
  reconstruction at junctions between {Weyl} semimetals},}\ }\href {\doibase
  10.1103/PhysRevB.97.064201} {\bibfield  {journal} {\bibinfo  {journal} {Phys.
  Rev. B}\ }\textbf {\bibinfo {volume} {97}},\ \bibinfo {pages} {064201}
  (\bibinfo {year} {2018})}\BibitemShut {NoStop}%
\bibitem [{\citenamefont {Murthy}\ \emph {et~al.}(2020)\citenamefont {Murthy},
  \citenamefont {Fertig},\ and\ \citenamefont {Shimshoni}}]{Murthy2020}%
  \BibitemOpen
  \bibfield  {author} {\bibinfo {author} {\bibfnamefont {Ganpathy}\
  \bibnamefont {Murthy}}, \bibinfo {author} {\bibfnamefont {H.~A.}\
  \bibnamefont {Fertig}}, \ and\ \bibinfo {author} {\bibfnamefont {Efrat}\
  \bibnamefont {Shimshoni}},\ }\bibfield  {title} {\enquote {\bibinfo {title}
  {Surface states and arcless angles in twisted {Weyl} semimetals},}\ }\href
  {\doibase 10.1103/PhysRevResearch.2.013367} {\bibfield  {journal} {\bibinfo
  {journal} {Phys. Rev. Res.}\ }\textbf {\bibinfo {volume} {2}},\ \bibinfo
  {pages} {013367} (\bibinfo {year} {2020})}\BibitemShut {NoStop}%
\bibitem [{\citenamefont {Ilan}\ \emph {et~al.}(2020)\citenamefont {Ilan},
  \citenamefont {Grushin},\ and\ \citenamefont {Pikulin}}]{ilan2020pseudo}%
  \BibitemOpen
  \bibfield  {author} {\bibinfo {author} {\bibfnamefont {Roni}\ \bibnamefont
  {Ilan}}, \bibinfo {author} {\bibfnamefont {Adolfo~G}\ \bibnamefont
  {Grushin}}, \ and\ \bibinfo {author} {\bibfnamefont {Dmitry~I}\ \bibnamefont
  {Pikulin}},\ }\bibfield  {title} {\enquote {\bibinfo {title}
  {Pseudo-electromagnetic fields in {3D} topological semimetals},}\ }\href
  {https://doi.org/10.1038/s42254-019-0121-8} {\bibfield  {journal} {\bibinfo
  {journal} {Nat. Rev. Phys.}\ }\textbf {\bibinfo {volume} {2}},\ \bibinfo
  {pages} {29--41} (\bibinfo {year} {2020})}\BibitemShut {NoStop}%
\bibitem [{\citenamefont {Blonder}\ \emph {et~al.}(1982)\citenamefont
  {Blonder}, \citenamefont {Tinkham},\ and\ \citenamefont
  {Klapwijk}}]{blonder1982transition}%
  \BibitemOpen
  \bibfield  {author} {\bibinfo {author} {\bibfnamefont {G.~E.}\ \bibnamefont
  {Blonder}}, \bibinfo {author} {\bibfnamefont {M.}~\bibnamefont {Tinkham}}, \
  and\ \bibinfo {author} {\bibfnamefont {T.~M.}\ \bibnamefont {Klapwijk}},\
  }\bibfield  {title} {\enquote {\bibinfo {title} {Transition from metallic to
  tunneling regimes in superconducting microconstrictions: Excess current,
  charge imbalance, and supercurrent conversion},}\ }\href {\doibase
  10.1103/PhysRevB.25.4515} {\bibfield  {journal} {\bibinfo  {journal} {Phys.
  Rev. B}\ }\textbf {\bibinfo {volume} {25}},\ \bibinfo {pages} {4515--4532}
  (\bibinfo {year} {1982})}\BibitemShut {NoStop}%
\bibitem [{\citenamefont {Gygi}\ and\ \citenamefont
  {Schl{\"u}ter}(1991)}]{gygi1991self}%
  \BibitemOpen
  \bibfield  {author} {\bibinfo {author} {\bibfnamefont {Fran{\c{c}}ois}\
  \bibnamefont {Gygi}}\ and\ \bibinfo {author} {\bibfnamefont {Michael}\
  \bibnamefont {Schl{\"u}ter}},\ }\bibfield  {title} {\enquote {\bibinfo
  {title} {Self-consistent electronic structure of a vortex line in a type-{II}
  superconductor},}\ }\href {\doibase 10.1103/PhysRevB.43.7609} {\bibfield
  {journal} {\bibinfo  {journal} {Phys. Rev. B}\ }\textbf {\bibinfo {volume}
  {43}},\ \bibinfo {pages} {7609} (\bibinfo {year} {1991})}\BibitemShut
  {NoStop}%
\bibitem [{Note3()}]{Note3}%
  \BibitemOpen
  \bibinfo {note} {We calculate values $\nu _{p_z}(E,r)$ on a relatively sparse
  grid of points in $(E,r)$-space, which does not include the point
  corresponding to the highest peak of $\nu _{p_z}(E,r)$, and then interpolate
  between the points. This results in that none of the peaks in the plot reach
  the value of 1.}\BibitemShut {Stop}%
\bibitem [{\citenamefont {Abrikosov}(1957)}]{abrikosov1957magnetic}%
  \BibitemOpen
  \bibfield  {author} {\bibinfo {author} {\bibfnamefont {A.A.}\ \bibnamefont
  {Abrikosov}},\ }\bibfield  {title} {\enquote {\bibinfo {title} {The magnetic
  properties of superconducting alloys},}\ }\href {\doibase
  https://doi.org/10.1016/0022-3697(57)90083-5} {\bibfield  {journal} {\bibinfo
   {journal} {J. Phys. Chem. Solids}\ }\textbf {\bibinfo {volume} {2}},\
  \bibinfo {pages} {199--208} (\bibinfo {year} {1957})}\BibitemShut {NoStop}%
\bibitem [{Note4()}]{Note4}%
  \BibitemOpen
  \bibinfo {note} {We restored $\hbar $ here}\BibitemShut {NoStop}%
\bibitem [{\citenamefont {Swartz}\ \emph {et~al.}(2018)\citenamefont {Swartz},
  \citenamefont {Inoue}, \citenamefont {Merz}, \citenamefont {Hikita},
  \citenamefont {Raghu}, \citenamefont {Devereaux}, \citenamefont {Johnston},\
  and\ \citenamefont {Hwang}}]{Swartz2018}%
  \BibitemOpen
  \bibfield  {author} {\bibinfo {author} {\bibfnamefont {Adrian~G.}\
  \bibnamefont {Swartz}}, \bibinfo {author} {\bibfnamefont {Hisashi}\
  \bibnamefont {Inoue}}, \bibinfo {author} {\bibfnamefont {Tyler~A.}\
  \bibnamefont {Merz}}, \bibinfo {author} {\bibfnamefont {Yasuyuki}\
  \bibnamefont {Hikita}}, \bibinfo {author} {\bibfnamefont {Srinivas}\
  \bibnamefont {Raghu}}, \bibinfo {author} {\bibfnamefont {Thomas~P.}\
  \bibnamefont {Devereaux}}, \bibinfo {author} {\bibfnamefont {Steven}\
  \bibnamefont {Johnston}}, \ and\ \bibinfo {author} {\bibfnamefont
  {Harold~Y.}\ \bibnamefont {Hwang}},\ }\bibfield  {title} {\enquote {\bibinfo
  {title} {Polaronic behavior in a weak-coupling superconductor},}\ }\href
  {\doibase 10.1073/pnas.1713916115} {\bibfield  {journal} {\bibinfo  {journal}
  {PNAS}\ }\textbf {\bibinfo {volume} {115}},\ \bibinfo {pages} {1475--1480}
  (\bibinfo {year} {2018})}\BibitemShut {NoStop}%
\bibitem [{\citenamefont {Lin}\ \emph {et~al.}(2014)\citenamefont {Lin},
  \citenamefont {Bridoux}, \citenamefont {Gourgout}, \citenamefont {Seyfarth},
  \citenamefont {Kr\"amer}, \citenamefont {Nardone}, \citenamefont {Fauqu\'e},\
  and\ \citenamefont {Behnia}}]{Lin2014}%
  \BibitemOpen
  \bibfield  {author} {\bibinfo {author} {\bibfnamefont {Xiao}\ \bibnamefont
  {Lin}}, \bibinfo {author} {\bibfnamefont {German}\ \bibnamefont {Bridoux}},
  \bibinfo {author} {\bibfnamefont {Adrien}\ \bibnamefont {Gourgout}}, \bibinfo
  {author} {\bibfnamefont {Gabriel}\ \bibnamefont {Seyfarth}}, \bibinfo
  {author} {\bibfnamefont {Steffen}\ \bibnamefont {Kr\"amer}}, \bibinfo
  {author} {\bibfnamefont {Marc}\ \bibnamefont {Nardone}}, \bibinfo {author}
  {\bibfnamefont {Beno\^{\i}t}\ \bibnamefont {Fauqu\'e}}, \ and\ \bibinfo
  {author} {\bibfnamefont {Kamran}\ \bibnamefont {Behnia}},\ }\bibfield
  {title} {\enquote {\bibinfo {title} {Critical doping for the onset of a
  two-band superconducting ground state in
  ${\mathrm{srtio}}_{3\ensuremath{-}\ensuremath{\delta}}$},}\ }\href {\doibase
  10.1103/PhysRevLett.112.207002} {\bibfield  {journal} {\bibinfo  {journal}
  {Phys. Rev. Lett.}\ }\textbf {\bibinfo {volume} {112}},\ \bibinfo {pages}
  {207002} (\bibinfo {year} {2014})}\BibitemShut {NoStop}%
\bibitem [{\citenamefont {Gor’kov}(1959)}]{gor1959microscopic}%
  \BibitemOpen
  \bibfield  {author} {\bibinfo {author} {\bibfnamefont {Lev~P.}\ \bibnamefont
  {Gor’kov}},\ }\bibfield  {title} {\enquote {\bibinfo {title} {Microscopic
  derivation of the {Ginzburg-Landau} equations in the theory of
  superconductivity},}\ }\href
  {http://www.jetp.ras.ru/cgi-bin/dn/e_009_06_1364.pdf} {\bibfield  {journal}
  {\bibinfo  {journal} {Sov. Phys. JETP}\ }\textbf {\bibinfo {volume} {9}},\
  \bibinfo {pages} {1364--1367} (\bibinfo {year} {1959})}\BibitemShut {NoStop}%
\bibitem [{\citenamefont {Schumann}\ \emph {et~al.}(2020)\citenamefont
  {Schumann}, \citenamefont {Galletti}, \citenamefont {Jeong}, \citenamefont
  {Ahadi}, \citenamefont {Strickland}, \citenamefont {Salmani-Rezaie},\ and\
  \citenamefont {Stemmer}}]{schumann2020possible}%
  \BibitemOpen
  \bibfield  {author} {\bibinfo {author} {\bibfnamefont {Timo}\ \bibnamefont
  {Schumann}}, \bibinfo {author} {\bibfnamefont {Luca}\ \bibnamefont
  {Galletti}}, \bibinfo {author} {\bibfnamefont {Hanbyeol}\ \bibnamefont
  {Jeong}}, \bibinfo {author} {\bibfnamefont {Kaveh}\ \bibnamefont {Ahadi}},
  \bibinfo {author} {\bibfnamefont {William~M.}\ \bibnamefont {Strickland}},
  \bibinfo {author} {\bibfnamefont {Salva}\ \bibnamefont {Salmani-Rezaie}}, \
  and\ \bibinfo {author} {\bibfnamefont {Susanne}\ \bibnamefont {Stemmer}},\
  }\bibfield  {title} {\enquote {\bibinfo {title} {Possible signatures of
  mixed-parity superconductivity in doped polar
  {$\mathrm{SrTi}{\mathrm{O}}_{3}$} films},}\ }\href {\doibase
  10.1103/PhysRevB.101.100503} {\bibfield  {journal} {\bibinfo  {journal}
  {Phys. Rev. B}\ }\textbf {\bibinfo {volume} {101}},\ \bibinfo {pages}
  {100503(R)} (\bibinfo {year} {2020})}\BibitemShut {NoStop}%
\bibitem [{Note5()}]{Note5}%
  \BibitemOpen
  \bibinfo {note} {For the curve $H_{c2}(T)$ we use approximate Gor'kov's
  formula~\cite {Gorkov1960} $H_{c2}(T)/H_{c}(T)\approx \chi (1.77 - 0.43
  (T/T_c)^2 + 0.07(T/T_c)^4)$, and we approximate $H_c(T)/H_c(0)\approx
  1-(T/T_c)^2$. Taking $H_{c2}(0) \simeq \protect \frac {\phi _0}{2 \pi \xi
  _0^2}$, where $\phi _0$ is the flux quantum, we find $\protect \frac {\xi
  (T)}{\xi _0} = \protect \sqrt {\protect \frac {H_{c2}(0)}{H_{c2}(T)}}$. Also,
  we have neglected the dependence of the parameter $K$ on
  temperature.}\BibitemShut {Stop}%
\bibitem [{\citenamefont {Chalker}\ and\ \citenamefont
  {Coddington}(1988)}]{chalker1988percolation}%
  \BibitemOpen
  \bibfield  {author} {\bibinfo {author} {\bibfnamefont {JT}~\bibnamefont
  {Chalker}}\ and\ \bibinfo {author} {\bibfnamefont {PD}~\bibnamefont
  {Coddington}},\ }\bibfield  {title} {\enquote {\bibinfo {title} {Percolation,
  quantum tunnelling and the integer {Hall} effect},}\ }\href
  {https://iopscience.iop.org/article/10.1088/0022-3719/21/14/008} {\bibfield
  {journal} {\bibinfo  {journal} {J. Phys. C: Solid State Phys.}\ }\textbf
  {\bibinfo {volume} {21}},\ \bibinfo {pages} {2665} (\bibinfo {year}
  {1988})}\BibitemShut {NoStop}%
\bibitem [{\citenamefont {Caroli}\ \emph {et~al.}(1964)\citenamefont {Caroli},
  \citenamefont {{De Gennes}},\ and\ \citenamefont {Matricon}}]{Caroli1964}%
  \BibitemOpen
  \bibfield  {author} {\bibinfo {author} {\bibfnamefont {C.}~\bibnamefont
  {Caroli}}, \bibinfo {author} {\bibfnamefont {P.G.}\ \bibnamefont {{De
  Gennes}}}, \ and\ \bibinfo {author} {\bibfnamefont {J.}~\bibnamefont
  {Matricon}},\ }\bibfield  {title} {\enquote {\bibinfo {title} {Bound fermion
  states on a vortex line in a type {II} superconductor},}\ }\href {\doibase
  https://doi.org/10.1016/0031-9163(64)90375-0} {\bibfield  {journal} {\bibinfo
   {journal} {Phys. Lett.}\ }\textbf {\bibinfo {volume} {9}},\ \bibinfo {pages}
  {307--309} (\bibinfo {year} {1964})}\BibitemShut {NoStop}%
\bibitem [{Note6()}]{Note6}%
  \BibitemOpen
  \bibinfo {note} {This is true only close to $H_{c2}$.}\BibitemShut {Stop}%
\bibitem [{\citenamefont {Gor'kov}\ and\ \citenamefont
  {Rashba}(2001)}]{Gorkov2001}%
  \BibitemOpen
  \bibfield  {author} {\bibinfo {author} {\bibfnamefont {Lev~P.}\ \bibnamefont
  {Gor'kov}}\ and\ \bibinfo {author} {\bibfnamefont {Emmanuel~I.}\ \bibnamefont
  {Rashba}},\ }\bibfield  {title} {\enquote {\bibinfo {title} {Superconducting
  {2D} system with lifted spin degeneracy: Mixed singlet-triplet state},}\
  }\href {\doibase 10.1103/PhysRevLett.87.037004} {\bibfield  {journal}
  {\bibinfo  {journal} {Phys. Rev. Lett.}\ }\textbf {\bibinfo {volume} {87}},\
  \bibinfo {pages} {037004} (\bibinfo {year} {2001})}\BibitemShut {NoStop}%
\bibitem [{\citenamefont {Gor'kov}(1960)}]{Gorkov1960}%
  \BibitemOpen
  \bibfield  {author} {\bibinfo {author} {\bibfnamefont {Lev~P.}\ \bibnamefont
  {Gor'kov}},\ }\bibfield  {title} {\enquote {\bibinfo {title} {The critical
  supercooling field in superconductivity theory},}\ }\href
  {http://www.jetp.ras.ru/cgi-bin/dn/e_010_03_0593.pdf} {\bibfield  {journal}
  {\bibinfo  {journal} {Sov. Phys. JETP}\ }\textbf {\bibinfo {volume} {10}},\
  \bibinfo {pages} {593--599} (\bibinfo {year} {1960})}\BibitemShut {NoStop}%
\end{thebibliography}%

\end{document}